June 28, 2015

# Optimal Equity Glidepaths in Retirement

CHRISTOPHER J. ROOK[*]

## ABSTRACT

Dynamic retirement glidepaths evolve over time based on some measure such as the retiree's funded status or current market valuations. Conversely, static glidepaths are fixed at a starting point and selected under the assumption that they will not change. In practice, new static glidepaths may be derived periodically making them more flexible. The optimal static retirement glidepath would be the one that performs better than all others with respect to some metric. When systematic withdrawals are made from a retirement portfolio, glidepaths are often assessed via the probability of ruin (or success). Our goal here is to derive the optimal static glidepath with respect to this metric. It is a result new to the literature and the shape will be of special interest to retirees, financial advisors, retirement researchers, and target-date fund providers.

[*] The author works as a statistical programming consultant and is concluding a degree in the Department of Systems Engineering at Stevens Institute of Technology. The attached Internet Appendix document contains proofs, derivations, and source code from a full C++ application.

The rise of defined contribution (DC) plans has coincided with the fall of defined benefit (DB) plans and, as a result, more retirements will be funded with personal accounts rather than traditional pensions. Coupled with low savings rates, this has led to a *retirement crisis* and mitigation strategies are in high demand by firms, advisors, and retirees. This research contributes to the literature by introducing a solution to the presently unsolved problem of finding the optimal equity glidepath in retirement. We assume the retiree withdraws a constant real amount and that the glidepath is fixed at some starting point, making it static. The model we develop applies to a retirement horizon of any length therefore can be rerun periodically to produce a new solution. This allows annual updating, for example, with a new static glidepath. To derive the optimal glidepath we express the probability of success (ruin) as a function of the equity ratios and maximize (minimize) it using traditional approaches.

This research is organized as follows. We begin Section I with a review of the literature, then in Section II introduce various concepts that will be needed later. In Section III we provide the optimization model in general terms and a specific implementation is given in Section IV. In Section V we extend the model using a random time of final withdrawal and summarize our findings in Section VI. A proofs appendix with full C++ application is included.

## I.  Literature Review

To generate retirement income, financial economists apply lifecycle models that select portfolios to maximize the expected utility of consumption[1]. These models are not restricted to the retirement decumulation phase. Research in this area uses Brownian motion calculus and is partly based on the works of Merton and Samuelson who published separate papers together in *The Review of Economics and Statistics*, August 1969. Samuelson worked on the discrete time case and Merton built the continuous time case by taking the limit as the interval width approached zero. A key finding of this research is that portfolio selection and consumption are independent of each other in certain special cases. Portfolio selection (amount to invest in the risky asset) is constant, based on risk aversion, whereas the consumption decision is based on wealth. More generally, consumption levels are a function of both risk aversion and risky asset volatility. The origins of these methods date back to 1827 when Scotsman Robert Brown published a paper describing the random motion of plant pollen when submersed in liquid. Seventy years later a math student named Louis Bachelier used Brown's methods to model prices on the Paris stock market. A quarter century later Norbert Wiener developed the full mathematical framework for this motion while at MIT (Wiersema (2008)). Merton's continuous-time model assumes risky asset returns are random and follow the Wiener process (Brownian motion) but discusses extensions, while Samuelson's is more general in this regard.

---

[1] The behavioral axioms and proposition that decision makers will choose among alternatives by maximizing the expected utility were developed by Von Neumann and Morgenstern (1947).

Research in this area of financial economics is active with Milevsky and Huang (2010) noting that over 50 such papers have appeared in top finance journals during the prior decade alone. The lifecycle model is often modified or extended in various ways. For example, Young (2004) determined the portfolio that minimizes the probability of lifetime ruin under constant force of mortality. Moore and Young (2006) followed up on this research using a more general mortality distribution. Huang, Milevsky and Salisbury (2012) incorporated a more realistic force of mortality into the utility maximization model and allowed the consumer to change spending as a function of their current health status. More recently, Bayraktar and Young (2015) derived the optimal lifecycle portfolio for a given bequest motive.

Despite their popularity in the academic literature, financial advisors routinely eschew these models for heuristics or rules-of-thumb that are easier to implement and explain. One such heuristic is the safe withdrawal rate (SWR) which suggests that the retiree withdraw a fixed amount from their savings each year, adjusted for inflation. Advisors prefer these heuristics because they more accurately reflect what retirees want, namely a constant and sustainable income stream akin to their prior paycheck (Fan, Murray and Pittman (2013)). Others, however, claim they are misguided and strongly discourage their use. For example, in a paper published in May 2014, Irlam and Tomlinson refer to rules-of-thumb, such as the SWR, as far from optimal suggesting there is no way of knowing when, or if, an optimal solution has been found. Rook (2014) preemptively raised doubts about these claims, developing an optimal decumulation model for the SWR heuristic using a dynamic glidepath strategy.[2] In this research we address the related problem of optimizing the SWR heuristic using a static glidepath strategy.

The glidepath for an SWR heuristic gives the equity ratio at each time point and it is primarily assessed via the probability of ruin (success). Terminal portfolio value or bequest wealth is another metric used to gauge an SWR's effectiveness. A glidepath is said to be *dynamic* when it can change over time and *static* when fixed at a starting point. Rook (2014) derived the dynamic glidepath that minimizes the probability of ruin and demonstrates that static glidepath strategies are suboptimal due to artificial constraints imposed on the retiree. Fan, Murray and Pittman (2013) derived the dynamic glidepath that maximizes bequest wealth subject to a shortfall penalty. Others advocating dynamic approaches include Guyton (2004), Stout and Mitchell (2006), and Frank, Mitchell and Blanchett (2011). Some approaches incorporate spending changes similar to lifecycle models. Despite the benefits of dynamic strategies, static glidepath research is an active field of study by practitioners and target-date fund providers.

Various static glidepath shapes have been studied in an attempt to find the best performing with respect to some metric. Here, "shape" refers to the equity ratio if plotted over time. The shapes attracting most attention are constant, declining, and rising. Bengen (1994) studied the question of how much a retiree could withdraw annually without depleting their portfolio in 30 years. The answer was slightly over 4% using U.S. historical stock/bond returns, leading to what is referred to as the "4% rule". Unlike some dynamic spending models, this rule imposes discipline when markets rise. Bengen (1994) found support for a constant glidepath using between 50% and 75% equities. Cooley, Hubbard, and Walz (1998) drew similar conclusions using constant glidepaths in the *Trinity Study*. Blanchett (2007) directly compared constant glidepaths to various types of declining glidepaths and found it outperforms. Cohen, Gardner, and Fan (2010) agree. Van Harlow (2011) also lends credence to constant glidepaths, but with lower equity ratios, and Estrada (2015) finds international evidence to support constant glidepaths compared to their declining and rising counterparts.

---

[2] A prior version of Rook (2014) can be found on SSRN at: http://ssrn.com/abstract=2420747. (Note that Version 1 of this Working Paper was published on April 6, 2014.)

Declining glidepaths mimic the "age-in-bonds" heuristic which posits that a retiree keep 100 or 120 minus their age in equities, as a percentage. This leads to a declining equity ratio over time. Many target-date (T-D) funds implement declining glidepaths, tacitly endorsing their use, and the industry now holds over $600 billion in assets (Yang and Lutton (2014)). Blanchett (2015) compared declining, rising, v-shaped and constant glidepaths under assumptions different from Blanchett (2007) and found that declining glidepaths are preferable. Estrada (2015) found that the international record favors declining over rising glidepaths in retirement. Fullmer (2014) notes that behavioral issues are a key reason why T-D funds use declining glidepaths, namely that retiree risk aversion increases with age therefore the equity ratio should decrease. Daverman and O'Hara (2015) disagree calling one's retirement day the riskiest of their life and conclude that constant glidepaths are more effective in T-D funds. This raises a practical concern with utility models. Two leading financial firms come to the exact opposite conclusion about retiree risk aversion and how it should impact the T-D fund glidepath.[3] On the topic, Young (2004) noted that maximizing expected utility is subjective and that a more objective metric such as minimizing the probability of financial ruin could be appropriate.

Rising glidepaths have attracted attention recently as being strategies that can reduce the probability of financial ruin in retirement. Pfau and Kitces (2014) demonstrated this using various return assumptions and note that rising glidepaths are actually more conservative on a weighted average basis because the account balance is larger at retirement start and becomes smaller with distributions. Bill Bernstein (2014) agrees and likens the rising glidepath to how a liability-matching portfolio evolves naturally over time. Fullmer (2014) notes that such glidepaths are not new and their benefits have been known, but that properly accounting for risk aversion limits their usefulness because a retiree is more likely to panic and sell as they age. Daverman and O'Hara (2015) lend credence to the rising glidepath from a lifecycle perspective finding that it is the optimal strategy when accounting for steady income streams such as those from Social Security or pensions, and Delorme (2015) arrives at a similar conclusion.

---

[3] Fullmer's main point that behavioral considerations should not be ignored when implementing mathematically optimized strategies bears repeating.

## II. Preliminary Discussion

In this section we present findings that will be required later. For all sections, some proof details will be placed into an appendix document that accompanies this research, as will source code from a full C++ application. The code requires independent user validation.

*A. Definitions and Notation*

We will adopt the same notation as used in Rook (2014). Namely, assume the retiree invests in stocks/bonds, and let α reflect the equity ratio at a given time point t.[4] The total return between times t-1 and t, for t=1, 2, …, $T_D$, using equity ratio α is denoted by $R_{(t,\alpha)}$. The quantity $R_{(t,\alpha)}$ is derived as:

$$R_{(t,\alpha)} = \frac{\text{(Time t Balance} - \text{Time t}-1\text{ Balance)}}{\text{Time t}-1\text{ Balance}} \quad (2.1)$$

It immediately follows that:

$$\text{Time t Balance} = (\text{Time t-1 Balance}) * (1 + R_{(t,\alpha)}) \quad (2.2)$$

Let $E_R$ reflect the expense ratio between times t-1 and t. Accounting for expenses yields:

$$\text{Actual Time t Balance} = (\text{Actual Time t-1 Balance}) * (1 + R_{(t,\alpha)}) * (1 - E_R) \quad (2.3)$$

The relationship between the total return, $R_{(t,\alpha)}$, the inflation rate, $I_t$, and the real return, $r_{(t,\alpha)}$, is:

$$(1 + R_{(t,\alpha)}) = (1 + I_t) * (1 + r_{(t,\alpha)}) \quad (2.4)$$

Solving for the compounding real return, $1 + r_{(t,\alpha)}$, yields:

$$1 + r_{(t,\alpha)} = \frac{(1 + R_{(t,\alpha)})}{(1 + I_t)} \quad (2.5)$$

In real terms, with expenses, the time t account balance is derived as: (2.6)

$$\text{Real Actual Time t Balance} = \frac{\text{Actual Time t Balance}}{\prod_{i=1}^{t}(1 + I_i)} = \frac{(\text{Actual Time t}-1\text{ Balance}) * (1 + R_{(t,\alpha)}) * (1 - E_R)}{\prod_{i=1}^{t-1}(1 + I_i) * (1 + I_t)}$$

$$\rightarrow \text{Real Actual Time t Balance} = (\text{Real Time t-1 Balance}) * (1 + r_{(t,\alpha)}) * (1 - E_R) \quad (2.7)$$

The quantity $(1 + r_{(t,\alpha)}) * (1 - E_R)$ will be denoted by $\hat{r}_{(t,\alpha)}$ and referred to as the inflation/expense-adjusted compounding return between times t-1 and t, using equity ratio α. Note that $\hat{r}_{(t,\alpha)}$ is a random variable (RV) since it is a function of the RV, $r_{(t,\alpha)}$, which is itself random since it is a function of the two RVs, $R_{(t,\alpha)}$ and $I_t$.

---

[4] The equity ratio α can differ by time point but we will avoid having subscripts on subscripts at this stage and only write α as $\alpha_t$ when it becomes necessary to do so.

## B. Mean and Variance of a Stock/Bond Portfolio

Let $r_{(s,t)}$ and $r_{(b,t)}$ be RVs reflecting real stock and bond returns between time points t-1 and t, respectively, with the following means, variances, and covariance:

$$E[r_{(s,t)}] = \mu_{(s,t)} \text{ and } V[r_{(s,t)}] = \sigma^2_{(s,t)} \tag{2.8}$$

$$E[r_{(b,t)}] = \mu_{(b,t)} \text{ and } V[r_{(b,t)}] = \sigma^2_{(b,t)} \tag{2.9}$$

$$Cov(r_{(s,t)}, r_{(b,t)}) = \sigma_{(s,b,t)} \tag{2.10}$$

The real return, $r_{(t,\alpha)}$, of a stock/bond portfolio between times t-1 and t with equity ratio $\alpha$ is:

$$r_{(t,\alpha)} = (\alpha)*r_{(s,t)} + (1-\alpha)*r_{(b,t)}, \tag{2.11}$$

with mean and variance:

$$E[r_{(t,\alpha)}] = (\alpha)\mu_{(s,t)} + (1-\alpha)\mu_{(b,t)} \tag{2.12}$$

$$V[r_{(t,\alpha)}] = (\alpha^2)\sigma^2_{(s,t)} + (1-\alpha)^2\sigma^2_{(b,t)} + 2(\alpha)(1-\alpha)\sigma_{(s,b,t)} \tag{2.13}$$

Finally, the inflation/expense-adjusted compounding return of a stock/bond portfolio between times t-1 and t, using equity ratio $\alpha$ is given by:

$$\hat{r}_{(t,\alpha)} = (1 - E_R)*(1 + (\alpha)r_{(s,t)} + (1-\alpha)r_{(b,t)}), \tag{2.14}$$

with mean and variance, denoted by $m_t(\alpha)$ and $v_t(\alpha)$, respectively:

$$m_t(\alpha) = E[\hat{r}_{(t,\alpha)}] = (1 - E_R)*(1 + (\alpha)\mu_{(s,t)} + (1-\alpha)\mu_{(b,t)}) \tag{2.15}$$

$$v_t(\alpha) = V[\hat{r}_{(t,\alpha)}] = (1 - E_R)^2*[(\alpha^2)\sigma^2_{(s,t)} + (1-\alpha)^2\sigma^2_{(b,t)} + 2(\alpha)(1-\alpha)\sigma_{(s,b,t)}] \tag{2.16}$$

The 1$^{st}$ and 2$^{nd}$ derivatives of these functions with respect to $\alpha$ are:

$$m_t'(\alpha) = (1 - E_R)*(\mu_{(s,t)} - \mu_{(b,t)}) \tag{2.17}$$

$$m_t''(\alpha) = 0 \tag{2.18}$$

$$v_t'(\alpha) = 2(1 - E_R)^2*[(\alpha)\sigma^2_{(s,t)} - (1-\alpha)\sigma^2_{(b,t)} + (1 - 2\alpha)\sigma_{(s,b,t)}] \tag{2.19}$$

$$v_t''(\alpha) = 2(1 - E_R)^2*[\sigma^2_{(s,t)} + \sigma^2_{(b,t)} - 2\sigma_{(s,b,t)}] \tag{2.20}$$

Note that $v_t''(\alpha) > 0$ since the bracketed term [·] in (2.20) is V(Z) where $Z = r_{(s,t)} - r_{(b,t)}$ and the variance of a continuous random variable is always positive. This implies that $v_t(\alpha)$ is a convex function so that its minimum value is the critical point satisfying $v_t'(\alpha) = 0$. Denoting the solution by $MV_t(\alpha)$ yields:

$$MV_t(\alpha) = \frac{\sigma_{(b,t)}^2 - \sigma_{(s,b,t)}}{\sigma_{(s,t)}^2 + \sigma_{(b,t)}^2 - 2\sigma_{(s,b,t)}}. \quad (2.21)$$

Here, $MV_t(\alpha)$ is the minimum variance $\alpha$ (Markowitz (1952)), derived analytically (Sigman (2005)) at time t. The situation is depicted graphically in Figure 1 below. Note that for any $\alpha > MV_t(\alpha)$, $v_t'(\alpha) > 0$. Further, we claim that a retiree will not benefit from any equity ratio $\alpha < MV_t(\alpha)$ since by drawing a horizontal line across the graph in Figure 1 we can build a portfolio with the same variance but higher expected return. This is because $m_t(\alpha)$ is an increasing function of $\alpha$ when $\mu_{(s,t)} > \mu_{(b,t)}$. We will therefore restrict the feasible region during optimization to $MV_t(\alpha) < \alpha \leq 1.00$ and rely on the fact that $v_t'(\alpha)$ is always positive.

### Figure 1
### The Portfolio Variance Function

This figure shows the portfolio variance as a function of the equity ratio $\alpha$ at time point t. The function is convex since $v_t''(\alpha) > 0$ and the minimum variance is achieved at the critical point which satisfies $v_t'(\alpha) = 0$, labeled $MV_t(\alpha)$. If the feasible region is restricted to equity ratios satisfying $MV_t(\alpha) < \alpha \leq 1.00$ then we can assume that $v_t'(\alpha) > 0$. We justify this restriction by noting that for any $\alpha < MV_t(\alpha)$, a portfolio with the same variance but higher expected return can be built since the expected return is an increasing function of $\alpha$. Draw a horizontal line across the graph, then straight down to find the preferred $\alpha$. This result is often used to conclude that diversification can lower risk.

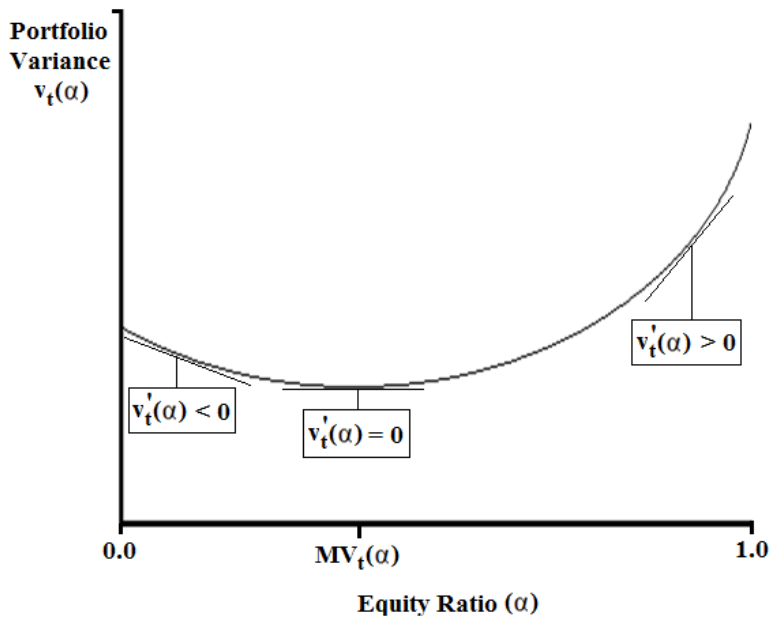

## C. Probability Distribution of Real Returns

We denote the inflation/expense-adjusted return at each time t by $\hat{r}_{(t,\alpha)}$ and assume it is a continuous RV. The multivariate probability density function (PDF) of all inflation/expense-adjusted returns across the retirement horizon will be represented by $f(\cdot)$, namely:

$$f(\hat{r}_{(1,\alpha)}, \hat{r}_{(2,\alpha)}, \ldots, \hat{r}_{(T_D,\alpha)}). \tag{2.22}$$

Note that $f(\cdot)$ is a function of the $T_D$ equity ratios $\vec{\alpha} = (\alpha_1, \alpha_2, \ldots, \alpha_{T_D})$, and $\vec{\alpha}$ represents a glidepath. Under the assumption of market efficiency this PDF can be expressed as:

$$f(\hat{r}_{(1,\alpha)}, \hat{r}_{(2,\alpha)}, \ldots, \hat{r}_{(T_D,\alpha)}) = \prod_{t=1}^{T_D} f^t(\hat{r}_{(t,\alpha)}), \tag{2.23}$$

where $f^t(\cdot)$ is the marginal PDF of $\hat{r}_{(t,\alpha)}$, which is a function of $\alpha_t$. The cumulative distribution function (CDF) of $f^t(\cdot)$ is $F^t(r) = P(\hat{r}_{(t,\alpha)} \leq r)$. The only restriction imposed on $f^t(\cdot)$ is that it be a valid PDF. For example, at times $t=t_1, t_2$, the PDFs $f^{t_1}(\cdot)$ and $f^{t_2}(\cdot)$ need not be identically distributed, nor must they originate from the same family of distributions.

## D. The Financial Ruin Event

A retiree who outlives his/her savings experiences financial ruin. Let Ruin(t) denote the event of financial ruin occurring at time t, and let $\text{Ruin}^C(t)$ denote the event of avoiding financial ruin at time t. Rook (2014) shows that, given ruin has been avoided at time points t=1, 2, …, t-1, the event Ruin(t) is equivalent to:

$$\text{Ruin}(t) \equiv \hat{r}_{(t,\alpha)} \leq \text{RF}(t-1), \tag{2.24}$$

where RF(t-1) is a quantity termed the *ruin factor* and defined as:

$$\text{RF}(0) = W_R, \text{ for } t=0, \text{ and,} \tag{2.25}$$

$$\text{RF}(t) = \frac{\text{RF}(t-1)}{\hat{r}_{(t,\alpha)} - \text{RF}(t-1)}, \text{ for } t=1, 2, \ldots, T_D-1. \tag{2.26}$$

If $\text{Ruin}^C(t)$ then $0 < \text{RF}(t) < \infty$, and if Ruin(t) then RF(t) < 0 (or = $\infty$). In terms of the portfolio balance at time t, the reciprocal of RF(t) equals the number of real withdrawals remaining (Rook (2014)). Therefore RF(t) is an indication of the retiree's funded status at time t. To avoid experiencing financial ruin the retiree must make successful withdrawals at all time points t=1, 2,

…, $T_D$. A successful withdrawal is made at time t *iff* $\hat{r}_{(t,\alpha)} >$ RF(t-1). Let Ruin($\leq$ t) denote the event of ruin occurring on or before time t, and let $\text{Ruin}^C(\leq t)$ denote the event of avoiding ruin at all time points up to and including time t. Since the terminal withdrawal is made at time $t=T_D$, $\text{Ruin}^C(\leq T_D)$ denotes the event of avoiding ruin at all time points. Here,

$$\text{Ruin}^C(\leq T_D) \equiv \text{Ruin}^C(1) \cap \text{Ruin}^C(2) \cap \ldots \cap \text{Ruin}^C(T_D), \quad (2.27)$$

and,

$$P(\text{Ruin}^C(\leq T_D)) = P(\text{Ruin}^C(1) \cap \text{Ruin}^C(2) \cap \ldots \cap \text{Ruin}^C(T_D)) \quad (2.28)$$

$$\rightarrow P(\text{Ruin}^C(\leq T_D)) = P(\hat{r}_{(1,\alpha)} > RF(0) \cap \hat{r}_{(2,\alpha)} > RF(1) \cap \ldots \cap \hat{r}_{(T_D,\alpha)} > RF(T_D-1)) \quad (2.29)$$

$$\rightarrow P(\text{Ruin}^C(\leq T_D)) =$$

$$\int_{RF(0)}^{\infty} \int_{RF(1)}^{\infty} \ldots \int_{RF(T_D-1)}^{\infty} f(\hat{r}_{(1,\alpha)}, \hat{r}_{(2,\alpha)}, \ldots, \hat{r}_{(T_D,\alpha)}) d\hat{r}_{(T_D,\alpha)} \ldots d\hat{r}_{(2,\alpha)} d\hat{r}_{(1,\alpha)}. \quad (2.30)$$

We can estimate $P(\text{Ruin}^C(\leq T_D))$ using simulation or a dynamic program (DP). Simulation uses the fact that ruin is avoided at time t *iff* $\hat{r}_{(t,\alpha)} >$ RF(t-1), which suggests the following algorithm. Iterate over times t=1, 2, …, $T_D$ (i.e., the retirement horizon) generating a random observation from the PDF $f^t(\cdot)$, say $\hat{r}_{(t,\alpha)}$, and if $\hat{r}_{(t,\alpha)} >$ RF(t-1) calculate RF(t), otherwise end the process and record a ruin event by incrementing a ruin counter (initialized to 0). Repeat the process N times and estimate $P(\text{Ruin}(\leq T_D))$ by (# of ruin events)/N. The quantity 1 - $P(\text{Ruin}(\leq T_D))$ estimates $P(\text{Ruin}^C(\leq T_D))$. If the PDF $f^t(\cdot)$ is not from a known family then rejection sampling can be used to generate random observations (Von Neumann (1951)).

To estimate $P(\text{Ruin}^C(\leq T_D))$ using a DP we simplify the model from Rook (2014) by removing the optimization at each (t, RF(t)). Namely, let:

$$V(t, RF(t)) = \text{Probability of ruin after time t given } RF(t) > 0. \quad (2.31)$$

Then it follows from Rook (2014) that:

$$V(t, RF(t)) = 1 - (1 - F^{t+1}(RF(t))) * (1 - E_{\hat{r}_{(t+1,\alpha)}}^+ \left[V(t+1, \frac{RF(t)}{\hat{r}_{(t+1,\alpha)} - RF(t)})\right]) \quad (2.32)$$

For: $0 \leq t \leq T_D-1$, $RF(t) > 0$, $V(T_D, RF(T_D)) = 0$.

This DP is solved in practice by discretizing the RF(t) dimension (Rook (2014)), with P(Ruin($\leq$ T$_D$)) = V(0, W$_R$). The RV $\hat{r}_{(t+1,\alpha)}^{+} = \hat{r}_{(t+1,\alpha)} | \hat{r}_{(t+1,\alpha)} >$ RF(t) from (2.32) has PDF:

$$\hat{r}_{(t+1,\alpha)}^{+} \sim \begin{cases} \frac{f(\hat{r}_{(t+1,\alpha)})}{\int_{RF(t)}^{\infty} f(\hat{r}_{(t+1,\alpha)}) \, d\hat{r}_{(t+1,\alpha)}} & , \text{ for } \hat{r}_{(t+1,\alpha)} > RF(t) \\ 0 & , \text{ O.W.} \end{cases} \quad (2.33)$$

Note that the DP is solved using $F^t(\cdot)$ not $f^t(\cdot)$, and 1 - P(Ruin($\leq$T$_D$)) estimates P(Ruin$^C$($\leq$T$_D$)).

*E. Testing Glidepaths for Equality*

Consider two retirement glidepaths GP$_1$ and GP$_2$ that generate binary outcomes (i.e., 0=Ruin($\leq$T$_D$), 1=Ruin$^C$($\leq$T$_D$)), and assign these to Bernoulli RVs X$_1$ and X$_2$ with P(X$_i$=1) = p$_i$ for i=1, 2. It follows trivially that E(X$_i$) = p$_i$ and V(X$_i$) = p$_i$(1-p$_i$). Interest is in testing the following hypothesis:

$$\begin{aligned} &H_o: p_1 = p_2 \\ \text{vs.} \quad &H_a: p_1 \neq p_2 \end{aligned} \quad (2.34)$$

If random samples of size N$_1$ and N$_2$ (both large) are drawn on RVs X$_1$ and X$_2$ then invoking the central limit theorem (CLT) yields (Ross (2009)):

$$\overline{X}_i \stackrel{.}{\sim} N\left(p_i, \frac{p_i(1-p_i)}{N_i}\right), \quad \text{for } i = 1, 2, \quad (2.35)$$

where $\overline{X}_i$ is the proportion of successes in sample i for i=1, 2. Under H$_o$ the variances are equal and we can pool both samples to estimate the normal variances as:

$$\frac{\hat{p}(1-\hat{p})}{N_i} \cong \frac{p_i(1-p_i)}{N_i}, \quad \text{for } i = 1, 2, \quad (2.36)$$

where,

$$\hat{p} = \frac{N_1 \overline{X}_1 + N_2 \overline{X}_2}{N_1 + N_2}, \quad (2.37)$$

is the proportion of successes in the combined sample, which is also the maximum likelihood estimator of both p$_1$ and p$_2$ under H$_o$ (Ross (2009)). Since the samples are independent, under $H_o$, the following statistic can be used for a large-sample test of $H_o$ with critical region: $|t_s| > Z_{\alpha/2}$, where $\alpha = P(\text{Type I Error})$:

$$t_s = \frac{(\bar{X}_1 - \bar{X}_2)}{\sqrt{\frac{\hat{p}(1-\hat{p})}{N_1} + \frac{\hat{p}(1-\hat{p})}{N_2}}} \stackrel{.}{\sim} N(0,1). \qquad (2.38)$$

*F. Testing Glidepaths for Non-Inferiority*

Consider the same 2 glidepath processes described in Section II.E, but interest is now in testing the hypothesis:

$$\begin{aligned} H_o: \ & p_1 \geq p_2 \\ \text{vs.} \quad H_a: \ & p_1 < p_2 \end{aligned} \qquad (2.39)$$

The pooled estimator $\hat{p}$ from Section II.E no longer applies (under $H_o$) but each variance can be estimated via maximum likelihood by (Ross (2009)):

$$\frac{\hat{p}_i(1-\hat{p}_i)}{N_i} \cong \frac{p_i(1-p_i)}{N_i}, \quad \text{for } i = 1, 2, \qquad (2.40)$$

where,

$$\hat{p}_i = \bar{X}_i \quad \text{for } i=1, 2. \qquad (2.41)$$

Invoking independence, and the CLT, for large samples, the quantity:

$$t_s^* = \frac{(\bar{X}_1 - \bar{X}_2)}{\sqrt{\frac{\hat{p}_1(1-\hat{p}_1)}{N_1} + \frac{\hat{p}_2(1-\hat{p}_2)}{N_2}}} \stackrel{.}{\sim} N(p_1 - p_2, 1). \qquad (2.42)$$

Small values of $t_s^*$ lead to rejecting $H_o$, and under $H_o$, $t_s^*$ is observed from a normal distribution with mean $> 0$ and variance $= 1$. Therefore, under $H_o$, $P(t_s^* < -Z_\alpha) \leq \alpha$. (Specifically, $P(t_s^* < -Z_\alpha) = \alpha$ when $p_1 = p_2$ and $P(t_s^* < -Z_\alpha) < \alpha$ when $p_1 > p_2$.) This suggests a large-sample test of $H_o$ with $P(\text{Type I Error})$ of at most $\alpha$ has critical region: $t_s^* < -Z_\alpha$.

*G. The Gradient of a Function*

The gradient of an n-dimensional continuous function $f(x_1, x_2, \ldots, x_n)$ is the vector of 1st order partial derivatives (assuming they exist), namely:

$$\vec{g} = \begin{pmatrix} \frac{\partial f(\cdot)}{\partial x_1} \\ \frac{\partial f(\cdot)}{\partial x_2} \\ \vdots \\ \frac{\partial f(\cdot)}{\partial x_n} \end{pmatrix} \quad (2.43)$$

When evaluated at a specific point, the gradient indicates the direction of steepest ascent (Hillier and Lieberman (2010)). Further, the critical points of f(·) are those satisfying $\vec{g} = \vec{0}$, and interior point optimums of f(·) can only exist at critical points (Anton (1988)).

*H. The Hessian of a Function*

The Hessian of an n-dimensional continuous function $f(x_1, x_2, \ldots, x_n)$ is the symmetric matrix of $2^{nd}$ order partial derivatives (assuming they exist), namely:

$$\bar{\bar{H}} = \begin{bmatrix} \frac{\partial^2 f(\cdot)}{\partial x_1^2} & \frac{\partial^2 f(\cdot)}{\partial x_1 \partial x_2} & \cdots & \frac{\partial^2 f(\cdot)}{\partial x_1 \partial x_n} \\ \frac{\partial^2 f(\cdot)}{\partial x_2 \partial x_1} & \frac{\partial^2 f(\cdot)}{\partial x_2^2} & \cdots & \frac{\partial^2 f(\cdot)}{\partial x_2 \partial x_n} \\ \vdots & \vdots & \ddots & \vdots \\ \frac{\partial^2 f(\cdot)}{\partial x_n \partial x_1} & \frac{\partial^2 f(\cdot)}{\partial x_n \partial x_2} & \cdots & \frac{\partial^2 f(\cdot)}{\partial x_n^2} \end{bmatrix} \quad (2.44)$$

When evaluated at a specific point, the Hessian matrix is negative semi-definite if the region around that point is concave (Jensen and Bard (2003)). Further, an *n*x*n* matrix is negative semi-definite *iff* its *n* eigenvalues are non-positive (Hillier and Lieberman (2010)). Note that the eigenvalues of a real symmetric matrix are guaranteed to both exist and be real.

*I. The Gradient Ascent Algorithm*

To maximize an n-dimensional continuous function $f(x_1, x_2, \ldots, x_n)$ using *gradient ascent* we compute the gradient $\vec{g}_0$ for an initial starting point $\vec{x}_0 = (x_{10}, x_{20}, \ldots, x_{n0})'$ and climb until no further progress is made, then repeat the process until the gradient is zero in all directions (i.e., $\vec{g} = \vec{0}$). We then compute the Hessian and confirm the stopping point is a maximum by showing that the region is concave. To climb, we build the gradient $\vec{g}_0$ for $\vec{x}_0$ then find the scalar t that maximizes f($\vec{x}_0$ + t*$\vec{g}_0$) (Hillier and Lieberman (2010)). In practice we choose an appropriate increment size then step in the direction of the gradient until progress ends, assigning $\vec{x}_1 = \vec{x}_0$ + t*$\vec{g}_0$. The vector $\vec{x}_1$ becomes the starting point for the next iteration (Hillier and Lieberman (2010)), and in this manner we zig-zag up the surface to the top. Iterations end when some desirable condition is met, for example when the largest absolute gradient element is smaller than a convergence threshold ε, see Figure 2. If different starting points, $\vec{x}_0$, lead to different local optimums then a metaheuristic such as tabu search, simulated annealing, or a genetic algorithm can be used to guide the procedure towards a global optimum. Some metaheuristics are guaranteed to converge if run long enough (Hillier and Lieberman (2010)).

**Figure 2**
**Climbing in the Direction of the Gradient**

This figure depicts the gradient climbing procedure where $\vec{g}$ is the gradient of an n-dimensional continuous function f($x_1$, $x_2$, …, $x_n$) evaluated at the point $\vec{x}$. The step size is t and the optimization reduces to maximizing the function f($\vec{x}$ + (t)$\vec{g}$) with respect to the unknown scalar t. That is, the n-dimensional optimization problem is reduced to a 1-dimensional optimization problem at each iteration. In practice we will choose an appropriate step size for t and climb in the direction of $\vec{g}$ until progress ends. Progress ends when f(·) begins to decrease. If t=t* is the value that maximizes f($\vec{x}$ + (t)$\vec{g}$) then $\vec{x}$ + (t*)$\vec{g}$ becomes the starting point for the next iteration. The procedure ends when $\vec{g}$ = $\vec{0}$, which implies that climbing in no direction will increase the objective, f(·). In practice we will set a threshold that all elements of $\vec{g}$ must not exceed in absolute terms as a stopping rule. (Hillier and Lieberman (2010))

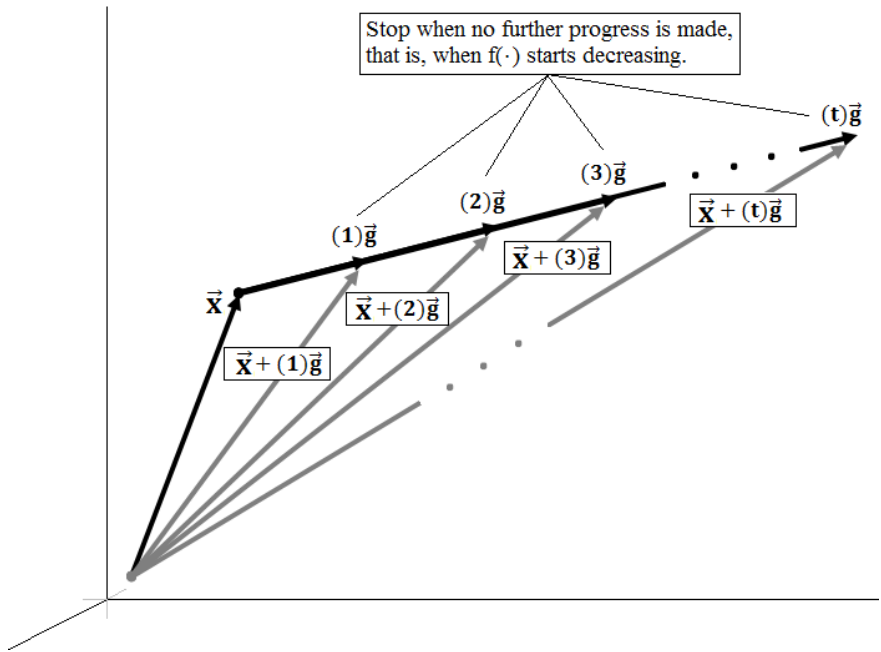

## J. Newton's Method

In practice the gradient ascent method often converges slowly. Alternatively, we can optimize an n-dimensional continuous function $f(x_1, x_2, \ldots, x_n)$ with *Newton's Method* by approximating its gradient $\vec{g}$ with a 1$^{st}$ order Taylor series in the neighborhood of some initial starting point $\vec{x}_0$. The 1$^{st}$ order Taylor expansion of $\vec{g}$ at $\vec{x}_0 = (x_{10}, x_{20}, \ldots, x_{n0})'$ is the RHS of:

$$\vec{g} \cong \vec{g}_0 + \bar{\bar{H}}_0(\vec{x} - \vec{x}_0). \tag{2.45}$$

Setting this approximation $= \vec{0}$ and solving yields the critical point, $\vec{x}_1 = (x_{11}, x_{21}, \ldots, x_{n1})'$, where:

$$\vec{x}_1 = \vec{x}_0 - \bar{\bar{H}}_0^{-1}\vec{g}_0, \tag{2.46}$$

which becomes the starting point for the next iteration. As before the procedure ends when some objective criteria is met, such as all gradient elements are $< \varepsilon$ in absolute value. Since Newton's method requires $\bar{\bar{H}}$ we can derive its eigenvalues at each step and confirm the region is concave when maximizing $f(\cdot)$. This is important because $\vec{g} = \vec{0}$ at local/global minimums also. As with gradient ascent, if different starting values lead to different local optimums a metaheuristic can be employed to guide the procedure towards a global optimum (Hillier and Lieberman (2010)).

## K. An Integral Evaluation

In upcoming sections we will need to evaluate integrals of the form:

$$\int_{-\infty}^{y} (x - \mu)^n \, e^{-\left(\frac{x-\mu}{\sqrt{2\sigma^2}}\right)^2} dx, \tag{2.47}$$

where n is a positive integer. A solution will be derived here and used throughout. The cases $y \leq \mu$ and $y > \mu$ will be considered separately.

<u>Case 1:</u>  $y \leq \mu$

$$\text{Let } u = \left(\frac{x - \mu}{\sqrt{2\sigma^2}}\right)^2 \tag{2.48}$$

$$\rightarrow \sqrt{u} = -\sqrt{\left(\frac{x-\mu}{\sqrt{2\sigma^2}}\right)^2} \rightarrow -\sqrt{u} = \left(\frac{x-\mu}{\sqrt{2\sigma^2}}\right) \tag{2.49}$$

$$\rightarrow \frac{du}{dx} = 2\left(\frac{x-\mu}{\sqrt{2\sigma^2}}\right)\left(\frac{1}{\sqrt{2\sigma^2}}\right) \rightarrow \left(\sqrt{\frac{\sigma^2}{2}}\right) du = \left(\frac{x-\mu}{\sqrt{2\sigma^2}}\right) dx \tag{2.50}$$

$$\text{As } x \to -\infty, u \to \infty, \text{ and,} \tag{2.51}$$

$$\text{as } x \to y, u \to \left(\frac{y-\mu}{\sqrt{2\sigma^2}}\right)^2 \tag{2.52}$$

Therefore, (2.47) can be written as:

$$\left(\sqrt{2\sigma^2}\right)^n \int_{-\infty}^{y} \left(\frac{x-\mu}{\sqrt{2\sigma^2}}\right)^n e^{-\left(\frac{x-\mu}{\sqrt{2\sigma^2}}\right)^2} dx = (-1)^n \left(\sqrt{2\sigma^2}\right)^n \left(\sqrt{\frac{\sigma^2}{2}}\right) \int_{\left(\frac{y-\mu}{\sqrt{2\sigma^2}}\right)^2}^{\infty} u^{\left(\frac{n-1}{2}\right)} e^{-u} du \tag{2.53}$$

$$= (-1)^n \left(\sqrt{\sigma^2}\right)^{n+1} \left(\sqrt{2}\right)^{n-1} \Gamma\left(\frac{n+1}{2}\right) \left[1 - F_{g\left[\frac{n+1}{2}, 1\right]}\left(\left(\frac{y-\mu}{\sqrt{2\sigma^2}}\right)^2\right)\right] \tag{2.54}$$

Where $\Gamma(\cdot)$ is the gamma function and $F_{g[\alpha,\beta]}(\cdot)$ represents the CDF of a gamma RV with shape and scale parameters $\alpha$ and $\beta$, respectively (Casella and Berger (1990)).

Case 2: $y > \mu$

Separate (2.47) into 2 parts as shown below:

$$\int_{-\infty}^{\mu} (x-\mu)^n e^{-\left(\frac{x-\mu}{\sqrt{2\sigma^2}}\right)^2} dx + \int_{\mu}^{y} (x-\mu)^n e^{-\left(\frac{x-\mu}{\sqrt{2\sigma^2}}\right)^2} dx. \tag{2.55} + (2.56)$$

Apply (2.54) from Case 1 to the 1st integral (2.55) which reduces to:

$$(-1)^n \left(\sqrt{\sigma^2}\right)^{n+1} \left(\sqrt{2}\right)^{n-1} \Gamma\left(\frac{n+1}{2}\right), \tag{2.57}$$

then make the same u-substitution from (2.48) in the 2nd integral (2.56), except now:

$$\sqrt{u} = \left(\frac{x-\mu}{\sqrt{2\sigma^2}}\right) \tag{2.58}$$

$$\text{And as } x \to \mu, u \to 0. \tag{2.59}$$

$$\rightarrow (2.56) = \left(\sqrt{2\sigma^2}\right)^n \int_{\mu}^{y} \left(\frac{x-\mu}{\sqrt{2\sigma^2}}\right)^n e^{-\left(\frac{x-\mu}{\sqrt{2\sigma^2}}\right)^2} dx = \left(\sqrt{2\sigma^2}\right)^n \left(\sqrt{\frac{\sigma^2}{2}}\right) \int_{0}^{\left(\frac{y-\mu}{\sqrt{2\sigma^2}}\right)^2} u^{\left(\frac{n-1}{2}\right)} e^{-u} du \tag{2.60}$$

$$= \left(\sqrt{\sigma^2}\right)^{n+1} \left(\sqrt{2}\right)^{n-1} \Gamma\left(\frac{n+1}{2}\right) F_{g\left[\frac{n+1}{2},1\right]}\left(\left(\frac{y-\mu}{\sqrt{2\sigma^2}}\right)^2\right). \tag{2.61}$$

Combining (2.57) and (2.61) yields (for Case 2):

$$(2.47) = (-1)^n \left(\sqrt{\sigma^2}\right)^{n+1} \left(\sqrt{2}\right)^{n-1} \Gamma\left(\frac{n+1}{2}\right)\left(1 + (-1)^n F_{g\left[\frac{n+1}{2},1\right]}\left(\left(\frac{y-\mu}{\sqrt{2\sigma^2}}\right)^2\right)\right). \tag{2.62}$$

And, combining *Cases* 1 and 2 results in the following piecewise definition for (2.47):

$$\begin{cases} (-1)^n \left(\sqrt{\sigma^2}\right)^{n+1} \left(\sqrt{2}\right)^{n-1} \Gamma\left(\frac{n+1}{2}\right)\left(1 - F_{g\left[\frac{n+1}{2},1\right]}\left(\left(\frac{y-\mu}{\sqrt{2\sigma^2}}\right)^2\right)\right) & , \text{for } y \leq \mu \\ \\ (-1)^n \left(\sqrt{\sigma^2}\right)^{n+1} \left(\sqrt{2}\right)^{n-1} \Gamma\left(\frac{n+1}{2}\right)\left(1 + (-1)^n F_{g\left[\frac{n+1}{2},1\right]}\left(\left(\frac{y-\mu}{\sqrt{2\sigma^2}}\right)^2\right)\right) & , \text{for } y > \mu \end{cases} \tag{2.63}$$

Using an indicator function we can express (2.63) more succinctly, namely, let $\mathbf{1}_A(x) = 1$ when $x \in A$, and 0 otherwise (Casella and Berger (1990)). Then, for $-\infty < y < \infty$, (2.47) is given by:

$$(2.64)$$
$$(-1)^n \left(\sqrt{\sigma^2}\right)^{n+1} \left(\sqrt{2}\right)^{n-1} \Gamma\left(\frac{n+1}{2}\right)\left(1 - (-1)^{(n-1)\mathbf{1}_{(\mu,\infty)}(y)} F_{g\left[\frac{n+1}{2},1\right]}\left(\left(\frac{y-\mu}{\sqrt{2\sigma^2}}\right)^2\right)\right).$$

Note that the expressions in (2.63) are identical $\forall$ odd integers $n > 0$. Further, $\Gamma(1/2) = \sqrt{\pi}$, $\Gamma(n) = (n-1)!$, and $\Gamma(\theta+1) = \theta\Gamma(\theta)$ for positive integers n and positive real numbers $\theta$ (Casella and Berger (1990)), thus:

$$\Gamma\left(\frac{n+1}{2}\right) = \begin{cases} 1 & , \text{for } n = 1 \\ \frac{\sqrt{\pi}}{2} & , \text{for } n = 2 \\ 1 & , \text{for } n = 3 \\ \frac{3\sqrt{\pi}}{4} & , \text{for } n = 4 \end{cases} \tag{2.65}$$

*L. Gaussian Moments*

There will also be a need to evaluate integrals of the following form:

$$\int_{-\infty}^{\infty} (x-\mu)^n \, e^{-\left(\frac{x-\mu}{\sqrt{2\sigma^2}}\right)^2} dx, \tag{2.66}$$

where n is a positive integer. Multiplying and dividing by the constant $\sigma^n\sqrt{2\pi}$ yields:

$$\sigma^n\sqrt{2\pi} \int_{-\infty}^{\infty} \frac{1}{\sqrt{2\pi}}\left(\frac{x-\mu}{\sigma}\right)^n e^{-\frac{1}{2}\left(\frac{x-\mu}{\sigma}\right)^2} dx = \begin{cases} 0 & \text{, for } n = 1 \\ \sigma^3\sqrt{2\pi} & \text{, for } n = 2 \\ 0 & \text{, for } n = 3 \\ 3\sigma^5\sqrt{2\pi} & \text{, for } n = 4 \end{cases} \quad (2.67)$$

For justification note that (2.66) = $\sigma\sqrt{2\pi}$ E[(X - $\mu$)$^n$] where X ~ N($\mu$, $\sigma^2$). Clearly, when n=1, (2.66) = 0 and when n=2 we invoke the definition of $\sigma^2$ so that (2.66) = $\sigma^3\sqrt{2\pi}$. Then let Z = (X - $\mu$)/$\sigma$ → dZ = (1/$\sigma$)dX → $\sigma$dZ = dX and (2.66) = $\sigma^{n+1}\sqrt{2\pi}$ E[Z$^n$], where E[Z$^n$] = (n-1)*(n-3) … 3*1 is the n$^{th}$ moment of a standardized normal RV (Johnson, Kotz and Balakrishnan (1994)). Using the formula, clearly (2.66) = 0 ∀ odd n and when n=4 (2.66) = $\sigma^5\sqrt{2\pi}$E[Z$^4$] = $3\sigma^5\sqrt{2\pi}$.

*M. Convex Sets*

A set S is convex if for any x, y ∈ S, $\lambda$x + (1-$\lambda$)y ∈ S, ∀ 0 ≤ $\lambda$ ≤ 1, that is, any convex combination of two set members is also a set member (Boyd and Vandenberghe (2004)). Let $\hat{R}$ denote the set of inflation/expense-adjusted returns that avoid financial ruin in retirement for a horizon of length $T_D$ and given initial withdrawal rate $W_R$=RF(0). Then from Section II.D,

$$\hat{R} = \{\hat{r}_{(1,\alpha)}, \hat{r}_{(2,\alpha)}, \ldots, \hat{r}_{(T_D,\alpha)} : \cap_{t=1}^{T_D}(\hat{r}_{(t,\alpha)} > RF(t-1))\}. \quad (2.68)$$

*Claim:* The set $\hat{R}$ is convex.

*Justification:* We must demonstrate that given any two vectors of returns that avoid financial ruin in retirement, say $\vec{\hat{r}}_1$ and $\vec{\hat{r}}_2 \in \hat{R}$, the vector of returns $\vec{\hat{r}}_c = \lambda\vec{\hat{r}}_1 + (1-\lambda)\vec{\hat{r}}_2$ also avoids financial ruin in retirement ∀ 0 ≤ $\lambda$ ≤ 1. That is we must show that $\vec{\hat{r}}_c \in \hat{R}$, where:

$$\vec{\hat{r}}_1 = \begin{pmatrix} \hat{r}_{1(1,\alpha)} \\ \hat{r}_{1(2,\alpha)} \\ \vdots \\ \hat{r}_{1(T_D,\alpha)} \end{pmatrix}, \vec{\hat{r}}_2 = \begin{pmatrix} \hat{r}_{2(1,\alpha)} \\ \hat{r}_{2(2,\alpha)} \\ \vdots \\ \hat{r}_{2(T_D,\alpha)} \end{pmatrix}, \vec{\hat{r}}_c = \begin{pmatrix} \hat{r}_{c(1,\alpha)} = \lambda\hat{r}_{1(1,\alpha)} + (1-\lambda)\hat{r}_{2(1,\alpha)} \\ \hat{r}_{c(2,\alpha)} = \lambda\hat{r}_{1(2,\alpha)} + (1-\lambda)\hat{r}_{2(2,\alpha)} \\ \vdots \\ \hat{r}_{c(T_D,\alpha)} = \lambda\hat{r}_{1(T_D,\alpha)} + (1-\lambda)\hat{r}_{2(T_D,\alpha)} \end{pmatrix}. \quad (2.69)$$

For $T_D$=1, a single-period retirement horizon, $\vec{\hat{r}}_1$ and $\vec{\hat{r}}_2 \in \hat{R}$ implies that $\hat{r}_{1(1,\alpha)}$>RF(0) and $\hat{r}_{2(1,\alpha)}$ > RF(0). Therefore, $\hat{r}_{c(1,\alpha)} = \lambda\hat{r}_{1(1,\alpha)} + (1-\lambda)\hat{r}_{2(1,\alpha)} > \lambda$RF(0)+(1-$\lambda$)RF(0) = RF(0) so that $\vec{\hat{r}}_c \in \hat{R}$.

For $T_D=2$, a two-period retirement horizon, $\vec{\hat{r}}_1$ and $\vec{\hat{r}}_2 \in \hat{R}$ implies that $\hat{r}_{1(1,\alpha)} > RF(0)$ and $\hat{r}_{2(1,\alpha)} > RF(0)$, but also $\hat{r}_{1(2,\alpha)} > RF_1(1)$ and $\hat{r}_{2(2,\alpha)} > RF_2(1)$. The condition at time t=2 states:

$$\hat{r}_{1(2,\alpha)} > RF_1(1) = \frac{RF(0)}{\hat{r}_{1(1,\alpha)} - RF(0)} \quad \text{and} \quad \hat{r}_{2(2,\alpha)} > RF_2(1) = \frac{RF(0)}{\hat{r}_{2(1,\alpha)} - RF(0)} \quad (2.70)$$

$$\rightarrow \lambda \hat{r}_{1(2,\alpha)} + (1-\lambda)\hat{r}_{2(2,\alpha)} > \lambda \left(\frac{RF(0)}{\hat{r}_{1(1,\alpha)} - RF(0)}\right) + (1-\lambda)\left(\frac{RF(0)}{\hat{r}_{2(1,\alpha)} - RF(0)}\right). \quad (2.71)$$

We must show that $\hat{r}_{c(2,\alpha)} > RF_c(1) = \frac{RF(0)}{\hat{r}_{c(1,\alpha)} - RF(0)}$, or:

$$\lambda \hat{r}_{1(2,\alpha)} + (1-\lambda)\hat{r}_{2(2,\alpha)} > \frac{RF(0)}{\lambda \hat{r}_{1(1,\alpha)} + (1-\lambda)\hat{r}_{2(1,\alpha)} - RF(0)}. \quad (2.72)$$

It suffices to show that the RHS of (2.71) is $\geq$ the RHS of (2.72), namely:

(2.73)
$$\lambda \left(\frac{RF(0)}{\hat{r}_{1(1,\alpha)} - RF(0)}\right) + (1-\lambda)\left(\frac{RF(0)}{\hat{r}_{2(1,\alpha)} - RF(0)}\right) \geq \frac{RF(0)}{\lambda \hat{r}_{1(1,\alpha)} + (1-\lambda)\hat{r}_{2(1,\alpha)} - RF(0)},$$

which holds,

$$\leftrightarrow \frac{\lambda\left(\hat{r}_{2(1,\alpha)} - RF(0)\right) + (1-\lambda)\left(\hat{r}_{1(1,\alpha)} - RF(0)\right)}{\left(\hat{r}_{1(1,\alpha)} - RF(0)\right)\left(\hat{r}_{2(1,\alpha)} - RF(0)\right)}$$

$$\geq \frac{1}{\lambda\left(\hat{r}_{1(1,\alpha)} - RF(0)\right) + (1-\lambda)\left(\hat{r}_{2(1,\alpha)} - RF(0)\right)} \quad (2.74)$$

$$\leftrightarrow [\lambda^2 + (1-\lambda)^2 - 1]\left(\hat{r}_{1(1,\alpha)} - RF(0)\right)\left(\hat{r}_{2(1,\alpha)} - RF(0)\right) + \lambda(1-\lambda)\left(\hat{r}_{1(1,\alpha)} - RF(0)\right)^2$$

$$+ \lambda(1-\lambda)\left(\hat{r}_{2(1,\alpha)} - RF(0)\right)^2 \geq 0, \quad (2.75)$$

but $\lambda^2 + (1-\lambda)^2 - 1 = 2\lambda^2 - 2\lambda = -2\lambda(1-\lambda)$ and $\lambda(1-\lambda)$ can be divided from each term on both sides leaving:

$$\leftrightarrow \left(\hat{r}_{1(1,\alpha)} - RF(0)\right)^2 - 2\left(\hat{r}_{1(1,\alpha)} - RF(0)\right)\left(\hat{r}_{2(1,\alpha)} - RF(0)\right) + \left(\hat{r}_{2(1,\alpha)} - RF(0)\right)^2 \geq 0$$

$$\leftrightarrow \left[\left(\hat{r}_{1(1,\alpha)} - RF(0)\right) - \left(\hat{r}_{2(1,\alpha)} - RF(0)\right)\right]^2 \geq 0, \quad (2.76)$$

which always holds, thus $\hat{r}_{c(2,\alpha)} > RF_c(1)$ and $\vec{\hat{r}}_c \in \hat{R}$ as was to be shown. For $T_D \geq 3$ we can justify our claim numerically by formulating and solving the following deterministic non-linear program (NLP):

| | | | |
|---|---|---|---|
| Minimize: | $Z = 1$ | | (2.77) |
| Subject to: | $RF_i(T_D) > 0$ and $RF_c(T_D) \leq 0$ | $i = 1, 2$ | (2.78) |
| For: | $RF_i(t) = \dfrac{RF_i(t-1)}{\hat{r}_{i(t,\alpha)} - RF_i(t-1)}$ | $i = 1, 2, c$ and $t = 1, 2, \ldots, T_D$ | (2.79a) |
| | $\hat{r}_{i(t,\alpha)} > 0$ | $i = 1, 2$ and $t = 1, 2, \ldots, T_D$ | (2.79b) |
| | $\hat{r}_{c(t,\alpha)} = \lambda \hat{r}_{1(t,\alpha)} + (1-\lambda)\hat{r}_{2(t,\alpha)}$ | $t = 1, 2, \ldots, T_D$ | (2.79c) |
| | $T_D > 0$ (integer), $RF(0) > 0$, $0 \leq \lambda \leq 1$ | | (2.79d) |

Any feasible solution to this NLP is a counter example that invalidates our claim. This NLP has no feasible solution and consequently the set $\hat{R}$ is convex. Since $\hat{R}$ does not include its boundary region it is also an open set. We can view $P(\text{Ruin}^C(\leq T_D))$ from (2.30) as the probability of observing a member of the set $\hat{R}$, and we seek the equity ratios that maximize this probability. Further, the set of inflation/expense-adjusted returns from a retirement portfolio that is successful when using an SWR strategy is independent of the glidepath. Once the withdrawal rate $W_R = RF(0)$ and horizon length $T_D$ are fixed the set $\hat{R}$ is completely determined. Elements of the set $\hat{R}$ are $T_D$-element vectors that avoid financial ruin for *any* and *every* glidepath. It is not possible for a vector of $T_D$ returns to succeed for one glidepath, but fail for another. Lastly, sequence of returns (SOR) risk emerges when using an SWR strategy in retirement (see, for example, Milevsky and Abaimova (2006)). This risk is often interpreted to mean that early decumulation years have outsized importance when determining portfolio success or failure. Using our notation SOR risk simply means that the set $\hat{R}$ is not closed under permutation. If we take a member of the set $\hat{R}$ and change its ordering, the resulting vector is not necessarily a member.

## N. The Concave Function Spectrum

The generalized mean of any two positive real numbers x, y is denoted by $M_\alpha^\lambda(x, y)$ and defined for $0 \leq \lambda \leq 1$ as (Kall and Mayer (2010))[5]:

$$M_\alpha^\lambda(x, y) = \begin{cases} [\lambda x^\alpha + (1-\lambda)y^\alpha]^{\frac{1}{\alpha}}, & \text{for } \alpha \notin \{0, -\infty, +\infty\} \\ x^\lambda y^{1-\lambda}, & \text{for } \alpha = 0 \\ \text{Min}(x, y), & \text{for } \alpha = -\infty \\ \text{Max}(x, y), & \text{for } \alpha = +\infty \end{cases} \quad (2.80)$$

A function $f(\cdot) > 0 \; \forall \; x, y \in (-\infty, +\infty)$ is said to be α-concave if $f(\lambda x + (1-\lambda)y) \geq M_\alpha^\lambda(f(x), f(y))$ $\forall \; 0 \leq \lambda \leq 1$. Cases of interest here are when $\alpha \in [-\infty, 1]$. Namely, a 1-concave function $f(\cdot)$ is just concave; a 0-concave function $f(\cdot)$ is log-concave which means that $\log[f(\cdot)]$ is concave; and a $-\infty$-concave function $f(\cdot)$ is quasi-concave (unimodal) (Kall and Mayer (2010)). If we replace "≥" with ">" in the definition above then for $\alpha = -\infty$, $f(\cdot)$ is said to be *strictly* quasi-concave. The domain of $f(\cdot)$ has been defined here as a scalar quantity, but this is not required and $f(\cdot)$ may accept a vector as its domain so long as it returns a single positive real number.

The spectrum of concave functions exhibits a natural ordering, meaning that a regular concave function is also log-concave, α-concave for $\alpha \in (-\infty, 0)$, and quasi-concave. Likewise, a function that is not quasi-concave is not α-concave for $\alpha \in (-\infty, 0)$, log-concave, or concave (Kall and Mayer (2010)). Concave functions are usually defined as the negative of convex functions and have no inflection points. Many (but not all) bell-shaped functions (including PDFs) are log-concave but not concave because they have inflection points near the bottom. A monotone step function, or one that does not decrease then increase, would be an example of a quasi-concave function that is not also strictly quasi-concave. Lastly, a function $f(\cdot)$ is said to be quasi-convex if $f(\lambda x + (1-\lambda)y) \leq \text{Max}(f(x), f(y))$ where x, y and λ are as defined above and $f(\cdot)$ is *strictly* quasi-convex if "≤" can be replaced by "<".

---

[5] In this section α refers to an arbitrary scalar not an equity ratio.

## III. Static Glidepath Optimization

As shown in (2.30) from Section II.D, $P(\text{Ruin}^C(\leq T_D))$, which we will denote by $P_{NR}(\vec{\alpha}) = P_{NR}(\vec{\alpha}, T_D)$ for a generic glidepath $\vec{\alpha} = (\alpha_1, \alpha_2, \ldots, \alpha_{T_D})'$ and horizon length $T_D$, refers to the probability of avoiding financial ruin at all time points and is computed as[6]:

$$P_{NR}(\vec{\alpha}) = P_{NR}(\vec{\alpha}, T_D) = \int_{RF(0)}^{\infty} \int_{RF(1)}^{\infty} \cdots \int_{RF(T_D-1)}^{\infty} f\left(\hat{r}_{(1,\alpha_1)}, \hat{r}_{(2,\alpha_2)}, \ldots, \hat{r}_{(T_D,\alpha_{T_D})}\right) d\vec{\hat{r}}, \quad (3.1)$$

where the differentials in (3.1), $d\hat{r}_{(T_D,\alpha_{T_D})} \ldots d\hat{r}_{(2,\alpha_2)} d\hat{r}_{(1,\alpha_1)}$, are abbreviated by $d\vec{\hat{r}}$. That is, $P_{NR}(\vec{\alpha})$ is a function of the unknown glidepath $\vec{\alpha}$ that we seek to optimize over given $T_D$. Recalling from Section II.B that only equity ratios $\alpha_t$ larger than the minimum variance portfolio will be considered, we can formulate this non-linear optimization problem as:

Maximize: $\quad Z = P_{NR}(\vec{\alpha}) = P_{NR}(\vec{\alpha}, T_D)$ \hfill (3.2)

Subject to: $\quad MV_t(\alpha) + \varepsilon \leq \alpha_t \leq 1.0, \text{ for } t = 1, 2, \ldots, T_D$ \hfill (3.3)

Here $\varepsilon$ is an arbitrarily small number that ensures each equity ratio exceeds $MV_t(\alpha)$ where $v_t'(\alpha) = 0$ because $v_t'(\alpha)$ will appear in the denominator of quantities to be derived. Since Z is constructed as a valid probability measure it is bounded by [0,1] and note that maximizing Z is equivalent to minimizing $(1.0 - Z)$, which is the probability of ruin. If $(1.0 - Z)$ is convex with respect to $\vec{\alpha}$ then the optimization has a convex objective with convex feasible region and therefore classifies as a *convex programming problem*. Such problems have the desirable property that local minimums are also global minimums (Jensen and Bard (2003)). The feasible region is convex since the collection of constraints forms a convex set. The entire set of constraints is: $\{-\alpha_t + (MV_t(\alpha) + \varepsilon) \leq 0: t = 1, 2, \ldots, T_D\} \cap \{\alpha_t - 1.0 \leq 0: t = 1, 2, \ldots, T_D\}$, which all have the general form $f(\alpha_t) \leq 0$, where $f(\alpha_t) = a(\alpha_t) + b$ is a linear function. This means they form a *polyhedron*, and thus represent a convex set (Boyd and Vandenberghe (2004))[7].

---

[6] Here it will become necessary to distinguish between equity ratios $\alpha$ at different time points, and $\alpha_t$ will reflect the equity ratio used between times t-1 and t. It is set at time t-1 and the resulting return using it is observed at time t.

[7] A constraint $f(x) \leq 0$ is convex if $f(x)$ is convex. By definition $f(x)$ is convex *iff* $f(\alpha x_1 + \beta x_2) \leq \alpha f(x_1) + \beta f(x_2)$ where $\alpha+\beta=1$. Linear functions $f(x) = ax + b$ satisfy $f(\alpha x_1 + \beta x_2) = \alpha f(x_1) + \beta f(x_2)$ thus are convex (Jensen and Bard (2003)). Finally, note that "∩" (the intersection operator) preserves convexity (Boyd and Vandenberghe (2004)).

Using Leibniz's Rule the gradient of $P_{NR}(\vec{\alpha})$ with respect to $\vec{\alpha}$ is the $T_D$-element vector $\vec{g}$ = $(g_1, g_2, ..., g_{T_D})'$ with $t^{th}$ term (t=1, 2, ..., $T_D$) given by (Flanders (1973)):

$$g_t = g_t|T_D = \int_{RF(0)}^{\infty} \int_{RF(1)}^{\infty} \cdots \int_{RF(T_D-1)}^{\infty} \frac{\partial}{\partial \alpha_t} f\left(\hat{r}_{(1,\alpha_1)}, \hat{r}_{(2,\alpha_2)}, ..., \hat{r}_{(T_D,\alpha_{T_D})}\right) d\vec{r}, \quad (3.4)$$

assuming the derivatives are Lebesgue measures as $\hat{r}_{(t,\alpha_t)} \to \{RF(t-1), \infty\}$. The $T_D \times T_D$ Hessian matrix $\overline{\overline{H}}$ for $P_{NR}(\vec{\alpha})$ has off-diagonal entries ($i \neq j \in 1, 2, ..., T_D$) as (Flanders (1973)):

$$H_{i,j} = H_{i,j}|T_D = \int_{RF(0)}^{\infty} \int_{RF(1)}^{\infty} \cdots \int_{RF(T_D-1)}^{\infty} \frac{\partial^2}{\partial \alpha_i \partial \alpha_j} f\left(\hat{r}_{(1,\alpha_1)}, \hat{r}_{(2,\alpha_2)}, ..., \hat{r}_{(T_D,\alpha_{T_D})}\right) d\vec{r}, \quad (3.5)$$

again assuming each 1$^{st}$ order derivative exists and the resulting integrand is a valid Lebesgue measure as $\hat{r}_{(t,\alpha_t)} \to \{RF(t-1), \infty\}$. Under the same assumptions for 2$^{nd}$ order derivatives, the diagonal Hessian elements (t=1, 2, ..., $T_D$) of $\overline{\overline{H}}$ take the form (Flanders (1973)):

$$H_{t,t} = H_{t,t}|T_D = \int_{RF(0)}^{\infty} \int_{RF(1)}^{\infty} \cdots \int_{RF(T_D-1)}^{\infty} \frac{\partial^2}{\partial \alpha_t^2} f\left(\hat{r}_{(1,\alpha_1)}, \hat{r}_{(2,\alpha_2)}, ..., \hat{r}_{(T_D,\alpha_{T_D})}\right) d\vec{r}. \quad (3.6)$$

Under efficient markets, applying (2.23) from Section II.C to (3.4), (3.5), and (3.6) yields:

$$g_t = g_t|T_D = \int_{RF(0)}^{\infty} \int_{RF(1)}^{\infty} \cdots \int_{RF(T_D-1)}^{\infty} \left(\prod_{\substack{k=1 \\ k \neq t}}^{T_D} f(\hat{r}_{(k,\alpha_k)})\right) \frac{\partial}{\partial \alpha_t} f(\hat{r}_{(t,\alpha_t)}) d\vec{r}, \quad (3.7)$$

$$H_{i,j} = H_{i,j}|T_D = \int_{RF(0)}^{\infty} \int_{RF(1)}^{\infty} \cdots \int_{RF(T_D-1)}^{\infty} \left(\prod_{\substack{k=1 \\ k \neq i,j}}^{T_D} f(\hat{r}_{(k,\alpha_k)})\right) \frac{\partial}{\partial \alpha_i} f(\hat{r}_{(i,\alpha_i)}) \frac{\partial}{\partial \alpha_j} f(\hat{r}_{(j,\alpha_j)}) d\vec{r}, \quad (3.8)$$

$$H_{t,t} = H_{t,t}|T_D = \int_{RF(0)}^{\infty} \int_{RF(1)}^{\infty} \cdots \int_{RF(T_D-1)}^{\infty} \left(\prod_{\substack{k=1 \\ k \neq t}}^{T_D} f(\hat{r}_{(k,\alpha_k)})\right) \frac{\partial^2}{\partial \alpha_t^2} f(\hat{r}_{(t,\alpha_t)}) d\vec{r}. \quad (3.9)$$

If the above derivatives exist as valid Lebesgue measures and can be estimated or approximated, then the techniques from Sections II.I and II.J can be used to optimize $P_{NR}(\vec{\alpha})$ with respect to the glidepath $\vec{\alpha} = (\alpha_1, \alpha_2, ..., \alpha_{T_D})'$. In the next section we show this to be the case.

## IV. An Implementation

We will assume that future real returns originate from the same probability distribution as past real returns and use S&P 500 total returns and 10-year Treasury Bond total returns as proxies for stocks and bonds. Under this assumption, future returns are identically distributed and the means/variances for real returns from an α-based portfolio between times t-1 and t simplify to (see Section II.B)[8]:

$$E[r_{(t,\alpha_t)}] = (\alpha_t)\mu_s + (1-\alpha_t)\mu_b \quad (4.1)$$

$$V[r_{(t,\alpha_t)}] = (\alpha_t^2)\sigma^2_s + (1-\alpha_t)^2\sigma^2_b + 2(\alpha_t)(1-\alpha_t)\sigma_{(s,b)} \quad (4.2)$$

The means and variances for inflation/expense-adjusted compounding returns $\forall\ t=1, 2, \ldots, T_D$, along with their derivatives with respect to $\alpha_t$, become (see Section II.B):

$$m(\alpha_t) = E[\hat{r}_{(t,\alpha_t)}] = (1 - E_R)*(1 + (\alpha_t)\mu_s + (1-\alpha_t)\mu_b) \quad (4.3)$$

$$m'(\alpha_t) = (1 - E_R)*(\mu_s - \mu_b) \quad (4.4)$$

$$m''(\alpha_t) = 0 \quad (4.5)$$

$$v(\alpha_t) = V[\hat{r}_{(t,\alpha_t)}] = (1 - E_R)^2*[(\alpha_t^2)\sigma^2_s + (1-\alpha_t)^2\sigma^2_b + 2(\alpha_t)(1-\alpha_t)\sigma_{(s,b)}] \quad (4.6)$$

$$v'(\alpha_t) = 2(1 - E_R)^2*[(\alpha_t)\sigma^2_s - (1-\alpha_t)\sigma^2_b + (1 - 2\alpha_t)\sigma_{(s,b)}] \quad (4.7)$$

$$v''(\alpha_t) = 2(1 - E_R)^2*[\sigma^2_s + \sigma^2_b - 2\sigma_{(s,b)}] \quad (4.8)$$

Further, the minimum variance equity ratio, $MV(\alpha)$, for identically distributed returns derived in Section II.B is constant with respect to time, and given by ($\forall\ t=1, 2, \ldots, T_D$):

$$MV(\alpha) = \frac{\sigma^2_b - \sigma_{(s,b)}}{\sigma^2_s + \sigma^2_b - 2\sigma_{(s,b)}}. \quad (4.9)$$

We will also assume that markets are efficient in the sense that any predictive capacity inherent in past patterns of returns is quickly accounted for, thereby removed as an arbitrage opportunity for retirees to exploit during their periodic rebalancing. The market efficiency assumption will be translated into the RVs, $\hat{r}_{(t,\alpha_t)}$, being independent between time points and the unit of time will be years. Further, the *iid* inflation/expense-adjusted returns are assumed to originate as normal RVs (see, Rook (2014)) with PDF $\forall\ t=1, 2, \ldots, T_D$ from Section II.C taking the form:

---

[8] The functions $m_t(\alpha)$ and $v_t(\alpha)$ become $m(\alpha_t)$ and $v(\alpha_t)$ since they are constant across time with respect to $\mu_s$, $\mu_b$, $\sigma^2_s$, $\sigma^2_b$, and $\sigma_{(s,b)}$. Their value changes with α, which is written as $\alpha_t$ reflecting the fact that it can change with time.

$$f(\hat{r}_{(t,\alpha_t)}) = \frac{1}{\sqrt{2\pi v(\alpha_t)}} e^{-\left(\frac{\hat{r}_{(t,\alpha_t)} - m(\alpha_t)}{\sqrt{2v(\alpha_t)}}\right)^2}, \quad \text{for} -\infty < \hat{r}_{(t,\alpha_t)} < \infty. \tag{4.10}$$

## A. *The Gradient of* $P_{NR}(\vec{\alpha}) = P(Ruin^C(\leq T_D))$

As shown in (3.1) the probability of avoiding financial ruin in retirement, $P_{NR}(\vec{\alpha})$, is a function of the glidepath $\vec{\alpha} = (\alpha_1, \alpha_2, \ldots, \alpha_{T_D})'$ where $MV(\alpha)+\varepsilon \leq \alpha_t \leq 1.00$ for $t=1, 2, \ldots, T_D$. By treating this glidepath as a vector of unknown variables we can optimize $P_{NR}(\vec{\alpha})$ over it subject to constraints as was formulated in (3.2) and (3.3). Using (3.7) and (4.10) the gradient is the $T_D$-element vector $\vec{g} = (g_1, g_2, \ldots, g_{T_D})'$ with $t^{th}$ term ($t=1, 2, \ldots, T_D$) given by:

$$g_t = \int_{RF(0)}^{\infty} \int_{RF(1)}^{\infty} \cdots \int_{RF(T_D-1)}^{\infty} \left( \prod_{\substack{k=1 \\ k \neq t}}^{T_D} f(\hat{r}_{(k,\alpha_k)}) \right) \frac{\partial}{\partial \alpha_t} \left[ \frac{1}{\sqrt{2\pi v(\alpha_t)}} e^{-\left(\frac{\hat{r}_{(t,\alpha_t)} - m(\alpha_t)}{\sqrt{2v(\alpha_t)}}\right)^2} \right] d\vec{r}. \tag{4.11}$$

Upon taking derivatives (see Appendix A), $g_t$ can be expressed as:

$$g_t = K_t * \left[ \int_{RF(0)}^{\infty} \int_{RF(1)}^{\infty} \cdots \int_{RF(T_D-1)}^{\infty} \left( \prod_{\substack{k=1 \\ k \neq t}}^{T_D} f(\hat{r}_{(k,\alpha_k)}) \right) g(\hat{r}_{(t,\alpha_t)}) d\vec{r} \right. \\ \left. - \int_{RF(0)}^{\infty} \int_{RF(1)}^{\infty} \cdots \int_{RF(T_D-1)}^{\infty} \prod_{k=1}^{T_D} f(\hat{r}_{(k,\alpha_k)}) d\vec{r} \right], \tag{4.12}$$

where,

$$K_t = \frac{v'(\alpha_t)}{2v(\alpha_t)} + \frac{m'(\alpha_t)^2}{2v'(\alpha_t)} \tag{4.13}$$

and,

$$g(\hat{r}_{(t,\alpha_t)}) = \frac{\left[v'(\alpha_t)\left(\hat{r}_{(t,\alpha_t)} - m(\alpha_t)\right) + m'(\alpha_t)v(\alpha_t)\right]^2}{[m'(\alpha_t)^2 v(\alpha_t)^2 + v(\alpha_t)v'(\alpha_t)^2]} f(\hat{r}_{(t,\alpha_t)}), \quad \text{for} -\infty < \hat{r}_{(t,\alpha_t)} < \infty. \tag{4.14}$$

Since $g(\hat{r}_{(t,\alpha_t)}) \geq 0$ for $-\infty < \hat{r}_{(t,\alpha_t)} < \infty$ and $\int_{-\infty}^{\infty} g(\hat{r}_{(t,\alpha_t)}) d\hat{r}_{(t,\alpha_t)} = 1.0$ (see Appendix B), $g(\hat{r}_{(t,\alpha_t)})$ is a valid PDF (Casella and Berger (1990)) and (4.12) is recognized as the difference of 2 success probabilities, multiplied by a constant. The 2$^{nd}$ probability uses the standard inflation/expense-adjusted return densities $f(\hat{r}_{(t,\alpha_t)})$ at all time points and thus equals $P_{NR}(\vec{\alpha})$, but the 1$^{st}$ probability replaces $f(\hat{r}_{(t,\alpha_t)})$ with $g(\hat{r}_{(t,\alpha_t)})$ when computing the t-th element of $\vec{g}$. We can therefore estimate (4.12) using simulation or a DP, see Section II.D. To estimate (4.12) with simulation, rejection sampling can be used to draw observations on $g(\hat{r}_{(t,\alpha_t)})$ (Von Neumann (1951)), and to estimate (4.12) with a DP we use the CDF $G(\hat{r}_{(t,\alpha_t)}) = P_g(\hat{r}_{(t,\alpha_t)} \leq r)$ for $\hat{r}_{(t,\alpha_t)} \sim g(\hat{r}_{(t,\alpha_t)})$, which takes the form (see Appendix C):

$$G(\hat{r}_{(t,\alpha_t)}) = P_g(\hat{r}_{(t,\alpha_t)} \leq r) = C_0 \left[ C_1 \left( 1 - (-1)^{\mathbf{1}_{(m(\alpha_t),\infty)}(r)} F_{g[\frac{3}{2},1]} \left( \left( \frac{r - m(\alpha_t)}{\sqrt{2v(\alpha_t)}} \right)^2 \right) \right) \right.$$
$$\left. + C_2 \left( 1 - F_{g[1,1]} \left( \left( \frac{r - m(\alpha_t)}{\sqrt{2v(\alpha_t)}} \right)^2 \right) \right) + C_3 \Phi \left( \frac{r - m(\alpha_t)}{\sqrt{v(\alpha_t)}} \right) \right],$$

(4.15)

where,

$$C_0 = \frac{1}{[m'(\alpha_t)^2 v(\alpha_t)^2 + v(\alpha_t)v'(\alpha_t)^2]}, \quad C_1 = \frac{v'(\alpha_t)^2 v(\alpha_t)}{2},$$
$$C_2 = -\frac{\sqrt{2}v'(\alpha_t)m'(\alpha_t)\sqrt{v(\alpha_t)}^3}{\sqrt{\pi}}, \quad \text{and } C_3 = m'(\alpha_t)^2 v(\alpha_t)^2.$$

(4.16)

Here, $\mathbf{1}_{(m(\alpha_t),\infty)}(r)$ is the indicator function that equals 1 when $m(\alpha_t) < r < \infty$, and 0 otherwise. Note that (4.15) is a linear combination of known CDF functions. We now have a means to estimate/approximate the gradient vector $\vec{g}$ for any glidepath $\vec{\alpha}$. The quantity $P_{NR}(\vec{\alpha})$ appears in each term but only needs to be estimated/approximated once. Further, since integration of the partial derivative in (4.11) was recognized as the difference of success probabilities, each of which is confined to [0,1], it must exist therefore satisfies the Lebesgue condition. Lastly, if simulation is used the resulting estimates are subject to *sampling error*, and if a discretized DP is used the estimates are subject to *approximation error*. In some sense, sampling error is more difficult to manage because it possesses randomness. When using simulation we can therefore test the hypothesis that the resulting gradient estimates are equal to zero with the method presented in Section II.E, since $K_t(p_1 - p_2) = 0$ *iff* $p_1 = p_2$ when $K_t \neq 0$, and each gradient element was shown to have this form in (4.12). Note that the constant $K_t$ is $> 0$ since $v(\alpha_t) > 0$ is the variance of a non-degenerate RV, and $v'(\alpha_t) > 0$ from Section II.B and (3.3). This also reveals an interesting condition required for optimality, namely that we seek a glidepath $\vec{\alpha}$ which results in all $t=1, 2, \ldots, T_D$ success probabilities using special density $g(\hat{r}_{(t,\alpha_t)})$ being equal to each other, and equal to $P_{NR}(\vec{\alpha})$.

## B. The Off-Diagonal Hessian Elements of $P_{NR}(\vec{\alpha}) = P(Ruin^C (\leq T_D))$

The off-diagonal elements of the $T_D \times T_D$ Hessian matrix, $\bar{\bar{H}}$, will reuse some of the quantities derived for $\vec{g}$. Namely, using (3.8) and (4.10) the Hessian matrix has i-j element ($i \neq j \in 1, 2, \ldots, T_D$) given by:

$$H_{i,j} = \int_{RF(0)}^{\infty} \int_{RF(1)}^{\infty} \cdots \int_{RF(T_D-1)}^{\infty} \left( \prod_{\substack{k=1 \\ k \neq i,j}}^{T_D} f(\hat{r}_{(k,\alpha_k)}) \right) \frac{\partial}{\partial \alpha_i} \left[ \frac{1}{\sqrt{2\pi v(\alpha_i)}} e^{-\left(\frac{\hat{r}_{(i,\alpha_i)} - m(\alpha_i)}{\sqrt{2v(\alpha_i)}}\right)^2} \right]$$
$$\frac{\partial}{\partial \alpha_j} \left[ \frac{1}{\sqrt{2\pi v(\alpha_j)}} e^{-\left(\frac{\hat{r}_{(j,\alpha_j)} - m(\alpha_j)}{\sqrt{2v(\alpha_j)}}\right)^2} \right] d\vec{r}.$$

(4.17)

From (A.1) in Appendix A, (4.17) can be written as:

$$H_{i,j} = \int_{RF(0)}^{\infty} \int_{RF(1)}^{\infty} \cdots \int_{RF(T_D-1)}^{\infty} \left( \prod_{\substack{k=1 \\ k \neq i,j}}^{T_D} f(\hat{r}_{(k,\alpha_k)}) \right) * K_i * [g(\hat{r}_{(i,\alpha_i)}) - f(\hat{r}_{(i,\alpha_i)})]$$
$$* K_j * [g(\hat{r}_{(j,\alpha_j)}) - f(\hat{r}_{(j,\alpha_j)})] d\vec{r},$$

(4.18)

where,

$$K_t = \frac{v'(\alpha_t)}{2v(\alpha_t)} + \frac{m'(\alpha_t)^2}{2v'(\alpha_t)}, \quad \text{for } t = i \neq j \in \{1, 2, \ldots, T_D\}$$

(4.19)

and,

$$g(\hat{r}_{(t,\alpha_t)}) = \frac{\left[v'(\alpha_t)\left(\hat{r}_{(t,\alpha_t)} - m(\alpha_t)\right) + m'(\alpha_t)v(\alpha_t)\right]^2}{[m'(\alpha_t)^2 v(\alpha_t)^2 + v(\alpha_t)v'(\alpha_t)^2]} f(\hat{r}_{(t,\alpha_t)}), \text{ for } -\infty < \hat{r}_{(t,\alpha_t)} < \infty \tag{4.20}$$

$$, \text{ and } t = i \neq j \in \{1, 2, \ldots, T_D\}$$

Here $g(\hat{r}_{(t,\alpha_t)})$ is the same valid PDF that was derived in (4.14). We can express (4.17) as:

$$K_i * K_j \left[ \int_{RF(0)}^{\infty} \int_{RF(1)}^{\infty} \cdots \int_{RF(T_D-1)}^{\infty} \left( \prod_{\substack{k=1 \\ k \neq i,j}}^{T_D} f(\hat{r}_{(k,\alpha_k)}) \right) \left[ g(\hat{r}_{(i,\alpha_i)}) g(\hat{r}_{(j,\alpha_j)}) - f(\hat{r}_{(i,\alpha_i)}) g(\hat{r}_{(j,\alpha_j)}) \right. \right.$$

$$\left. \left. - g(\hat{r}_{(i,\alpha_i)}) f(\hat{r}_{(j,\alpha_j)}) + f(\hat{r}_{(i,\alpha_i)}) f(\hat{r}_{(j,\alpha_j)}) \right] d\vec{r} \right], \tag{4.21}$$

which is equivalent to:

$$K_i * K_j \left[ \int_{RF(0)}^{\infty} \int_{RF(1)}^{\infty} \cdots \int_{RF(T_D-1)}^{\infty} \left( \prod_{\substack{k=1 \\ k \neq i,j}}^{T_D} f(\hat{r}_{(k,\alpha_k)}) \right) g(\hat{r}_{(i,\alpha_i)}) g(\hat{r}_{(j,\alpha_j)}) \, d\vec{r} \right. \tag{4.22a}$$

$$- \int_{RF(0)}^{\infty} \int_{RF(1)}^{\infty} \cdots \int_{RF(T_D-1)}^{\infty} \left( \prod_{\substack{k=1 \\ k \neq j}}^{T_D} f(\hat{r}_{(k,\alpha_k)}) \right) g(\hat{r}_{(j,\alpha_j)}) \, d\vec{r} \tag{4.22b}$$

$$- \int_{RF(0)}^{\infty} \int_{RF(1)}^{\infty} \cdots \int_{RF(T_D-1)}^{\infty} \left( \prod_{\substack{k=1 \\ k \neq i}}^{T_D} f(\hat{r}_{(k,\alpha_k)}) \right) g(\hat{r}_{(i,\alpha_i)}) \, d\vec{r} \tag{4.22c}$$

$$\left. + \int_{RF(0)}^{\infty} \int_{RF(1)}^{\infty} \cdots \int_{RF(T_D-1)}^{\infty} \left( \prod_{k=1}^{T_D} f(\hat{r}_{(k,\alpha_k)}) \right) d\vec{r} \right]. \tag{4.22d}$$

We recognize (4.22[a-d]) as consisting of valid success probabilities, this time a simple linear combination multiplied by the constant term $K_i * K_j$. The last expression (4.22d) is the standard computation for $P_{NR}(\vec{\alpha})$ using only the inflation/expense-adjusted PDFs. In practice we will compute $P_{NR}(\vec{\alpha})$ once and reuse it where needed when building $\bar{\bar{H}}$ (and $\vec{g}$). Further, note that the probabilities in (4.22b) and (4.22c) have already been computed when building $\vec{g}$, that is:

$$(4.22b) = -\left(\frac{g_j}{K_j} + P_{NR}(\vec{\alpha})\right), \tag{4.23}$$

and,

$$(4.22c) = -\left(\frac{g_i}{K_i} + P_{NR}(\vec{\alpha})\right), \tag{4.24}$$

with

$$K_t = \frac{v'(\alpha_t)}{2v(\alpha_t)} + \frac{m'(\alpha_t)^2}{2v'(\alpha_t)}, \quad \text{for } t = i \neq j \in \{1, 2, \ldots, T_D\} \tag{4.25}$$

and, (4.22d) equals $P_{NR}(\vec{\alpha})$. Consequently, $H_{i,j}$ from (4.17) can be expressed as:

$$H_{i,j} = K_i * K_j \left[ \int_{RF(0)}^{\infty} \int_{RF(1)}^{\infty} \cdots \int_{RF(T_D-1)}^{\infty} \left( \prod_{\substack{k=1 \\ k \neq i,j}}^{T_D} f(\hat{r}_{(k,\alpha_k)}) \right) g(\hat{r}_{(i,\alpha_i)}) g(\hat{r}_{(j,\alpha_j)}) d\vec{\hat{r}} - \frac{g_j}{K_j} - \frac{g_i}{K_i} - P_{NR}(\vec{\alpha}) \right]. \tag{4.26}$$

Once $\vec{g}$ has been built there is a single success probability needed to compute each i-j off-diagonal Hessian element. It is the probability of avoiding ruin when inflation/expense-adjusted returns at *both* times i and j follow PDF $g(\cdot)$, and the returns at all other time points follow the standard PDF $f(\cdot)$. As before, this probability can be estimated/approximated with rejection sampling using $g(\cdot)$ or with a DP using $G(\cdot)$ for *both* time points i, j. Lastly, the application of Leibniz's rule in (3.8) again appears valid here since the integrals in (4.22[a-d]) exist and as valid probabilities are bounded by [0,1]. The constant $K_i * K_j$ in (4.22[a-d]) is $> 0$ everywhere.

C. *The Diagonal Hessian Elements of $P_{NR}(\vec{\alpha}) = P(Ruin^C(\leq T_D))$*

From (3.9) and (4.10), the diagonal elements of $\bar{\bar{H}}$ take the form:

$$H_{t,t} = \int_{RF(0)}^{\infty} \int_{RF(1)}^{\infty} \cdots \int_{RF(T_D-1)}^{\infty} \left( \prod_{\substack{k=1 \\ k \neq t}}^{T_D} f(\hat{r}_{(k,\alpha_k)}) \right) \frac{\partial^2}{\partial \alpha_t^2} \left[ \frac{1}{\sqrt{2\pi v(\alpha_t)}} e^{-\left(\frac{\hat{r}_{(t,\alpha_t)} - m(\alpha_t)}{\sqrt{2v(\alpha_t)}}\right)^2} \right] d\vec{\hat{r}}. \tag{4.27}$$

Upon taking derivatives (see Appendix D), $H_{t,t}$ can be written as:

$$H_{t,t} = Q_t^{(1)} * \left[ \int_{RF(0)}^{\infty} \int_{RF(1)}^{\infty} \cdots \int_{RF(T_D-1)}^{\infty} \left( \prod_{\substack{k=1 \\ k \neq t}}^{T_D} f(\hat{r}_{(k,\alpha_k)}) \right) h_1(\hat{r}_{(t,\alpha_t)}) \, d\vec{\hat{r}} \right] \quad (4.28a)$$

$$+ Q_t^{(2)} * \left[ \int_{RF(0)}^{\infty} \int_{RF(1)}^{\infty} \cdots \int_{RF(T_D-1)}^{\infty} \left( \prod_{\substack{k=1 \\ k \neq t}}^{T_D} f(\hat{r}_{(k,\alpha_k)}) \right) h_2(\hat{r}_{(t,\alpha_t)}) \, d\vec{\hat{r}} \right] \quad (4.28b)$$

$$+ Q_t^{(3)} * \left[ \int_{RF(0)}^{\infty} \int_{RF(1)}^{\infty} \cdots \int_{RF(T_D-1)}^{\infty} \left( \prod_{k=1}^{T_D} f(\hat{r}_{(k,\alpha_k)}) \right) d\vec{\hat{r}} \right] \quad (4.28c)$$

where,

$$Q_t^{(1)} = \frac{\theta_t}{2v(\alpha_t)^2} + \frac{2v'(\alpha_t)^2 m'(\alpha_t)^2}{v(\alpha_t) * \theta_t}, \quad \text{for } t = 1, 2, \ldots, T_D, \quad (4.29)$$

$$h_1(\hat{r}_{(t,\alpha_t)}) = \frac{\left[ \left( \hat{r}_{(t,\alpha_t)} - m(\alpha_t) \right) - \left( \frac{2v(\alpha_t)v'(\alpha_t)m'(\alpha_t)}{\theta_t} \right) \right]^2}{v(\alpha_t) + \left[ \frac{2v(\alpha_t)v'(\alpha_t)m'(\alpha_t)}{\theta_t} \right]^2} f(\hat{r}_{(t,\alpha_t)}), \quad \text{for } -\infty < \hat{r}_{(t,\alpha_t)} < \infty \quad (4.30)$$

and,

$$Q_t^{(2)} = \frac{v'(\alpha_t)^2 + 2v(\alpha_t)m'(\alpha_t)^2}{2v(\alpha_t)^2}, \quad \text{for } t = 1, 2, \ldots, T_D, \quad (4.31)$$

$$h_2(\hat{r}_{(t,\alpha_t)}) = \frac{2 \left[ \frac{v'(\alpha_t)}{2v(\alpha_t)} \left( \hat{r}_{(t,\alpha_t)} - m(\alpha_t) \right)^2 + m'(\alpha_t) \left( \hat{r}_{(t,\alpha_t)} - m(\alpha_t) \right) - \frac{v'(\alpha_t)}{2} \right]^2}{v'(\alpha_t)^2 + 2v(\alpha_t)m'(\alpha_t)^2} f(\hat{r}_{(t,\alpha_t)}), \quad (4.32)$$

$$\text{for } -\infty < \hat{r}_{(t,\alpha_t)} < \infty$$

and,

$$Q_t^{(3)} = -\left[ \frac{v''(\alpha_t)v(\alpha_t) - v'(\alpha_t)^2 + 2v(\alpha_t)m'(\alpha_t)^2}{2v(\alpha_t)^2} + \frac{2v'(\alpha_t)^2 m'(\alpha_t)^2}{v(\alpha_t) * \theta_t} \right], \quad \text{for } t = 1, 2, \ldots, T_D, \quad (4.33)$$

$$\theta_t = v(\alpha_t)v''(\alpha_t) - 2v'(\alpha_t)^2, \quad \text{for } t = 1, 2, \ldots, T_D. \quad (4.34)$$

The quantity $\theta_t$ has been defined separately because initial inspection suggests $H_{t,t}$ is undefined at $\alpha_t$ where $\theta_t = 0$, see (4.29), (4.30), and (4.33). Fortunately this is not the case (however, we still cannot divide by zero in the code). The term $h_1(\hat{r}_{(t,\alpha_t)})$ is indeterminate ($\infty/\infty$) when $\theta_t \to 0$. Since the numerator and denominator approach $\infty$ at the same polynomial rate, L'Hôpital's rule indicates that the limit approaches the ratio of coefficients on the highest ordered terms, which here is 1, leaving $h_1(\hat{r}_{(t,\alpha_t)}) = f(\hat{r}_{(t,\alpha_t)})$. Further, as $\theta_t \to 0$, select terms from (4.28[a,c]) approach the following indeterminate form (0/0), which ultimately vanishes:

$$\frac{2v'(\alpha_t)^2 m'(\alpha_t)^2}{v(\alpha_t) * \theta_t} * [P_{NR}(\vec{\alpha}) - P_{NR}(\vec{\alpha})]. \tag{4.35}$$

The diagonal Hessian elements, $H_{t,t}$, in (4.28[a-c]) have been expressed in terms of $h_1(\hat{r}_{(t,\alpha_t)})$ and $h_2(\hat{r}_{(t,\alpha_t)})$ because they are valid PDFs (see Appendices E and F). Consequently, $H_{t,t}$ for $t=1, 2, \ldots, T_D$ is a linear function of valid success probabilities, therefore it exists. This validates the application of Leibniz's rule in (3.6). As before, we can estimate/approximate these probabilities using simulation or a DP. The term [·] in (4.28c) is $P_{NR}(\vec{\alpha})$, which we assume has already been calculated leaving 2 new glidepath success probabilities to compute for each of the $T_D$ elements, $H_{t,t}$. Using a DP will require the following CDFs for $h_1(\hat{r}_{(t,\alpha_t)})$ and $h_2(\hat{r}_{(t,\alpha_t)})$, denoted by $H_1(\hat{r}_{(t,\alpha_t)}) = P_{h_1}(\hat{r}_{(t,\alpha_t)} \le r)$ for $\hat{r}_{(t,\alpha_t)} \sim h_1(\hat{r}_{(t,\alpha_t)})$ and $H_2(\hat{r}_{(t,\alpha_t)}) = P_{h_2}(\hat{r}_{(t,\alpha_t)} \le r)$ for $\hat{r}_{(t,\alpha_t)} \sim h_2(\hat{r}_{(t,\alpha_t)})$, respectively (see Appendices G and H):

$$H_1(\hat{r}_{(t,\alpha_t)}) = P_{h_1}(\hat{r}_{(t,\alpha_t)} \le r) = H_{1,t}^{(0)}\left[H_{1,t}^{(1)}\left(1 - (-1)^{\mathbf{1}_{(m(\alpha_t),\infty)}(r)} F_{g[\frac{3}{2},1]}\left(\left(\frac{r - m(\alpha_t)}{\sqrt{2v(\alpha_t)}}\right)^2\right)\right)\right.$$
$$\left. + H_{1,t}^{(2)}\left(1 - F_{g[1,1]}\left(\left(\frac{r - m(\alpha_t)}{\sqrt{2v(\alpha_t)}}\right)^2\right)\right) + H_{1,t}^{(3)} \Phi\left(\frac{r - m(\alpha_t)}{\sqrt{v(\alpha_t)}}\right)\right], \tag{4.36}$$

where,

$$H_{1,t}^{(0)} = \left[v(\alpha_t) + \left(\frac{2v(\alpha_t)v'(\alpha_t)m'(\alpha_t)}{\theta_t}\right)^2\right]^{-1}, \quad H_{1,t}^{(1)} = \frac{v(\alpha_t)}{2},$$

$$H_{1,t}^{(2)} = \left(\frac{\sqrt{2v(\alpha_t)}^3 v'(\alpha_t)m'(\alpha_t)}{\theta_t\sqrt{\pi}}\right), \text{ and } H_{1,t}^{(3)} = \left(\frac{2v(\alpha_t)v'(\alpha_t)m'(\alpha_t)}{\theta_t}\right)^2,$$

(4.37)

and,

$$H_2(\hat{r}_{(t,\alpha_t)}) = P_{h_2}(\hat{r}_{(t,\alpha_t)} \leq r) = H_{2,t}^{(0)} \Bigg[ H_{2,t}^{(1)}\left(1 - (-1)^{\mathbf{1}_{(m(\alpha_t),\infty)}(r)} F_{g[\frac{5}{2},1]}\left(\left(\frac{r - m(\alpha_t)}{\sqrt{2v(\alpha_t)}}\right)^2\right)\right)$$

$$+ H_{2,t}^{(2)}\left(1 - F_{g[2,1]}\left(\left(\frac{r - m(\alpha_t)}{\sqrt{2v(\alpha_t)}}\right)^2\right)\right)$$

$$+ H_{2,t}^{(3)}\left(1 - (-1)^{\mathbf{1}_{(m(\alpha_t),\infty)}(r)} F_{g[\frac{3}{2},1]}\left(\left(\frac{r - m(\alpha_t)}{\sqrt{2v(\alpha_t)}}\right)^2\right)\right)$$

$$+ H_{2,t}^{(4)}\left(1 - F_{g[1,1]}\left(\left(\frac{r - m(\alpha_t)}{\sqrt{2v(\alpha_t)}}\right)^2\right)\right) + H_{2,t}^{(5)} \Phi\left(\frac{r - m(\alpha_t)}{\sqrt{v(\alpha_t)}}\right)\Bigg],$$

(4.38)

where,

$$H_{2,t}^{(0)} = \left(\frac{2}{v'(\alpha_t)^2 + 2v(\alpha_t)m'(\alpha_t)^2}\right), \quad H_{2,t}^{(1)} = \left(\frac{3v'(\alpha_t)^2}{8}\right),$$

$$H_{2,t}^{(2)} = -\frac{m'(\alpha_t)v'(\alpha_t)\sqrt{2v(\alpha_t)}}{\sqrt{\pi}}, \quad H_{2,t}^{(3)} = \left(\frac{2v(\alpha_t)m'(\alpha_t)^2 - v'(\alpha_t)^2}{4}\right)$$

(4.39)

$$H_{2,t}^{(4)} = \frac{m'(\alpha_t)v'(\alpha_t)\sqrt{v(\alpha_t)}}{\sqrt{2\pi}}, \text{ and } H_{2,t}^{(5)} = \frac{v'(\alpha_t)^2}{4}.$$

As before, the quantity $\mathbf{1}_{(m(\alpha_t),\infty)}(r)$ is an indicator function that equals 1 when $m(\alpha_t) < r < \infty$, and 0 otherwise. The CDFs $H_1(\hat{r}_{(t,\alpha_t)})$ and $H_2(\hat{r}_{(t,\alpha_t)})$ for PDFs $h_1(\hat{r}_{(t,\alpha_t)})$ and $h_2(\hat{r}_{(t,\alpha_t)})$ reveal themselves as linear combinations of known CDF calls that are trivial to implement, in practice.

## D. Characteristics of a Solution

In Section 3 we noted that minimizing a convex function Z over a convex feasible region is considered a convex programming problem, and that in such problems local optimums are global optimums. Since minimizing Z is equivalent to maximizing –Z, maximizing a concave function over a convex feasible region is itself a convex programming problem. With a simple transformation we can show that maximizing a log-concave function over a convex feasible region is also a convex programming problem (Lovász and Vempala (2006)). The same can be said for maximizing a strictly quasi-concave function over a convex feasible region. Maximizing just a quasi-concave function over a convex feasible region is almost, but not quite, a convex programming problem yet still has desirable properties. This is intuitive since quasi-concave functions can have plateaus with zero gradient, but the function increases again after. The problem defined in (3.1) is to maximize $Z=P_{NR}(\vec{\alpha})$ over a convex region and the goal here is to determine whether or not $P_{NR}(\vec{\alpha})$ falls on the concave function spectrum for retirement horizons of length $T_D$ and withdrawal rates $W_R=RF(0)$. The withdrawal rate must be reasonable since $P_{NR}(\vec{\alpha}) \to 1.0$ as $RF(0) \to 0.0$ and $P_{NR}(\vec{\alpha}) \to 0.0$ as $RF(0) \to \infty$. At these limits, all glidepaths either succeed or fail. Lastly, if we can demonstrate that $P_{NR}(\vec{\alpha})$ is not always quasi-concave then it does not always fall on the spectrum defined in Section II.N and solutions found using the technique proposed here may reflect local optimums. From (2.68), (3.1), and (4.10), we have:

$$P_{NR}(\vec{\alpha}) = \int_{\hat{R}} \left(\frac{1}{\sqrt{2\pi}}\right)^{T_D} \prod_{t=1}^{T_D} \left(\frac{1}{\sqrt{v(\alpha_t)}}\right) e^{-\frac{1}{2}\sum_{t=1}^{T_D}\left(\frac{\hat{r}_{(t,\alpha_t)} - m(\alpha_t)}{\sqrt{v(\alpha_t)}}\right)^2} d\vec{\hat{r}}, \qquad (4.40)$$

for $-\infty < \hat{r}_{(t,\alpha_t)} < \infty$, $MV(\alpha) < \alpha_t \leq 1.0$, and t=1, 2, …, $T_D$. Gardner (2002) details the following consequence of Prékopa-Leindler's inequality. In words, if a log-concave function of many variables has some integrated out over an open convex set then the resulting function in the remaining variables is itself log-concave. Therefore, if we can show that the kernel of (4.40) is log-concave in both $\vec{\hat{r}}$ and $\vec{\alpha}$, we are done since it was shown in Section II.M that $\hat{R}$ is an open convex set.[9] Unfortunately, it is not. In fact, we can demonstrate via counter-example that the kernel of (4.40) is not even quasi-concave for all reasonable $T_D$ and $RF(0)$, therefore cannot be log-concave. This is not surprising since a plot of $\log[P_{NR}(\vec{\alpha})]$ for select $RF(0)$ when $T_D=1$ reveals that it has inflection points. From (4.10) we assume $\hat{r}_{(t,\alpha_t)} \sim iid\ N(m(\alpha_t), v(\alpha_t))$, therefore $\hat{r}_{(t,\alpha_t)} = \sqrt{v(\alpha_t)}*z_t + m(\alpha_t)$ where $z_t \sim iid\ N(0, 1)$, for $t=1, 2, \ldots, T_D$. Further, the set of inflation/expense-adjusted returns $\hat{R}$ from (2.68) that avoid financial ruin in retirement can be expressed in terms of standardized returns as $Z$ where:

$$Z = \{z_1, z_2, \ldots, z_{T_D} : \cap_{t=1}^{T_D}(z_t > ZF(t-1))\} \tag{4.41}$$

with,

$$ZF(t-1) = \frac{RF(t-1) - m(\alpha_t)}{\sqrt{v(\alpha_t)}} \tag{4.42}$$

and,

$$RF(t) = \frac{RF(t-1)}{\left[\sqrt{v(\alpha_t)} * z_t + m(\alpha_t)\right] - RF(t-1)}, \tag{4.43}$$

for $t=1, 2, \ldots, T_D$. Under this formulation, (4.40) can be expressed as:

$$P_{NR}(\vec{\alpha}) = \int_Z \left(\frac{1}{\sqrt{2\pi}}\right)^{T_D} e^{-\frac{1}{2}\sum_{t=1}^{T_D} z_t^2}\, d\vec{z}. \tag{4.44}$$

The kernel in (4.44) is now log-concave since $z_t^2/2$ is convex therefore $-z_t^2/2$ is concave, and the product of log-concave functions is log-concave (Boyd and Vandenberghe (2004)). The set $Z$ however, is not convex therefore this reformulation does not help us apply Prékopa's theorems, but it will allow sufficient progress to be made. The function $P_{NR}(\vec{\alpha})$ from (4.44) will be analyzed case-by-case for increasing retirement horizon lengths, starting with $T_D=1$.

---

[9] Not quite done, but almost. For some withdrawal rates $W_R=RF(0)$ and horizon lengths $T_D$, the function could be log-concave but with peak outside of our feasible region. In this case we are left operating along a boundary using either gradient ascent or Newton's method. At the boundary the gradient will continue to point in the direction of steepest ascent but the constraints prohibit us from climbing in that direction along certain dimensions. (See Section IV.E.8.)

Case 1:  $T_D=1$

For a single-period retirement, financial ruin is avoided if $z_1>ZF(0)$ with probability:

$$P_{NR}(\alpha_1) = \int_{ZF(0)}^{\infty} \left(\frac{1}{\sqrt{2\pi}}\right) e^{-\frac{1}{2}z_1^2} dz_1. \tag{4.45}$$

To show that $P_{NR}(\alpha_1)$ is strictly quasi-concave it suffices to show that $ZF(0)$ is strictly quasi-convex in $\alpha_1$. This is because we are computing the probability of being to the right of a function of $\alpha_1$. If this function evaluated at any convex combination of two $\alpha$'s is not the largest value, then the corresponding probability cannot be the smallest value. Let $\alpha_{11}$ and $\alpha_{21}$ be two equity ratios between $MV(\alpha_1)$ and 1.0, and let $\alpha_{c1}$ reflect the convex combination $\lambda\alpha_{11} + (1-\lambda)\alpha_{21}$ $\forall\ 0 \leq \lambda \leq 1$. Further, let $ZF_1(0)$, $ZF_2(0)$, and $ZF_c(0)$ be the respective standardized withdrawal rates. By definition, $ZF(0)$ is strictly quasi-convex if $ZF_c(0) < \text{Max}\{ZF_1(0), ZF_2(0)\}$ $\forall\ \alpha_{11}, \alpha_{21}, \lambda$, and $RF(0)$. This condition can only be violated if $ZF(0)$ has a local maximum between $MV(\alpha_1)$ and 1.0 for some $RF(0) > 0$. A plot of $ZF(0)$ for various $RF(0)=W_R$ is shown below in Figure 3.

Local optimums of $ZF(0)$ will occur at critical points where the first derivative equals zero and to prove our conjecture we must show that these points always reflect local minimums between $MV(\alpha_1)$ and 1.0. The first derivative of $ZF(0)$ with respect to $\alpha_1$ is given by:

$$\frac{d}{d\alpha_1}[ZF(0)] = \frac{[m(\alpha_1) - RF(0)]v'(\alpha_1)}{2\sqrt{v(\alpha_1)}^3} - \frac{m'(\alpha_1)}{\sqrt{v(\alpha_1)}}, \tag{4.46}$$

which equals zero when:

$$[m(\alpha_1) - RF(0)]v'(\alpha_1) - 2v(\alpha_1)m'(\alpha_1) = 0. \tag{4.47}$$

Solving this equation for $\alpha_1$ using the expressions derived in (4.3), (4.4), (4.6), and (4.7) yields the following single critical point, $\alpha_1^*$:

$$\alpha_1^* = \frac{(\mu_s - \mu_b)\sigma_b^2 - \left[1 + \mu_b - \frac{RF(0)}{(1-E_R)}\right](\sigma_{(s,b)} - \sigma_b^2)}{\left[1 + \mu_b - \frac{RF(0)}{(1-E_R)}\right](\sigma_s^2 + \sigma_b^2 - 2\sigma_{(s,b)}) - (\mu_s - \mu_b)(\sigma_{(s,b)} - \sigma_b^2)}. \tag{4.48}$$

# Figure 3
# The Standardized Withdrawal Rate for a Single-Period Retirement Horizon

This figure depicts the standardized withdrawal rate ZF(0) as a function of the equity ratio $\alpha_1$ at time t=1 for various initial withdrawal rates RF(0) (=$W_R$). By showing that ZF(0) is strictly quasi-convex we show that $P_{NR}(\alpha_1)$ is strictly quasi-concave for a single-period retirement. The large dots represent local minimums and importantly there are no local maximums between MV($\alpha_1$) and 1.0. Since there are no local maximums $ZF_c(0) < \text{Max}\{ZF_1(0), ZF_2(0)\}$ and ZF(0) is strictly quasi-convex as a function of $\alpha_1$ for a single-period retirement ($T_D$=1).

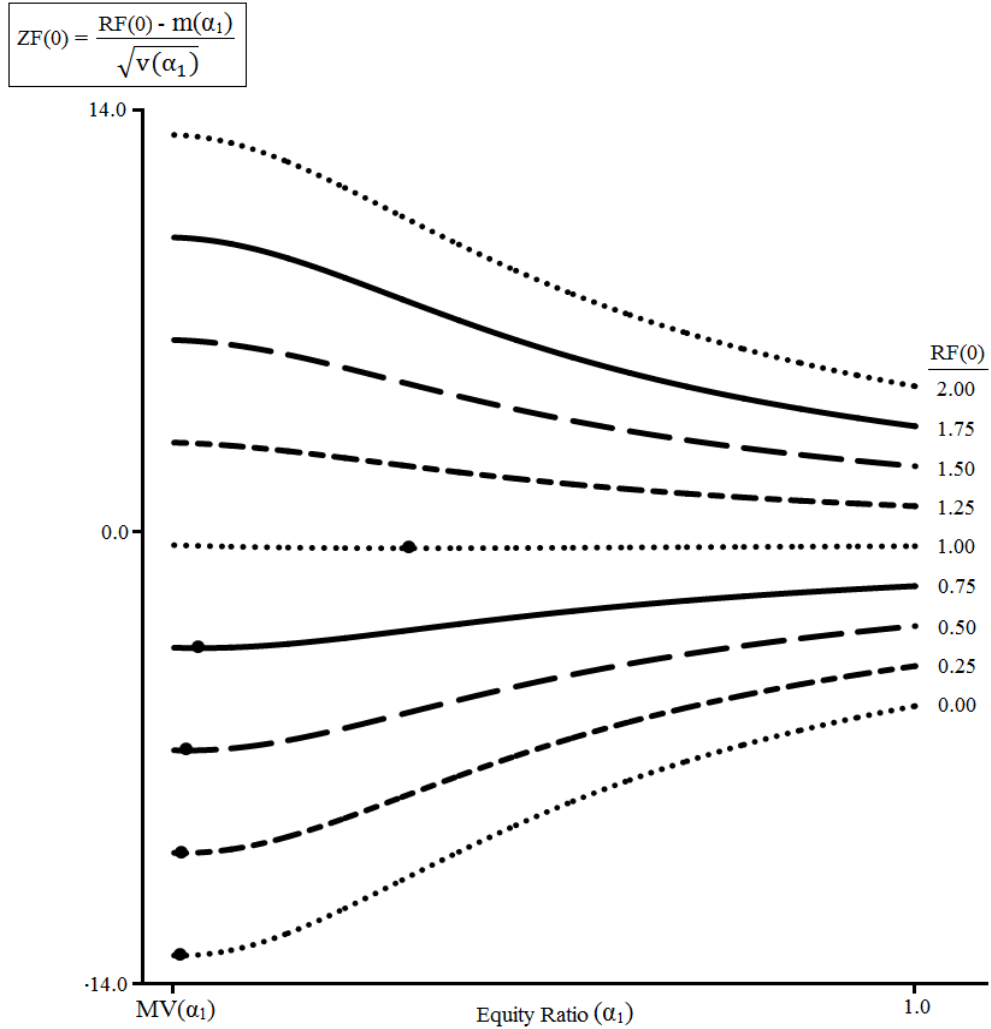

The critical points $\alpha_1^*$ are represented by the large dots in Figure 3. The 2nd derivative of $ZF(0)$ with respect to $\alpha_1$ is given by:

$$\frac{d^2}{d\alpha_1^2}[ZF(0)] = \frac{[m(\alpha_1) - RF(0)]\left[v(\alpha_1)v''(\alpha_1) - \frac{3}{2}v'(\alpha_1)^2\right] + 2m'(\alpha_1)v(\alpha_1)v'(\alpha_1)}{2\sqrt{v(\alpha_1)}^5}, \quad (4.49)$$

which is always $> 0$ when evaluated at $\alpha_1^*$ between $MV(\alpha_1)$ and 1.0 (confirmed numerically). Therefore, all local optimums of $ZF(0)$ between $MV(\alpha_1)$ and 1.0 are minimums and $ZF(0)$ is strictly quasi-convex making $P_{NR}(\alpha_1)$ strictly quasi-concave. Consequently, for a single-period retirement, any locally optimal glidepath is also globally optimal.

Case 2: $T_D=2$

For a 2-period retirement, financial ruin is avoided if $(z_1 > ZF(0) \cap z_2 > ZF(1))$, which occurs with probability:

$$P_{NR}(\vec{\alpha}) = \int_{ZF(0)}^{\infty} \int_{ZF(1)}^{\infty} \left(\frac{1}{\sqrt{2\pi}}\right)^2 e^{-\frac{1}{2}\sum_{t=1}^{2} z_t^2} \, dz_2 dz_1. \quad (4.50)$$

This probability falls on the concave function spectrum if it is at least quasi-concave in $\vec{\alpha}$. The situation is depicted in Figure 4 below. For reasonable $T_D$ and $RF(0)$, $P_{NR}(\vec{\alpha}_c) \geq \text{Min}\{P_{NR}(\vec{\alpha}_1), P_{NR}(\vec{\alpha}_2)\}$, $\forall \ 0 \leq \lambda \leq 1$ must hold where:

$$\vec{\alpha}_1 = \begin{pmatrix}\alpha_{11}\\\alpha_{12}\end{pmatrix}, \ \vec{\alpha}_2 = \begin{pmatrix}\alpha_{21}\\\alpha_{22}\end{pmatrix}, \ \vec{\alpha}_c = \begin{pmatrix}\alpha_{c1}\\\alpha_{c2}\end{pmatrix} = \begin{pmatrix}\lambda\alpha_{11} + (1-\lambda)\alpha_{21}\\\lambda\alpha_{12} + (1-\lambda)\alpha_{22}\end{pmatrix}. \quad (4.51)$$

**Figure 4**
**Comparing Glidepaths for a Two-Period Retirement Horizon**

This figure depicts (4.50) graphically for two glidepaths $\vec{\alpha}_1$, $\vec{\alpha}_2$ and a given convex combination $\vec{\alpha}_c$. Retirement ruin is avoided when $(z_1, z_2)$ falls above and to the right of each curve which reflects the equation $z_2 = ZF_i(1)$ for $i = 1, 2, c$. The corresponding probability is the volume of the joint density $\phi(\cdot)$ above and to the right of each curve. The curve associated with the largest volume represents the glidepath with the greatest probability of avoiding ruin. The circles represent density contours out to 5 standard deviations and over 99.99% of the total probability is contained within the lightly drawn box around the largest circle. When comparing glidepaths it suffices to focus on the region between the curves. We can approximate this region's volume by constructing a grid of narrow rectangles as depicted. We then formulate a deterministic NLP with the goal of finding $\vec{\alpha}_1$, $\vec{\alpha}_2$ and $\vec{\alpha}_c$ such that $P_{NR}(\vec{\alpha}_c) < \text{Min}\{P_{NR}(\vec{\alpha}_1), P_{NR}(\vec{\alpha}_2)\}$. The formal objective is to minimize $Z = P_{NR}(\vec{\alpha}_c) - P_{NR}(\vec{\alpha}_2)$ subject to $P_{NR}(\vec{\alpha}_1) - P_{NR}(\vec{\alpha}_c) > 0$. A solution $Z < 0$ would refute the claim that $P_{NR}(\vec{\alpha})$ is quasi-concave. The black dots indicate intersection points and when comparing only 2 glidepaths there may be zero, one or two points of intersection. The blue dots demonstrate that the set $Z$ is not convex, as was stated above. That is, there exist standardized returns $(z_{11}, z_{12})$ and $(z_{21}, z_{22})$ that succeed for glidepaths $\vec{\alpha}_1$ and $\vec{\alpha}_2$, respectively, but a convex combination $(z_{c1}, z_{c2})$ fails for glidepath $\vec{\alpha}_c$.

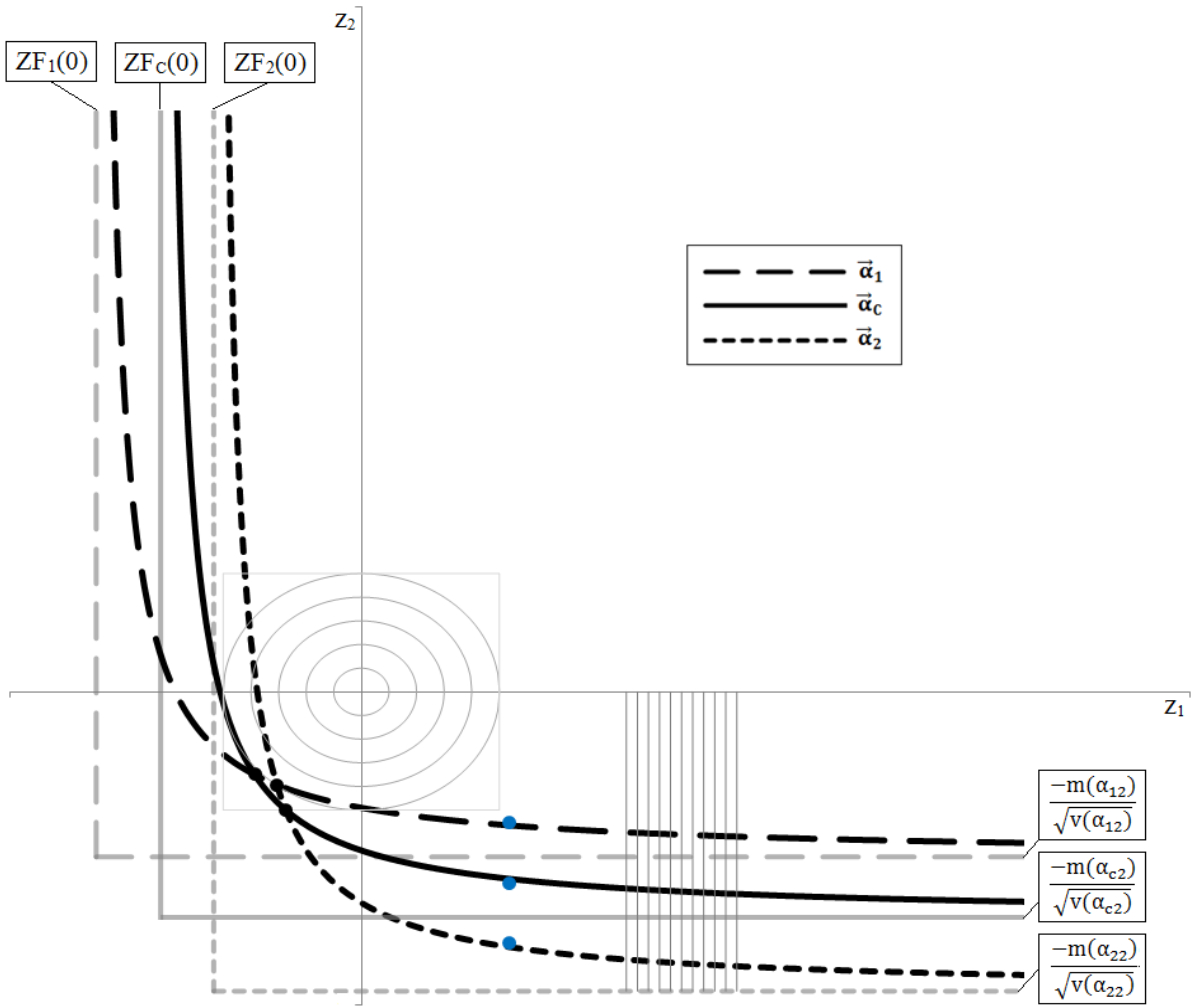

Our goal is to express the difference in glidepath success probabilities compactly then formulate a deterministic NLP with the objective of minimizing Z = $P_{NR}(\vec{\alpha}_c)$ - $P_{NR}(\vec{\alpha}_2)$ subject to $P_{NR}(\vec{\alpha}_1)$ - $P_{NR}(\vec{\alpha}_c)$ > 0. A solution with Z < 0 invalidates the claim that $P_{NR}(\vec{\alpha})$ is quasi-concave as it would imply that the probability of avoiding ruin in retirement using the convex combination glidepath $\vec{\alpha}_c$ is lower than the probability with either $\vec{\alpha}_1$ or $\vec{\alpha}_2$. This would imply that, as a surface, $P_{NR}(\vec{\alpha})$ has a dip or valley and therefore cannot be unimodal. The following compact expression for $P_{NR}(\vec{\alpha}_1)$ - $P_{NR}(\vec{\alpha}_2)$ applies in general to the difference in probabilities between any two glidepaths when $T_D=2$. Define the function $F_i(z_1)$, for i=1,2 as:

$$F_i(z_1) = \begin{cases} 1 & \text{for } z_1 \leq ZF_i(0) \\ \Phi(ZF_i(1)) & \text{for } z_1 > ZF_i(0) \end{cases}, \quad (4.52)$$

recalling that $ZF_i(1) = \frac{RF_i(1) - m(\alpha_{i2})}{\sqrt{v(\alpha_{i2})}}$, where $RF_i(1) = \frac{RF(0)}{\sqrt{v(\alpha_{i1})} \ast Z_1 + m(\alpha_{i1}) - RF(0)}$ is a function of $z_1$.

For any 2 glidepaths $\vec{\alpha}_1$ and $\vec{\alpha}_2$, $P_{NR}(\vec{\alpha}_1)$ - $P_{NR}(\vec{\alpha}_2)$ is given by:

$$P_{NR}(\vec{\alpha}_1) - P_{NR}(\vec{\alpha}_2) = \int_{ZF_1(0)}^{\infty}\int_{ZF_1(1)}^{\infty} \phi(z_1)\phi(z_2)\,dz_2 dz_1 - \int_{ZF_2(0)}^{\infty}\int_{ZF_2(1)}^{\infty} \phi(z_1)\phi(z_2)\,dz_2 dz_1 \quad (4.53)$$

$$= \int_{-\infty}^{Min(ZF_1(0), ZF_2(0))} \phi(z_1)[F_2(z_1) - F_1(z_1)]\,dz_1 \quad (4.54a)$$

$$+ \int_{Min(ZF_1(0), ZF_2(0))}^{Max(ZF_1(0), ZF_2(0))} \phi(z_1)[F_2(z_1) - F_1(z_1)]\,dz_1 \quad (4.54b)$$

$$+ \int_{Max(ZF_1(0), ZF_2(0))}^{\infty} \phi(z_1)[F_2(z_1) - F_1(z_1)]\,dz_1 \quad (4.54c)$$

$$= \int_{-\infty}^{\infty} \phi(z_1)[F_2(z_1) - F_1(z_1)]\,dz_1 \quad (4.55)$$

$$= E_{z_1}[F_2(z_1) - F_1(z_1)]. \quad (4.56)$$

To justify this development, split the $z_1$-axis into the following 3 sections: (a) $z_1 \leq \text{Min}\{ZF_1(0), ZF_2(0)\}$, (b) $\text{Min}\{ZF_1(0), ZF_2(0)\} < z_1 < \text{Max}\{ZF_1(0), ZF_2(0)\}$, and (c) $z_1 \geq \text{Max}\{ZF_1(0), ZF_2(0)\}$. In section (a) there is no volume of interest and $F_2(z_1) - F_1(z_1) = 0$ as desired (4.54a). Section (c) reflects (4.54c) and contains both curves. The outer integral is with respect to $z_1$ for $z_1 \in [\text{Max}\{ZF_1(0), ZF_2(0)\}, \infty)$ and the inner integral takes the form:

$$\int_{ZF_1(1)}^{\infty} \phi(z_2)\, dz_2 - \int_{ZF_2(1)}^{\infty} \phi(z_2)\, dz_2 = [1 - \Phi(ZF_1(1))] - [1 - \Phi(ZF_2(1))] \quad (4.57)$$

$$= \Phi(ZF_2(1)) - \Phi(ZF_1(1)) = F_2(z_1) - F_1(z_1). \quad (4.58)$$

Lastly, section (b) reflects (4.54b) and contains only one curve. If that curve is for glidepath $\vec{\alpha}_1$ then $F_2(z_1) = 1$ and the extra term in (4.53) is on the left side of the minus sign. The outer integral is with respect to $z_1$ for $z_1 \in (ZF_1(0), ZF_2(0))$ and the inner integral takes the form:

$$\int_{ZF_1(1)}^{\infty} \phi(z_2)\, dz_2 = 1 - \Phi(ZF_1(1)) = F_2(z_1) - F_1(z_1). \quad (4.59)$$

Conversely, if that curve is for glidepath $\vec{\alpha}_2$ then $F_1(z_1) = 1$ and the extra term in (4.53) is on the right side of the minus sign. The outer integral is with respect to $z_1$ for $z_1 \in (ZF_2(0), ZF_1(0))$ and the inner integral takes the form:

$$-\int_{ZF_2(1)}^{\infty} \phi(z_2)\, dz_2 = -[1 - \Phi(ZF_2(1))] = [\Phi(ZF_2(1)) - 1] = F_2(z_1) - F_1(z_1). \quad (4.60)$$

The expected value of $[F_2(z_1) - F_1(z_1)]$ with respect to $z_1$ can be approximated using the grid technique shown in Figure 4. If k rectangles are being drawn between fixed points $Z_L < 0$ and $Z_U > 0$ then the rectangle width is $w = \frac{Z_U - Z_L}{k}$ and:

$$E_{z_1}[F_2(z_1) - F_1(z_1)] \cong \sum_{r=1}^{k} (P_r) * [F_2(z_{1r}^*) - F_1(z_{1r}^*)] \quad (4.61)$$

where $P_r = P[Z_L + (r-1)w < z_1 < Z_L + (r)w]$ is the probability that $z_1$ falls in the $r^{th}$ rectangle and $z_{1r}^*$

is the mid-point of each rectangle, namely, $z_{1r}^* = Z_L+(r-1)w + \frac{w}{2}$, for r=1, 2, …, k. In vector notation, we can define $\vec{P}$ and $\vec{F}_{(1:2)}$ as:

$$\vec{P} = \begin{pmatrix} \Phi(Z_L + (1)w) - \Phi(Z_L + (0)w) \\ \Phi(Z_L + (2)w) - \Phi(Z_L + (1)w) \\ \vdots \\ \Phi(Z_L + (k)w) - \Phi(Z_L + (k-1)w) \end{pmatrix}, \tag{4.62}$$

and,

$$\vec{F}_{(1:2)} = \begin{pmatrix} F_2(z_{11}^*) - F_1(z_{11}^*) \\ F_2(z_{12}^*) - F_1(z_{12}^*) \\ \vdots \\ F_2(z_{1k}^*) - F_1(z_{1k}^*) \end{pmatrix}. \tag{4.63}$$

Then,

$$E_{z_1}[F_2(z_1) - F_1(z_1)] \cong \vec{P}^T * \vec{F}_{(1:2)}. \tag{4.64}$$

Note that $\vec{P}$ is a vector of constant probabilities that does not change once the grid has been drawn, and it should be constructed first. The deterministic NLP below can then be formulated in any non-linear solver, for example, we used Excel:

Minimize:     $Z = \vec{P}^T * \vec{F}_{(C:2)}$ (4.65)

Subject to:     $\vec{P}^T * \vec{F}_{(1:C)} > 0$ (4.66)

For:     $MV(\alpha) < \alpha_{ij} \leq 1.0,$     i=1,2 and j=1,2 (4.67a)

$\alpha_{cj} = \lambda\alpha_{1j} + (1-\lambda)\alpha_{2j},$     j=1,2 (4.67b)

$RF(0) > 0, 0 \leq \lambda \leq 1$ (4.67c)

Any solution $Z < 0$ invalidates the claim that $P_{NR}(\vec{\alpha})$ from (4.40) is quasi-concave. Such solutions with reasonable $W_R=RF(0)$ do exist and a specific example is detailed in Appendix I. The probability of avoiding ruin in retirement is therefore not necessarily unimodal as a function of the glidepath. Consequently, the optimization problem in (3.1) is not necessarily convex and we cannot guarantee that local optimums are always global optimums.

*E. Examples*

A full C++ implementation of the optimization technique presented in this research is included in Appendix J. There are 4 ways to perform the optimization: (1) Newton's method using a DP, (2) Newton's method using simulation, (3) gradient ascent using a DP, and, (4) gradient ascent using simulation. Refer to Sections II.I and II.J for more details on the two optimization methods, and Section II.D for details on the two estimation methods. Approach (1) usually converges fastest and produces the most accurate estimates, but is not well defined when operating along a boundary region. It reflects the approach used for all scenarios presented in this section except Scenario 8 which uses approach (3). Additional details are provided in the appendix. In this section we will analyze 8 different scenarios as described in Table 1 below.

**Table 1**
**Assumptions used for Fixed TD Glidepath Optimization Scenarios**

This table defines the assumptions for each of the 8 scenarios presented in this section. Historical returns were derived from NYU Professor Aswath Damodaran's website of stock (S&P 500) and bond (10-year Treasury) total returns from 1928-2013, see Rook (2014). The parameters for *Historical* real returns are: $\mu_s = 0.0825$, $\mu_b = 0.0214$, $\sigma^2_s = 0.0403$, $\sigma^2_b = 0.0070$, $\sigma_{(s,b)} = 0.0007$. Inflation adjustments were made using CPI-U data retrieved from the Federal Reserve's Bank of Minneapolis website. The *Evensky* assumptions reflect lower returns and were drawn from Scenario A of Pfau and Kitces (2014) who retrieved these assumptions from a 2013 version of the MoneyGuidePro™ software package. (<u>Note:</u> They assume lognormal returns and we assume normal.) Parameters for these real returns are: $\mu_s = 0.0550$, $\mu_b = 0.0175$, $\sigma^2_s = 0.0428$, $\sigma^2_b = 0.0042$, $\sigma_{(s,b)} = 0.0040$. The procedure requires that the user specify a starting point and each scenario will start at the following 5 general glidepaths ($\vec{\alpha}$): (1) rising, (2) declining, (3) constant, (4) random #1, and (5) random #2. Each scenario then reports the optimal glidepath the procedure converges to. Different solutions indicate the presence of multiple local optimums.

| Scenario # | Real Return Assumptions | Retirement Length ($T_D$) | Withdrawal Rate ($W_R$=RF(0)) | Expense Ratio ($E_R$) |
|---|---|---|---|---|
| 1 | Historical | 30 yrs | 4.0% | 0.0% |
| 2 | Historical | 30 yrs | 4.0% | 1.0% |
| 3 | Evensky | 30 yrs | 4.0% | 0.0% |
| 4 | Evensky | 30 yrs | 4.0% | 1.0% |
| 5 | Historical | 30 yrs | 5.0% | 0.0% |
| 6 | Historical | 30 yrs | 5.0% | 1.0% |
| 7 | Evensky | 30 yrs | 5.0% | 0.0% |
| 8 | Evensky | 30 yrs | 5.0% | 1.0% |

The 5 starting glidepaths used for each of the 8 scenarios did not change per scenario and are shown below in Figure 5.

**Figure 5**
**Starting Glidepaths for all 8 Optimization Scenarios**

This figure depicts the starting glidepaths used for each of the 8 scenarios described in Table 1 above. The rising glidepath begins at 30.5% equities and increases by 1% per year. The declining glidepath begins at 59.5% equities and declines by 1% each year. The constant glidepath is fixed at 45% equities and the final 2 starting glidepaths are completely random but generated so that they are greater than the MV($\alpha$) for both sets of return assumptions. <u>Note</u>: We start the procedure at various glidepaths in an attempt to find different local optimums (i.e., ending glidepaths).

$$\vec{\alpha}_1 = \begin{pmatrix} 0.305 \\ 0.315 \\ 0.325 \\ 0.335 \\ 0.345 \\ 0.355 \\ 0.365 \\ 0.375 \\ 0.385 \\ 0.395 \\ 0.405 \\ 0.415 \\ 0.425 \\ 0.435 \\ 0.445 \\ 0.455 \\ 0.465 \\ 0.475 \\ 0.485 \\ 0.495 \\ 0.505 \\ 0.515 \\ 0.525 \\ 0.535 \\ 0.545 \\ 0.555 \\ 0.565 \\ 0.575 \\ 0.585 \\ 0.595 \end{pmatrix}_{\text{Rising GP}}, \vec{\alpha}_2 = \begin{pmatrix} 0.595 \\ 0.585 \\ 0.575 \\ 0.565 \\ 0.555 \\ 0.545 \\ 0.535 \\ 0.525 \\ 0.515 \\ 0.505 \\ 0.495 \\ 0.485 \\ 0.475 \\ 0.465 \\ 0.455 \\ 0.445 \\ 0.435 \\ 0.425 \\ 0.415 \\ 0.405 \\ 0.395 \\ 0.385 \\ 0.375 \\ 0.365 \\ 0.355 \\ 0.345 \\ 0.335 \\ 0.325 \\ 0.315 \\ 0.305 \end{pmatrix}_{\text{Declining GP}}, \vec{\alpha}_3 = \begin{pmatrix} 0.450 \\ 0.450 \\ 0.450 \\ 0.450 \\ 0.450 \\ 0.450 \\ 0.450 \\ 0.450 \\ 0.450 \\ 0.450 \\ 0.450 \\ 0.450 \\ 0.450 \\ 0.450 \\ 0.450 \\ 0.450 \\ 0.450 \\ 0.450 \\ 0.450 \\ 0.450 \\ 0.450 \\ 0.450 \\ 0.450 \\ 0.450 \\ 0.450 \\ 0.450 \\ 0.450 \\ 0.450 \\ 0.450 \\ 0.450 \end{pmatrix}_{\text{Constant GP}}, \vec{\alpha}_4 = \begin{pmatrix} 0.636 \\ 0.214 \\ 0.193 \\ 0.637 \\ 0.626 \\ 0.597 \\ 0.943 \\ 0.877 \\ 0.254 \\ 0.823 \\ 0.903 \\ 0.294 \\ 0.444 \\ 0.513 \\ 0.529 \\ 0.160 \\ 0.564 \\ 0.293 \\ 0.698 \\ 0.228 \\ 0.311 \\ 0.776 \\ 0.689 \\ 0.764 \\ 0.596 \\ 0.793 \\ 0.911 \\ 0.624 \\ 0.709 \\ 0.205 \end{pmatrix}_{\text{Random \#1 GP}}, \vec{\alpha}_5 = \begin{pmatrix} 0.813 \\ 0.886 \\ 0.227 \\ 0.684 \\ 0.328 \\ 0.379 \\ 0.484 \\ 0.145 \\ 0.763 \\ 0.284 \\ 0.690 \\ 0.476 \\ 0.876 \\ 0.649 \\ 0.147 \\ 0.643 \\ 0.521 \\ 0.662 \\ 0.161 \\ 0.864 \\ 0.867 \\ 0.332 \\ 0.281 \\ 0.224 \\ 0.471 \\ 0.777 \\ 0.922 \\ 0.880 \\ 0.295 \\ 0.860 \end{pmatrix}_{\text{Random \#2 GP}}$$

## E.1 Scenario #1: Historical Real Returns, $T_D=30$ yrs, $W_R=4\%$, and $E_R=0.0\%$

Let $\vec{\alpha}_{S_1}$ be the optimal static glidepath, and $P_{NR}(\vec{\alpha}_{S_1})$ be the corresponding probability of avoiding financial ruin for Scenario #1. We used a DP to estimate the probabilities and Newton's method to perform the optimization. All starting glidepaths from Figure 5 converge to the same solution at $\varepsilon = (0.1)^{11}$, which is in a region of $P_{NR}(\vec{\alpha})$ where the Hessian is negative definite, thus concave. This is empirical evidence of an interior point optimum for this scenario. (See Figure 6.1 below.) The optimal glidepath is rising and appears to be slightly convex but almost linear. The first equity ratio at time (year) t=1 is about 36.85% and the last equity ratio at time (year) t=30 is about 77.66%. The success rate using this glidepath is approximately 91.97%.

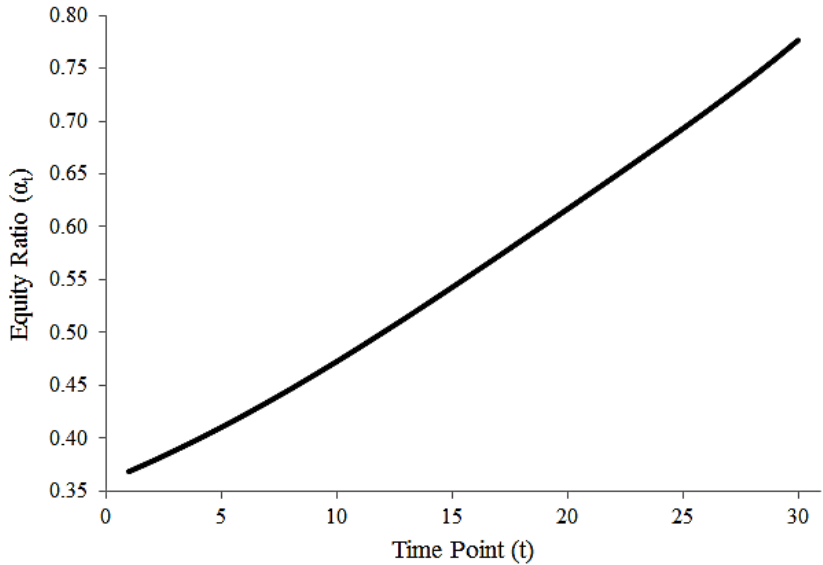

**Figure 6.1**
**Scenario #1: Optimal Glidepath**

Optimal GP
$P_{NR}(\vec{\alpha}_{S_1}) = 0.9196892347$

$$\vec{\alpha}_{S_1} = \begin{pmatrix} 0.3684826900 \\ 0.3782146268 \\ 0.3884336499 \\ 0.3991332015 \\ 0.4103025899 \\ 0.4219268946 \\ 0.4339869745 \\ 0.4464596018 \\ 0.4593177397 \\ 0.4725309778 \\ 0.4860661326 \\ 0.4998880110 \\ 0.5139603274 \\ 0.5282467547 \\ 0.5427120841 \\ 0.5573234631 \\ 0.5720516836 \\ 0.5868724955 \\ 0.6017679310 \\ 0.6167276424 \\ 0.6317502766 \\ 0.6468449369 \\ 0.6620328225 \\ 0.6773491818 \\ 0.6928457918 \\ 0.7085942800 \\ 0.7246907811 \\ 0.7412627184 \\ 0.7584790253 \\ 0.7765660632 \end{pmatrix}$$

This glidepath was derived using Newton's method with DP approximation having discretization level of 5,000. (Increase this value to reduce approximation error.) Convergence is achieved at $\varepsilon = (0.1)^{11}$ in 3 or 4 iterations, which takes 1 or 2 hours. This $\varepsilon$-level provides accuracy to about 10 decimal places in the glidepath, which is unnecessary in practice. All starting glidepaths converge to the same solution which exists in a concave region of the function and is empirical evidence of an interior point optimum. As shown below, the optimal glidepath is rising and appears slightly convex but nearly linear.

## E.2 Scenario #2: Historical Real Returns, $T_D$=30 yrs, $W_R$=4%, and $E_R$=1.0%

Let $\vec{\alpha}_{S_2}$ be the optimal static glidepath, and $P_{NR}(\vec{\alpha}_{S_2})$ be the corresponding probability of avoiding financial ruin for Scenario #2. We used a DP to estimate the probabilities and Newton's method to perform the optimization. All starting glidepaths from Figure 5 converge to the same solution at $\varepsilon = (0.1)^{11}$, which is in a region of $P_{NR}(\vec{\alpha})$ where the Hessian is negative definite, thus concave. This is empirical evidence of an interior point optimum for this scenario. (See Figure 6.2 below.) The optimal glidepath is rising and starts out linear, then becomes concave. The first equity ratio at time (year) t=1 is about 48.16% and the last equity ratio at time (year) t=30 is about 79.03%. The success rate using this glidepath is approximately 83.82%.

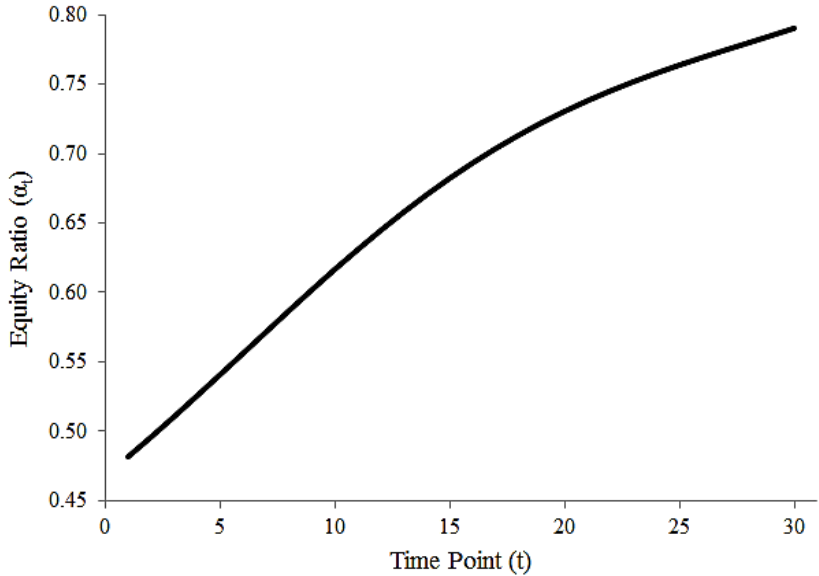

**Figure 6.2
Scenario #2: Optimal Glidepath**

Optimal GP

$P_{NR}(\vec{\alpha}_{S_2}) = 0.8382288556$

$$\vec{\alpha}_{S_2} = \begin{pmatrix} 0.4815660184 \\ 0.4957842116 \\ 0.5104349622 \\ 0.5254321640 \\ 0.5406767663 \\ 0.5560591499 \\ 0.5714623484 \\ 0.5867659540 \\ 0.6018504457 \\ 0.6166016113 \\ 0.6309147086 \\ 0.6446980372 \\ 0.6578756689 \\ 0.6703891904 \\ 0.6821984320 \\ 0.6932812645 \\ 0.7036326286 \\ 0.7132630093 \\ 0.7221965828 \\ 0.7304692477 \\ 0.7381267277 \\ 0.7452228904 \\ 0.7518183982 \\ 0.7579797780 \\ 0.7637789865 \\ 0.7692935574 \\ 0.7746074485 \\ 0.7798127805 \\ 0.7850127956 \\ 0.7903265702 \end{pmatrix}$$

This glidepath was derived using Newton's method with DP approximation having discretization level of 5,000. (Increase this value to reduce approximation error.) Convergence is achieved at $\varepsilon = (0.1)^{11}$ in 3 or 4 iterations, which takes 1 or 2 hours. This $\varepsilon$-level provides accuracy to about 10 decimal places in the glidepath, which is unnecessary in practice. All starting glidepaths converge to the same solution which exists in a concave region of the function and is empirical evidence of an interior point optimum. As shown below, the optimal glidepath is rising and mostly concave.

## E.3 Scenario #3: Evensky Real Returns, $T_D$=30 yrs, $W_R$=4%, and $E_R$=0.0%

Let $\vec{\alpha}_{S_3}$ be the optimal static glidepath, and $P_{NR}(\vec{\alpha}_{S_3})$ be the corresponding probability of avoiding financial ruin for Scenario #3. We used a DP to estimate the probabilities and Newton's method to perform the optimization. All starting glidepaths from Figure 5 converge to the same solution at $\varepsilon = (0.1)^{11}$, which is in a region of $P_{NR}(\vec{\alpha})$ where the Hessian is negative definite, thus concave. This is empircal evidence of an interior point optimum for this scenario. (See Figure 6.3 below.) The optimal glidepath is rising and starts out linear, then becomes concave. The first equity ratio at time (year) t=1 is about 28.12% and the last equity ratio at time (year) t=30 is about 48.10%. The success rate using this glidepath is approximately 74.80%.

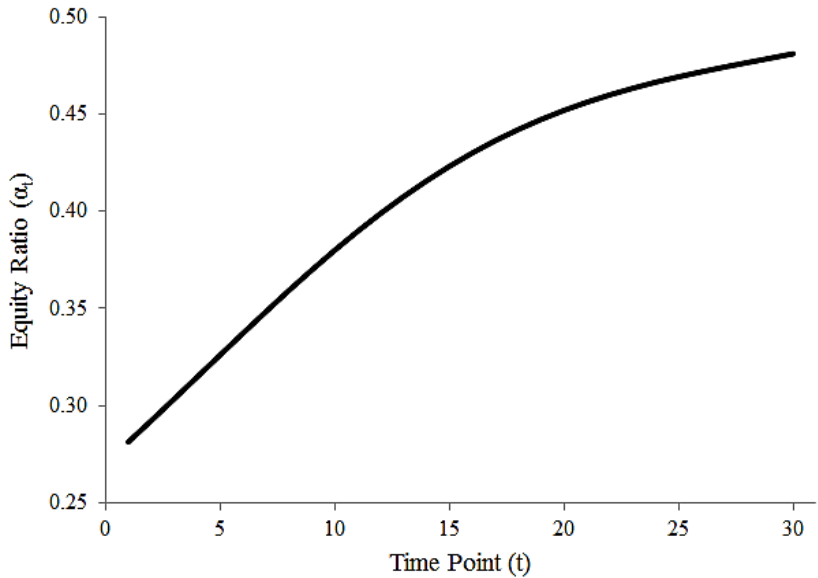

Figure 6.3
Scenario #3: Optimal Glidepath

This glidepath was derived using Newton's method with DP approximation having discretization level of 5,000. (Increase this value to reduce approximation error.) Convergence is achieved at $\varepsilon = (0.1)^{11}$ in 3 or 4 iterations, which takes 1 or 2 hours. This $\varepsilon$-level provides accuracy to about 10 decimal places in the glidepath, which is unnecessary in practice. All starting glidepaths converge to the same solution which exists in a concave region of the function and is empirical evidence of an interior point optimum. As shown below, the optimal glidepath is rising and mostly concave with lower equity ratios than found using historical returns.

Optimal GP
$P_{NR}(\vec{\alpha}_{S_3}) = 0.7480382844$

$$\vec{\alpha}_{S_3} = \begin{pmatrix} 0.2812123778 \\ 0.2922071207 \\ 0.3033747908 \\ 0.3146440730 \\ 0.3259369786 \\ 0.3371708360 \\ 0.3482607393 \\ 0.3591223192 \\ 0.3696746507 \\ 0.3798430801 \\ 0.3895617560 \\ 0.3987756762 \\ 0.4074421188 \\ 0.4155313969 \\ 0.4230269477 \\ 0.4299248317 \\ 0.4362327616 \\ 0.4419688049 \\ 0.4471599053 \\ 0.4518403570 \\ 0.4560503411 \\ 0.4598346080 \\ 0.4632413612 \\ 0.4663213786 \\ 0.4691273889 \\ 0.4717137107 \\ 0.4741361609 \\ 0.4764522373 \\ 0.4787215939 \\ 0.4810068389 \end{pmatrix}$$

### E.4 Scenario #4: Evensky Real Returns, $T_D$=30 yrs, $W_R$=4%, and $E_R$=1.0%

Let $\vec{\alpha}_{S_4}$ be the optimal static glidepath, and $P_{NR}(\vec{\alpha}_{S_4})$ be the corresponding probability of avoiding financial ruin for Scenario #4. We used a DP to estimate the probabilities and Newton's method to perform the optimization. All starting glidepaths from Figure 5 converge to the same solution at $\varepsilon = (0.1)^{11}$, which is in a region of $P_{NR}(\vec{\alpha})$ where the Hessian is negative definite, thus concave. This is empircal evidence of an interior point optimum for this scenario. (See Figure 6.4 below.) The optimal glidepath rises for about the first 9 years, then declines. The first equity ratio at time (year) t=1 is about 57.06% and the last equity ratio at time (year) t=30 is about 49.15%. The success rate using this glidepath is approximately 59.99%.

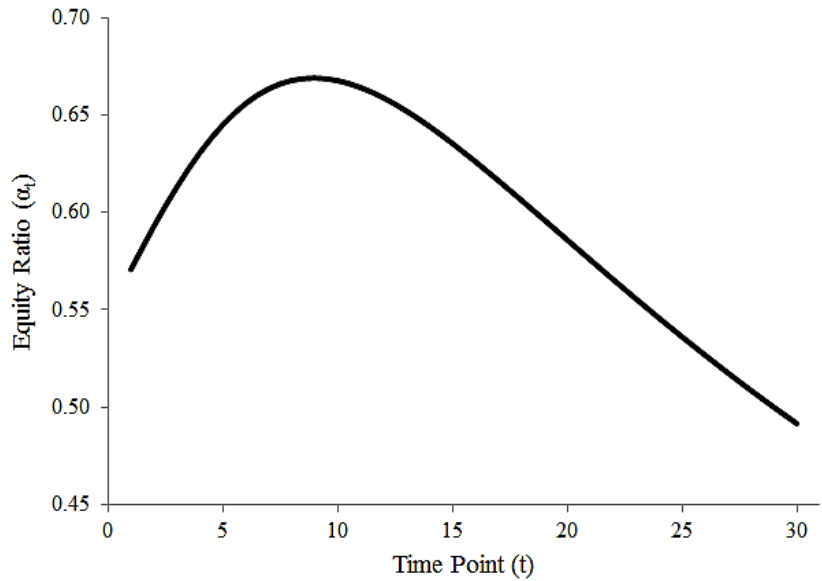

Figure 6.4
Scenario #4: Optimal Glidepath

This glidepath was derived using Newton's method with DP approximation having discretization level of 5,000. (Increase this value to reduce approximation error.) Convergence is achieved at $\varepsilon = (0.1)^{11}$ in 3 or 4 iterations, which takes 1 or 2 hours. This $\varepsilon$-level provides accuracy to about 10 decimal places in the glidepath, which is unnecessary in practice. All starting glidepaths converge to the same solution which exists in a concave region of the function and is empirical evidence of an interior point optimum. As shown below, the optimal glidepath rises for the first 9 years then declines for the remainder of retirement.

Optimal GP
$P_{NR}(\vec{\alpha}_{S_4}) = 0.5998654133$

$\vec{\alpha}_{S_4} = \begin{pmatrix} 0.5706001278 \\ 0.5927784900 \\ 0.6129556047 \\ 0.6304850727 \\ 0.6448803856 \\ 0.6558717793 \\ 0.6634184339 \\ 0.6676782655 \\ 0.6689517784 \\ 0.6676200346 \\ 0.6640914236 \\ 0.6587636517 \\ 0.6520009401 \\ 0.6441232226 \\ 0.6354034439 \\ 0.6260696330 \\ 0.6163093765 \\ 0.6062751824 \\ 0.5960898717 \\ 0.5858515575 \\ 0.5756380295 \\ 0.5655105044 \\ 0.5555167779 \\ 0.5456938428 \\ 0.5360700508 \\ 0.5266668922 \\ 0.5175004623 \\ 0.5085826720 \\ 0.4999222574 \\ 0.4915256292 \end{pmatrix}$

## E.5  Scenario #5: Historical Real Returns, $T_D=30$ yrs, $W_R=5\%$, and $E_R=0.0\%$

Let $\vec{\alpha}_{S_5}$ be the optimal static glidepath, and $P_{NR}(\vec{\alpha}_{S_5})$ be the corresponding probability of avoiding financial ruin for Scenario #5. We used a DP to estimate the probabilities and Newton's method to perform the optimization. All starting glidepaths from Figure 5 converge to the same solution at $\varepsilon = (0.1)^{11}$, which is in a region of $P_{NR}(\vec{\alpha})$ where the Hessian is negative definite, thus concave. This is empirical evidence of an interior point optimum for this scenario. (See Figure 6.5 below.) The optimal glidepath rises quickly then starts to decline very slowly midway through retirement. The first equity ratio at time (year) $t=1$ is about 56.55% and the last equity ratio at time (year) $t=30$ is about 79.7%. The success rate is approximately 77.52%.

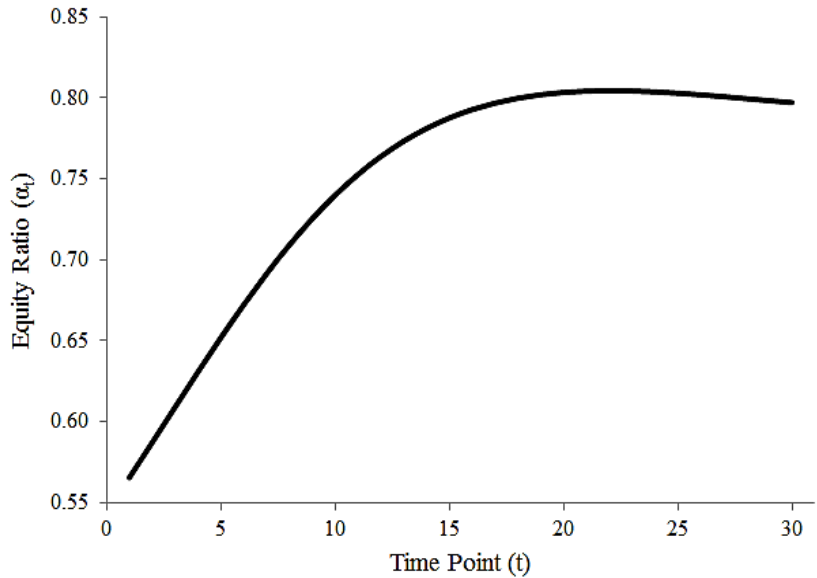

### Figure 6.5
### Scenario #5: Optimal Glidepath

**Optimal GP**

$P_{NR}(\vec{\alpha}_{S_5}) = 0.7752227003$

$$\vec{\alpha}_{S_5} = \begin{pmatrix} 0.5654643330 \\ 0.5871486400 \\ 0.6089818917 \\ 0.6306405078 \\ 0.6517869464 \\ 0.6720922620 \\ 0.6912589009 \\ 0.7090401592 \\ 0.7252534149 \\ 0.7397856956 \\ 0.7525918144 \\ 0.7636866473 \\ 0.7731338052 \\ 0.7810329517 \\ 0.7875075640 \\ 0.7926942856 \\ 0.7967344104 \\ 0.7997675761 \\ 0.8019274624 \\ 0.8033391506 \\ 0.8041177747 \\ 0.8043681211 \\ 0.8041848929 \\ 0.8036534230 \\ 0.8028506784 \\ 0.8018464592 \\ 0.8007047386 \\ 0.7994851393 \\ 0.7982445802 \\ 0.7970391309 \end{pmatrix}$$

This glidepath was derived using Newton's method with DP approximation having discretization level of 5,000. (Increase this value to reduce approximation error.) Convergence is achieved at $\varepsilon = (0.1)^{11}$ in 3 or 4 iterations, which takes 1 or 2 hours. This $\varepsilon$-level provides accuracy to about 10 decimal places in the glidepath, which is unnecessary in practice. All starting glidepaths converge to the same solution which exists in a concave region of the function and is empirical evidence of an interior point optimum. As shown below, the optimal glidepath rises quickly at first, then begins to decline very slowly about halfway through retirement.

## E.6 Scenario #6: Historical Real Returns, $T_D=30$ yrs, $W_R=5\%$, and $E_R=1.0\%$

Let $\vec{\alpha}_{S_6}$ be the optimal static glidepath, and $P_{NR}(\vec{\alpha}_{S_6})$ be the corresponding probability of avoiding financial ruin for Scenario #6. We used a DP to estimate the probabilities and Newton's method to perform the optimization. All starting glidepaths from Figure 5 converge to the same solution at $\varepsilon = (0.1)^{11}$, which is in a region of $P_{NR}(\vec{\alpha})$ where the Hessian is negative definite, thus concave. This is empirical evidence of an interior point optimum for this scenario. (See Figure 6.6 below.) The optimal glidepath rises quickly until year 11, then declines. The first equity ratio at time (year) t=1 is about 78.14% and the last equity ratio at time (year) t=30 is about 80.35%. The success rate is approximately 67.93%.

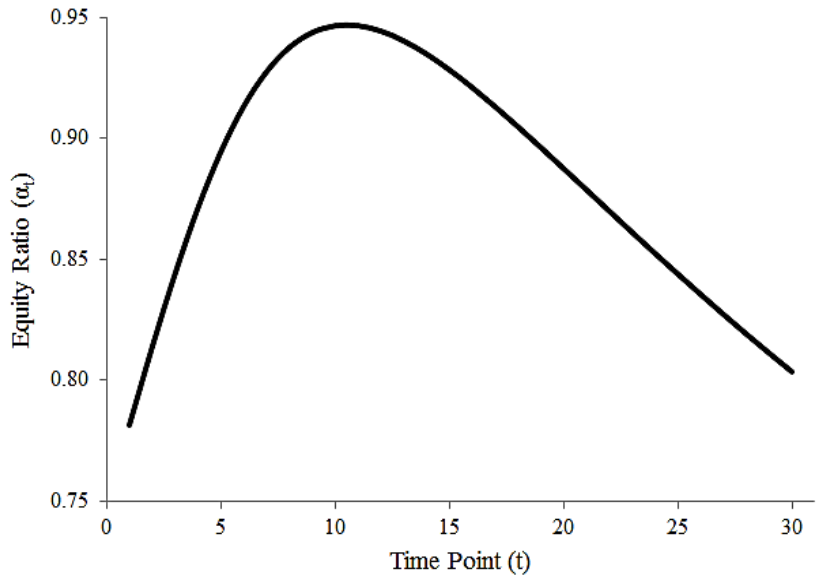

Figure 6.6
Scenario #6: Optimal Glidepath

Optimal GP
$P_{NR}(\vec{\alpha}_{S_6}) = 0.6793316432$

$$\vec{\alpha}_{S_6} = \begin{pmatrix} 0.7813938642 \\ 0.8137162909 \\ 0.8439703289 \\ 0.8711046081 \\ 0.8943111775 \\ 0.9131277585 \\ 0.9274556971 \\ 0.9375025200 \\ 0.9436839991 \\ 0.9465229636 \\ 0.9465683560 \\ 0.9443419618 \\ 0.9403098222 \\ 0.9348713868 \\ 0.9283595508 \\ 0.9210463895 \\ 0.9131512258 \\ 0.9048490994 \\ 0.8962786551 \\ 0.8875490457 \\ 0.8787457540 \\ 0.8699353978 \\ 0.8611696437 \\ 0.8524883742 \\ 0.8439222481 \\ 0.8354947786 \\ 0.8272240381 \\ 0.8191240878 \\ 0.8112062132 \\ 0.8034799987 \end{pmatrix}$$

This glidepath was derived using Newton's method with DP approximation having discretization level of 5,000. (Increase this value to reduce approximation error.) Convergence is achieved at $\varepsilon = (0.1)^{11}$ in 3 or 4 iterations, which takes 1 or 2 hours. This $\varepsilon$-level provides accuracy to about 10 decimal places in the glidepath, which is unnecessary in practice. All starting glidepaths converge to the same solution which exists in a concave region of the function and is empirical evidence of an interior point optimum. As shown below, the optimal glidepath rises very quickly until year 11 then declines for the remainder of retirement at a slower rate.

## E.7 Scenario #7: Evensky Real Returns, $T_D=30$ yrs, $W_R=5\%$, and $E_R=0.0\%$

Let $\vec{\alpha}_{S_7}$ be the optimal static glidepath, and $P_{NR}(\vec{\alpha}_{S_7})$ be the corresponding probability of avoiding financial ruin for Scenario #7. We used a DP to estimate the probabilities and Newton's method to perform the optimization. All starting glidepaths from Figure 5 converge to the same solution at $\varepsilon = (0.1)^{11}$, which is in a region of $P_{NR}(\vec{\alpha})$ where the Hessian is negative definite, thus concave. This is empirical evidence of an interior point optimum for this scenario. (See Figure 6.7 below.) The optimal glidepath rises quickly for the first 4 years, then declines for the remainder of retirement. The first equity ratio at time t=1 is about 82.35% and the last equity ratio at time t=30 is about 49.23%. The success rate is approximately 52.80%.

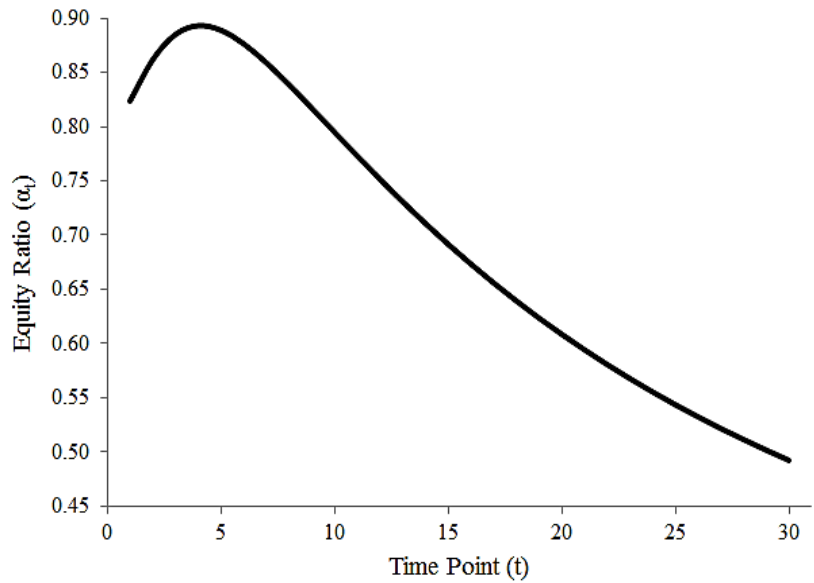

**Figure 6.7**
**Scenario #7: Optimal Glidepath**

This glidepath was derived using Newton's method with DP approximation having discretization level of 5,000. (Increase this value to reduce approximation error.) Convergence is achieved at $\varepsilon = (0.1)^{11}$ in 3 or 4 iterations, which takes 1 or 2 hours. This $\varepsilon$-level provides accuracy to about 10 decimal places in the glidepath, which is unnecessary in practice. All starting glidepaths converge to the same solution which exists in a concave region of the function and is empirical evidence of an interior point optimum. As shown below, the optimal glidepath rises quickly for 4 years then declines for the remainder of retirement.

Optimal GP
$P_{NR}(\vec{\alpha}_{S_7}) = 0.5279521553$

$\vec{\alpha}_{S_7} = \begin{pmatrix} 0.8235272966 \\ 0.8617503896 \\ 0.8850732670 \\ 0.8931568162 \\ 0.8890061069 \\ 0.8765350143 \\ 0.8590020928 \\ 0.8386754281 \\ 0.8170174723 \\ 0.7949412393 \\ 0.7730074379 \\ 0.7515547956 \\ 0.7307821069 \\ 0.7107993614 \\ 0.6916598009 \\ 0.6733802825 \\ 0.6559544071 \\ 0.6393611079 \\ 0.6235703362 \\ 0.6085468596 \\ 0.5942528104 \\ 0.5806493950 \\ 0.5676980327 \\ 0.5553611039 \\ 0.5436024264 \\ 0.5323875470 \\ 0.5216839028 \\ 0.5114608960 \\ 0.5016899088 \\ 0.4923442815 \end{pmatrix}$

## E.8 Scenario #8: Evensky Real Returns, $T_D=30$ yrs, $W_R=5\%$, and $E_R=1.0\%$

Let $\vec{\alpha}_{S_8}$ be the optimal static glidepath, and $P_{NR}(\vec{\alpha}_{S_8})$ be the corresponding probability of avoiding financial ruin for Scenario #8. We used a DP to estimate the probabilities and gradient ascent to perform the optimization. All starting glidepaths from Figure 5 converge to the same solution at $\varepsilon = (0.13)^{10}$, which is in a region of $P_{NR}(\vec{\alpha})$ where the Hessian is negative definite, thus concave. The optimum for this scenario exists in a concave border region of the function. (See Figure 6.8 below.) The optimal glidepath is constant at 100% equities initially, then declines for the remainder of retirement. The first equity ratio at time t=1 is 100% and the last equity ratio at time t=30 is about 49.61%. The success rate is approximately 43.23%.

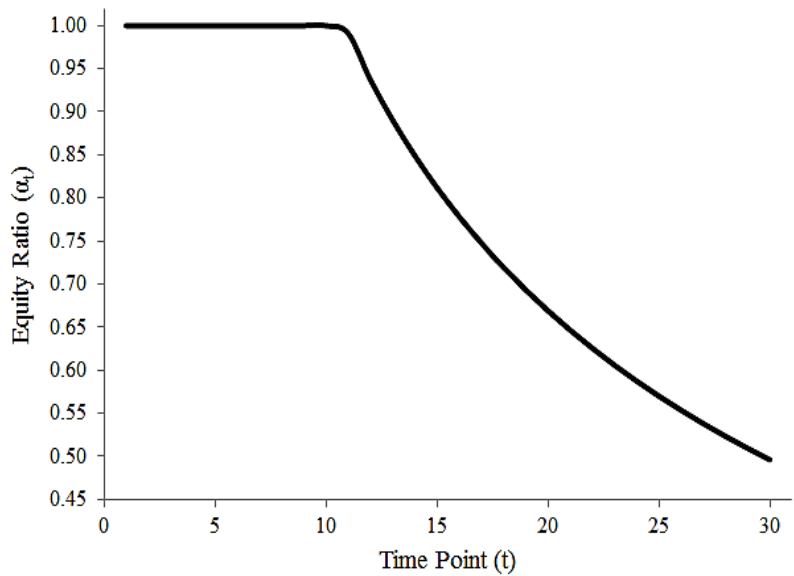

Figure 6.8
Scenario #8: Optimal Glidepath

Optimal GP
$P_{NR}(\vec{\alpha}_{S_8}) = 0.4322869545$

$$\vec{\alpha}_{S_8} = \begin{pmatrix} 1.0000000000 \\ 1.0000000000 \\ 1.0000000000 \\ 1.0000000000 \\ 1.0000000000 \\ 1.0000000000 \\ 1.0000000000 \\ 1.0000000000 \\ 1.0000000000 \\ 1.0000000000 \\ 0.9911850486 \\ 0.9373566848 \\ 0.8906630236 \\ 0.8493134032 \\ 0.8121883772 \\ 0.7785260350 \\ 0.7477740007 \\ 0.7195131648 \\ 0.6934139924 \\ 0.6692102999 \\ 0.6466825215 \\ 0.6256465423 \\ 0.6059459405 \\ 0.5874464078 \\ 0.5700316138 \\ 0.5536000608 \\ 0.5380626395 \\ 0.5233406936 \\ 0.5093644632 \\ 0.4960727971 \end{pmatrix}$$

This glidepath was derived using gradient ascent with DP approximation having discretization level of 10,000. (Increase this value to reduce approximation error.) Convergence is achieved at $\varepsilon = (0.13)^{10}$ in a few hours. This $\varepsilon$-level provides accuracy to about 8 decimal places in the glidepath, which is unnecessary in practice. All starting glidepaths converge to the same solution which exists in a concave border region of the function. The actual optimum requires more than 100% equities during the first few years, which violates the constraints. As shown below, the optimal glidepath starts and stays at 100% equities, then declines for the remainder of retirement.

## V. Random Time of Final Withdrawal

In general, the last withdrawal occurs at time $t=T_D$, and in the above examples $T_D=30$ (years). The first equity ratio, $\alpha_1$, is set at time $t=0$ and the corresponding first return, $\hat{r}_{(1,\alpha_1)}$, is observed at time $t=1$. When $T_D=30$, the last equity ratio, $\alpha_{30}$, is applied at time $t=29$ and the corresponding last return, $\hat{r}_{(30,\alpha_{30})}$, is observed at time $t=30$. When $T_D$ is random the last withdrawal is based on longevity and $T_D = 0, 1, 2, \ldots, S_{Max}$, where $S_{Max}$ is the largest possible value of $T_D$ based on the current age of the retiree(s). Let $P(T_D=t) = p_t$, for $t=0, 1, 2, \ldots, S_{Max}$. Here, the glidepath $\vec{\alpha}$ contains $S_{Max}$ equity ratios, one for each possible time point $t > 0$. No withdrawal is attempted at time $t=0$, and $T_D=0$ indicates that death occurs prior to the first withdrawal attempt. Let $P_{NR}^{(r)}(\vec{\alpha})$ represent the probability of avoiding ruin in retirement when $T_D$ is random, and let $P_{NR}(\vec{\alpha}, T_D)$ be as defined in (3.1). Then:

$$P_{NR}^{(r)}(\vec{\alpha}) = P\big[(T_D = 0 \cap \text{Ruin}^c(\leq 0)) \cup (T_D = 1 \cap \text{Ruin}^c(\leq 1)) \\ \cup \ldots \cup (T_D = S_{Max} \cap \text{Ruin}^c(\leq S_{Max}))\big] \quad (5.1)$$

Since these events are mutually exclusive, their probabilities can be added:

$$\rightarrow P_{NR}^{(r)}(\vec{\alpha}) = \sum_{t=0}^{S_{Max}} P(T_D = t \cap \text{Ruin}^c(\leq t)) \quad (5.2)$$

$$= \sum_{t=0}^{S_{Max}} P(T_D = t) * P(\text{Ruin}^c(\leq t) \mid T_D = t) \quad (5.3)$$

$$= \sum_{t=0}^{S_{Max}} p_t * P_{NR}(\vec{\alpha}, t) \quad (5.4)$$

$$= p_0 + \sum_{t=1}^{S_{Max}} p_t * P_{NR}(\vec{\alpha}, t). \quad (5.5)$$

To optimize this function with respect to the glidepath, $\vec{\alpha}$, using the techniques proposed here requires the $S_{Max}$-element gradient vector, $\vec{g}^{(r)}$, and $(S_{Max}) \times (S_{Max})$ Hessian matrix, $\overline{\overline{H}}^{(r)}$, for the function $P_{NR}^{(r)}(\vec{\alpha})$. Since the derivative of a sum equals the sum of the derivatives, the $t^{th}$ gradient element, $g_t^{(r)}$, of $\vec{g}^{(r)} = \left(g_1^{(r)}, g_2^{(r)}, \ldots, g_{S_{Max}}^{(r)}\right)$ is given by:

$$g_t^{(r)} = \frac{\partial}{\partial \alpha_t}\left[P_{NR}^{(r)}(\vec{\alpha})\right] = \sum_{k=1}^{S_{Max}} p_k * \frac{\partial}{\partial \alpha_t}[P_{NR}(\vec{\alpha}, k)] = \sum_{k=1}^{S_{Max}} p_k * (g_t | T_D = k), \quad (5.6)$$

where $g_t | T_D = k$ is as defined in (3.4) for $t \leq k$, and 0, O.W. Here, $g_t^{(r)}$ is defined for t=1, 2, …, $S_{Max}$. The off-diagonal elements, $H_{i,j}^{(r)}$, of the corresponding Hessian matrix, $\bar{\bar{H}}^{(r)}$, are given by:

$$H_{i,j}^{(r)} = \frac{\partial^2}{\partial \alpha_i \partial \alpha_j}\left[P_{NR}^{(r)}(\vec{\alpha})\right] = \sum_{k=1}^{S_{Max}} p_k * \frac{\partial^2}{\partial \alpha_i \partial \alpha_j}[P_{NR}(\vec{\alpha}, k)] = \sum_{k=1}^{S_{Max}} p_k * (H_{i,j} | T_D = k), \quad (5.7)$$

where $H_{i,j} | T_D = k$ is as defined in (3.5) for $i \neq j \in \{1, 2, \ldots, k\}$, and it is 0, O.W. Here, $H_{i,j}^{(r)}$ is defined for $i \neq j = 1, 2, \ldots, S_{Max}$. The diagonal elements, $H_{t,t}^{(r)}$, of the Hessian matrix, $\bar{\bar{H}}^{(r)}$, are:

$$H_{t,t}^{(r)} = \frac{\partial^2}{\partial \alpha_t^2}\left[P_{NR}^{(r)}(\vec{\alpha})\right] = \sum_{k=1}^{S_{Max}} p_k * \frac{\partial^2}{\partial \alpha_t^2}[P_{NR}(\vec{\alpha}, k)] = \sum_{k=1}^{S_{Max}} p_k * (H_{t,t} | T_D = k), \quad (5.8)$$

where $H_{t,t} | T_D = k$ is as defined in (3.6) for $t \leq k$, and t=1, 2, …, $S_{Max}$. The problem of finding the optimal static glidepath in retirement for random time of final withdrawal ($T_D$) is therefore tractable. The results presented in this section may apply to the mortality of a person or group.

### Table 2
### Assumptions used for Random TD Glidepath Optimization Scenarios

This table defines the assumptions for the two random TD scenarios presented in this section. The results are for a same-age male/female couple with age 65 reflecting time t=0. The corresponding probabilities were derived using lifetables from SSA.gov and the maximum male/female ages are 111 and 113, respectively. Therefore this glidepath requires 48 equity ratios, one for each possible time point and the initial glidepaths from Figure 5 used for Scenarios 1-8 do not apply. Historical and Evensky returns are exactly as described in Table 1. Each scenario will start at varying glidepaths of different shapes to determine whether or not there are local optimums. These 2 scenarios can be directly compared against Scenarios #1 and #4 from Section IV.E which are the fixed $T_D$=30 year counterparts.

| Scenario # (Compare vs.) | Real Return Assumptions | Retirement Length ($S_{Max}$) | Withdrawal Rate ($W_R$=RF(0)) | Expense Ratio ($E_R$) |
|---|---|---|---|---|
| 9 (vs. 1) | Historical | 48 yrs | 4.0% | 0.0% |
| 10 (vs. 4) | Evensky | 48 yrs | 4.0% | 1.0% |

## A.1 Scenario #9: Historical Real Returns, $S_{Max}$=48 yrs, $W_R$=4%, and $E_R$=0.0%

Let $\vec{\alpha}_{S_9}$ be the optimal static glidepath, and $P_{NR}(\vec{\alpha}_{S_9})$ be the corresponding probability of avoiding financial ruin for Scenario #9. We used a DP to estimate the probabilities and Newton's method to perform the optimization. All starting glidepaths tested converge to the same solution at $\varepsilon = (0.1)^8$, which is in a region of $P_{NR}(\vec{\alpha})$ where the Hessian is negative definite, thus concave. This is empirical evidence of an interior point optimum for this scenario. (See Figure 7.1 below.) The optimal glidepath rises then levels off at about (year) time t=30. The first equity ratio at time (year) t=1 is about 34.19% and the last equity ratio at time (year) t=48 is about 79.77%. The success rate using this glidepath is approximately 96.06%.

$P_{NR}(\vec{\alpha}_{S_9}) = 0.9606338340$

$$\vec{\alpha}_{S_9} = \begin{pmatrix} 0.3418935494 \\ 0.3507076211 \\ 0.3601130892 \\ 0.3701360132 \\ 0.3808017293 \\ 0.3921343765 \\ 0.4041562743 \\ 0.4168871147 \\ 0.4303429142 \\ 0.4445346538 \\ 0.4594665065 \\ 0.4751335044 \\ 0.4915184412 \\ 0.5085877681 \\ 0.5262863679 \\ 0.5445315421 \\ 0.5632073261 \\ 0.5821610426 \\ 0.6012040338 \\ 0.6201177741 \\ 0.6386653181 \\ 0.6566066845 \\ 0.6737158883 \\ 0.6897963971 \\ 0.7046935759 \\ 0.7183016244 \\ 0.7305650025 \\ 0.7414750256 \\ 0.7510641237 \\ 0.7593986857 \\ 0.7665739362 \\ 0.7727005656 \\ 0.7778937520 \\ 0.7822699079 \\ 0.7859336835 \\ 0.7889775284 \\ 0.7914821787 \\ 0.7935205750 \\ 0.7951554269 \\ 0.7964409013 \\ 0.7974227038 \\ 0.7981400499 \\ 0.7986073987 \\ 0.7988880537 \\ 0.7989330710 \\ 0.7986744386 \\ 0.7984614566 \\ 0.7976847814 \end{pmatrix}$$

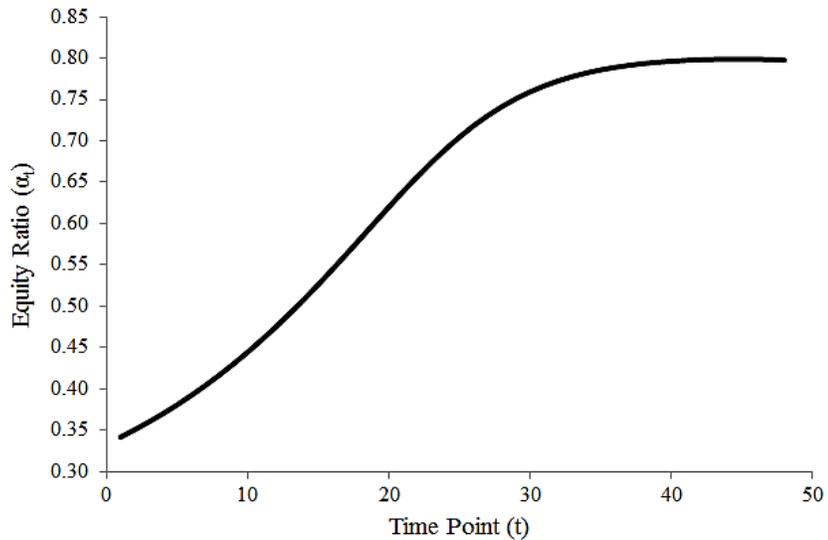

**Figure 7.1**
**Scenario #9: Optimal Glidepath**

This glidepath was derived using Newton's method with DP approximation having discretization level of 5,000. (Increase this value to reduce approximation error.) Convergence is achieved at $\varepsilon = (0.1)^8$. This $\varepsilon$-level provides accuracy to about 7 decimal places in the glidepath, which is unnecessary in practice. All starting glidepaths tested converge to the same solution which exists in a concave region of the function and is empirical evidence of an interior point optimum. As seen below, the optimal glidepath rises initially and then levels off at a relatively high equity exposure for the remainder of retirement.

## A.2 Scenario #10: Evensky Real Returns, $S_{Max}$=48 yrs, $W_R$=4%, and $E_R$=1.0%

Let $\vec{\alpha}_{S_{10}}$ be the optimal static glidepath, and $P_{NR}(\vec{\alpha}_{S_{10}})$ be the corresponding probability of avoiding financial ruin for Scenario #10. We used a DP to estimate the probabilities and Newton's method to perform the optimization. All starting glidepaths tested converge to the same solution at $\varepsilon = (0.1)^{10}$, which is in a region of $P_{NR}(\vec{\alpha})$ where the Hessian is negative definite, thus concave. This is empirical evidence of an interior point optimum for this scenario. (See Figure 7.2 below.) The optimal glidepath rises then begins to decline at a slower rate about halfway through retirement. The first equity ratio at time (year) t=1 is about 31.37% and the last equity ratio at time (year) t=48 is about 49.72%. The success rate is approximately 77.87%.

$P_{NR}(\vec{\alpha}_{S_{10}}) = 0.7786867264$

$$\vec{\alpha}_{S_{10}} = \begin{pmatrix} 0.3136699500 \\ 0.3263162327 \\ 0.3394973455 \\ 0.3531827075 \\ 0.3673301932 \\ 0.3818851629 \\ 0.3967797125 \\ 0.4119321725 \\ 0.4272468277 \\ 0.4426137453 \\ 0.4579084827 \\ 0.4729913050 \\ 0.4877055506 \\ 0.5018754195 \\ 0.5153051225 \\ 0.5277830135 \\ 0.5390939382 \\ 0.5490398111 \\ 0.5574639344 \\ 0.5642718264 \\ 0.5694427623 \\ 0.5730299416 \\ 0.5751506361 \\ 0.5759707255 \\ 0.5756853495 \\ 0.5745020780 \\ 0.5726265617 \\ 0.5702513293 \\ 0.5675440916 \\ 0.5646366343 \\ 0.5616059028 \\ 0.5584906130 \\ 0.5553031695 \\ 0.5520162449 \\ 0.5486320251 \\ 0.5451797171 \\ 0.5417106785 \\ 0.5382424320 \\ 0.5348079700 \\ 0.5313789145 \\ 0.5279252704 \\ 0.5245091433 \\ 0.5207791996 \\ 0.5173003822 \\ 0.5133594248 \\ 0.5080167217 \\ 0.5042683421 \\ 0.4972053385 \end{pmatrix}$$

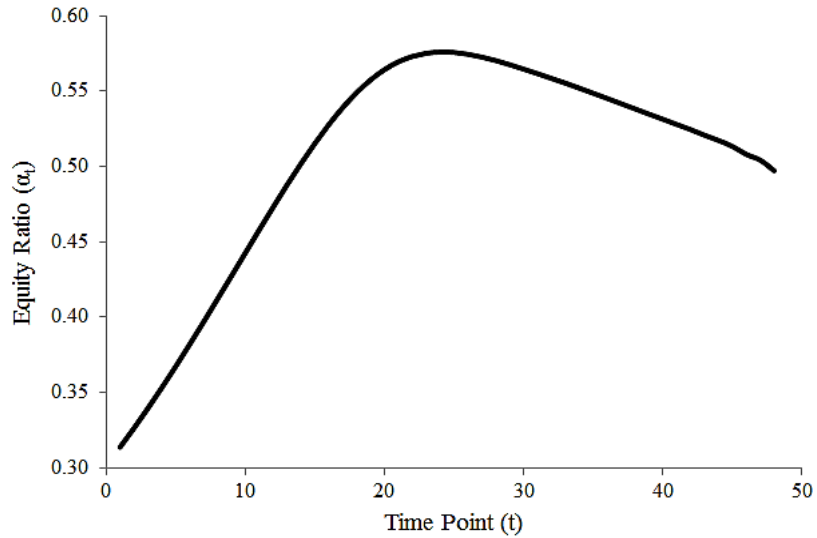

### Figure 7.2
### Scenario #10: Optimal Glidepath

This glidepath was derived using Newton's method with DP approximation having discretization level of 5,000. (Increase this value to reduce approximation error.) Convergence is achieved at $\varepsilon = (0.1)^{10}$. This $\varepsilon$-level provides accuracy to about 9 decimal places in the glidepath, which is unnecessary in practice. All starting glidepaths tested converge to the same solution which exists in a concave region of the function and is empirical evidence of an interior point optimum. As seen below, the optimal glidepath rises initially and then declines for the remainder of retirement.

## VI. Summary/Conclusions and Future Research

Dynamic glidepaths change with time as market returns are consumed, whereas static glidepaths are predetermined at some starting point. In general, static glidepaths are suboptimal strategies because they impose artificial constraints on the retiree (Rook (2014)). They are easier to implement and understand, however. Static glidepaths also form the basis of most T-D funds, thus are of interest to practitioners and retirement researchers. We have introduced a technique to derive the optimal static glidepath in retirement. Since the portfolio success probability is not always quasi-concave as a function of the equity ratios, the procedure may find a local optimum. In practice we have not encountered local optimums for typical retirement horizons and initial withdrawal rates. If the user finds multiple local optimums in their particular scenario, a metaheuristic can be used to guide the procedure towards a global optimum (see Section II.I).

We have examined 10 scenarios with varying assumptions about withdrawal rates, underlying asset class returns, expense ratios, and fixed vs. random times of final withdrawal. We have found that the optimal static glidepath in retirement changes as a function of these parameters, which is not surprising. We have also found that the sequence of portfolio returns can impact the optimal solution in different ways. When the strategy used has a high probability of success, SOR becomes a risk and it is mitigated in optimal solutions with a rising glidepath (see Scenario #1). When the strategy has a low probability of success, SOR becomes a potential reward and is seen in optimal solutions as heavy equity investments early in retirement (see Scenario #8). T-D funds with declining glidepaths therefore bias in favor of SOR reward, at the expense of SOR risk. This approach is consistent with strategies that have a high withdrawal rate, high expenses, and lower success probability. A costly T-D fund that lacks knowledge of the retiree's withdrawal rate may justify a declining glidepath, perhaps for the wrong reasons.

Consistent with the findings of Young (2004), Moore and Young (2006), Bayraktar and Young (2007), and Rook (2014), we find in this research that the optimal equity ratio is not constant across time when minimizing the probability of ruin using a static glidepath. This conflicts with many lifecycle models that maximize expected utility. The results of this research differ somewhat from the existing literature on static glidepaths. The optimal glidepath's shape changes based on underlying assumptions making it difficult to advocate for one over another. Lacking a consensus on the parameters of future returns, each user is left to choose the optimal glidepath that is built on the foundation of their assumptions. Further, these results are not directly comparable to many findings reviewed in Section I. For example, models built by backtesting or bootstrapping historical data, with expected utility as the objective, or with serially correlated asset class returns will naturally differ from the results presented here.

Future research on this topic will take two directions. First, we will introduce techniques that extend the models developed in this paper and in Rook (2014). We will also study general classes of retirement decumulation problems that can be solved using these approaches. Second, we will attempt to develop new models that optimize practitioner retirement heuristics.

**Acknowledgements:**

I am deeply grateful to my past and present instructors, all of whom have selflessly shared so much of their knowledge. In particular, to Dr. Irwin Guttman, who was my graduate instructor and thesis advisor while studying statistics at the University of Buffalo, and to Dr. Guillermo Gallego who was my graduate instructor in statistics and revenue management courses while studying at Columbia. From each I learned a great deal and, in fact, the tools from *Minimizing the Probability of Ruin in Retirement* are nothing more than dynamic pricing models applied to retirement decumulation.[10] The interested reader can find all details in *The Theory and Practice of Revenue Management* by Talluri and van Ryzin. Alternatively, they can take classes with Dr. Gallego where they will learn to build, decompose, code, and analyze the runtime complexity of several such models. I would also like to thank Dr. Mitchell Kerman of Stevens Institute who I am presently working with on a related research project that can be used to extend these models. Please note that all errors and mistakes in this and past research are solely my own and reflect a failure to correctly apply what I have been taught.

---

[10] For example, commercial airlines build a grid for their flights with time-to-departure on the vertical axis and remaining capacity on the horizontal axis. At each cell in the grid the fare price is chosen to maximize expected revenue under a random demand function. Downward pressure is placed on price as time approaches departure and upward pressure as capacity diminishes. In retirement we draw the exact same grid but with time-to-death on the vertical axis and funded status on the horizontal axis. At each cell in the grid the equity ratio is chosen to minimize the probability of ruin under a random return function. Downward pressure is placed on the equity ratio as time approaches death and upward pressure as the funded status worsens. Both models are solved backwards and then lived forward.

June 27, 2015

**Optimal Equity Glidepaths in Retirement**

CHRISTOPHER J. ROOK

**INTERNET APPENDIX**

# Table of Contents

## Appendix A. Derivation of the Gradient $\vec{g}$ for $P_{NR}(\vec{\alpha}) = P(Ruin^C(\leq T_D))$

Each element $g_t$ of the gradient vector $\vec{g} = (g_1, g_2, \ldots, g_{TD})'$ requires evaluation of the following derivative:

$$\frac{\partial}{\partial \alpha_t}\left[\frac{1}{\sqrt{2\pi v(\alpha_t)}} e^{-\left(\frac{\hat{r}_{(t,\alpha_t)} - m(\alpha_t)}{\sqrt{2v(\alpha_t)}}\right)^2}\right], \tag{A.1}$$

which, by repeated applications of the product and chain rules (Anton (1988)), is given by:

$$= \frac{1}{\sqrt{2\pi}}\left\{\frac{1}{\sqrt{v(\alpha_t)}} e^{-\left(\frac{\hat{r}_{(t,\alpha_t)} - m(\alpha_t)}{\sqrt{2v(\alpha_t)}}\right)^2}\left[\frac{\left(\hat{r}_{(t,\alpha_t)} - m(\alpha_t)\right) m'(\alpha_t)}{v(\alpha_t)} + \frac{\left(\hat{r}_{(t,\alpha_t)} - m(\alpha_t)\right)^2 v'(\alpha_t)}{2v(\alpha_t)^2}\right]\right.$$

$$\left. - \frac{v'(\alpha_t)}{2\sqrt{v(\alpha_t)}^3} e^{-\left(\frac{\hat{r}_{(t,\alpha_t)} - m(\alpha_t)}{\sqrt{2v(\alpha_t)}}\right)^2}\right\} \tag{A.2}$$

$$= \frac{1}{\sqrt{2\pi v(\alpha_t)}} e^{-\left(\frac{\hat{r}_{(t,\alpha_t)} - m(\alpha_t)}{\sqrt{2v(\alpha_t)}}\right)^2}\left[\frac{v'(\alpha_t)\left(\hat{r}_{(t,\alpha_t)} - m(\alpha_t)\right)^2}{2v(\alpha_t)^2} + \frac{m'(\alpha_t)\left(\hat{r}_{(t,\alpha_t)} - m(\alpha_t)\right)}{v(\alpha_t)} - \frac{v'(\alpha_t)}{2v(\alpha_t)}\right]. \tag{A.3}$$

Upon factoring, adding terms needed to complete the square with respect to $(\hat{r}_{(t,\alpha_t)} - m(\alpha_t))$ inside $[\cdot]$, and using $f(\hat{r}_{(t,\alpha_t)})$ to represent its PDF, (A.3) yields:

$$= \frac{v'(\alpha_t) f(\hat{r}_{(t,\alpha_t)})}{2v(\alpha_t)^2}\left[\left(\hat{r}_{(t,\alpha_t)} - m(\alpha_t)\right)^2 + \frac{2v(\alpha_t)m'(\alpha_t)\left(\hat{r}_{(t,\alpha_t)} - m(\alpha_t)\right)}{v'(\alpha_t)} + \left(\frac{v(\alpha_t)m'(\alpha_t)}{v'(\alpha_t)}\right)^2\right.$$

$$\left. - v(\alpha_t) - \left(\frac{v(\alpha_t)m'(\alpha_t)}{v'(\alpha_t)}\right)^2\right] \tag{A.4}$$

$$= \frac{v'(\alpha_t) f(\hat{r}_{(t,\alpha_t)})}{2v(\alpha_t)^2}\left\{\left[\left(\hat{r}_{(t,\alpha_t)} - m(\alpha_t)\right) + \frac{v(\alpha_t)m'(\alpha_t)}{v'(\alpha_t)}\right]^2 - \left[v(\alpha_t) + \left(\frac{v(\alpha_t)m'(\alpha_t)}{v'(\alpha_t)}\right)^2\right]\right\} \tag{A.5}$$

We now create the difference of 2 valid PDFs by factoring the right most term $[\cdot]$ inside $\{\cdot\}$ from the entire expression and simplifying the result:

$$= \left[v(\alpha_t) + \left(\frac{v(\alpha_t)m'(\alpha_t)}{v'(\alpha_t)}\right)^2\right]\frac{v'(\alpha_t) f(\hat{r}_{(t,\alpha_t)})}{2v(\alpha_t)^2}\left\{\frac{\left[\left(\hat{r}_{(t,\alpha_t)} - m(\alpha_t)\right) + \frac{v(\alpha_t)m'(\alpha_t)}{v'(\alpha_t)}\right]^2}{\left[v(\alpha_t) + \left(\frac{v(\alpha_t)m'(\alpha_t)}{v'(\alpha_t)}\right)^2\right]} - 1\right\} \tag{A.6}$$

$$= \left[\frac{v'(\alpha_t)}{2v(\alpha_t)} + \frac{m'(\alpha_t)^2}{2v'(\alpha_t)}\right] \left\{\frac{\left[v'(\alpha_t)\left(\hat{r}_{(t,\alpha_t)} - m(\alpha_t)\right) + m'(\alpha_t)v(\alpha_t)\right]^2}{[m'(\alpha_t)^2 v(\alpha_t)^2 + v(\alpha_t)v'(\alpha_t)^2]} f\bigl(\hat{r}_{(t,\alpha_t)}\bigr) - f\bigl(\hat{r}_{(t,\alpha_t)}\bigr)\right\} \quad (A.7)$$

$$= K * \left[g\bigl(\hat{r}_{(t,\alpha_t)}\bigr) - f\bigl(\hat{r}_{(t,\alpha_t)}\bigr)\right], \quad (A.8)$$

where $f(\hat{r}_{(t,\alpha_t)})$, K, and $g(\hat{r}_{(t,\alpha_t)})$ are as defined in (4.10), (4.13), and (4.14), respectively. The final step, which is trivial and not shown, is to replace (A.1) in (4.11) with the term just derived in (A.8) to yield (4.12). The quantity in (4.12) is recognized as the difference of 2 glidepath success probabilities, therefore it can be estimated/approximated to any required degree of accuracy.

## Appendix B. Proof that $g(\hat{r}_{(t,\alpha_t)})$ is a Valid PDF

A valid PDF is any function $w(x) \geq 0$ for $-\infty < x < \infty$ where $\int_{-\infty}^{\infty} w(x) dx = 1.0$ (Casella and Berger (1990)). Consider the function $g(\hat{r}_{(t,\alpha_t)})$, where:

$$g(\hat{r}_{(t,\alpha_t)}) = \frac{\left[v'(\alpha_t)\left(\hat{r}_{(t,\alpha_t)} - m(\alpha_t)\right) + m'(\alpha_t)v(\alpha_t)\right]^2}{[m'(\alpha_t)^2 v(\alpha_t)^2 + v(\alpha_t)v'(\alpha_t)^2]} f(\hat{r}_{(t,\alpha_t)}), \qquad \text{for } -\infty < \hat{r}_{(t,\alpha_t)} < \infty. \tag{B.1}$$

Clearly $g(\hat{r}_{(t,\alpha_t)}) \geq 0$ for all $\hat{r}_{(t,\alpha_t)}$ since the numerator of the ratio is always $\geq 0$ and the denominator is $\geq 0$ if $v(\alpha_t) > 0$. By definition $v(\alpha_t)$ represents the variance of the RV $\hat{r}_{(t,\alpha_t)} \sim f(\cdot)$ from (4.6), which is non-degenerate. Finally, $f(\cdot) \geq 0$ everywhere since it is a valid PDF. The 2nd condition requires that $g(\hat{r}_{(t,\alpha_t)})$ integrate to 1.0 over all real numbers, and:

$$\int_{-\infty}^{\infty} \frac{\left[v'(\alpha_t)\left(\hat{r}_{(t,\alpha_t)} - m(\alpha_t)\right) + m'(\alpha_t)v(\alpha_t)\right]^2}{[m'(\alpha_t)^2 v(\alpha_t)^2 + v(\alpha_t)v'(\alpha_t)^2]} f(\hat{r}_{(t,\alpha_t)}) \, d(\hat{r}_{(t,\alpha_t)}) \tag{B.2}$$

$$= \frac{1}{[m'(\alpha_t)^2 v(\alpha_t)^2 + v(\alpha_t)v'(\alpha_t)^2]} \Bigg[ v'(\alpha_t)^2 \int_{-\infty}^{\infty} \left(\hat{r}_{(t,\alpha_t)} - m(\alpha_t)\right)^2 f(\hat{r}_{(t,\alpha_t)}) \, d(\hat{r}_{(t,\alpha_t)})$$

$$+ 2v'(\alpha_t)m'(\alpha_t)v(\alpha_t) \int_{-\infty}^{\infty} \left(\hat{r}_{(t,\alpha_t)} - m(\alpha_t)\right) f(\hat{r}_{(t,\alpha_t)}) \, d(\hat{r}_{(t,\alpha_t)}) \tag{B.3}$$

$$+ m'(\alpha_t)^2 v(\alpha_t)^2 \int_{-\infty}^{\infty} f(\hat{r}_{(t,\alpha_t)}) \, d(\hat{r}_{(t,\alpha_t)}) \Bigg],$$

which, upon invoking (2.67) from Section II.L, yields:

$$= \frac{1}{[m'(\alpha_t)^2 v(\alpha_t)^2 + v(\alpha_t)v'(\alpha_t)^2]} [v'(\alpha_t)^2 v(\alpha_t) + 0 + m'(\alpha_t)^2 v(\alpha_t)^2] \tag{B.4}$$

$$= 1. \tag{B.5}$$

Since $g(\hat{r}_{(t,\alpha_t)})$ satisfies both conditions given it is a valid PDF.

## Appendix C. Derivation of the CDF $G(\hat{r}_{(t,\alpha_t)})$ for $g(\hat{r}_{(t,\alpha_t)})$

An RV $\hat{r}_{(t,\alpha_t)} \sim g(\hat{r}_{(t,\alpha_t)})$ has CDF $G(\hat{r}_{(t,\alpha_t)}) = P_g(\hat{r}_{(t,\alpha_t)} \leq r)$ defined as:

$$P_g(\hat{r}_{(t,\alpha_t)} \leq r) = \int_{-\infty}^{r} \frac{\left[v'(\alpha_t)\left(\hat{r}_{(t,\alpha_t)} - m(\alpha_t)\right) + m'(\alpha_t)v(\alpha_t)\right]^2}{[m'(\alpha_t)^2 v(\alpha_t)^2 + v(\alpha_t)v'(\alpha_t)^2]} f(\hat{r}_{(t,\alpha_t)})\, d(\hat{r}_{(t,\alpha_t)}) \tag{C.1}$$

$$= \frac{1}{[m'(\alpha_t)^2 v(\alpha_t)^2 + v(\alpha_t)v'(\alpha_t)^2]} \left[ \frac{v'(\alpha_t)^2}{\sqrt{2\pi v(\alpha_t)}} \int_{-\infty}^{r} \left(\hat{r}_{(t,\alpha_t)} - m(\alpha_t)\right)^2 e^{-\left(\frac{\hat{r}_{(t,\alpha_t)} - m(\alpha_t)}{\sqrt{2v(\alpha_t)}}\right)^2} d(\hat{r}_{(t,\alpha_t)}) \right.$$

$$+ \frac{2v'(\alpha_t)m'(\alpha_t)v(\alpha_t)}{\sqrt{2\pi v(\alpha_t)}} \int_{-\infty}^{r} \left(\hat{r}_{(t,\alpha_t)} - m(\alpha_t)\right) e^{-\left(\frac{\hat{r}_{(t,\alpha_t)} - m(\alpha_t)}{\sqrt{2v(\alpha_t)}}\right)^2} d(\hat{r}_{(t,\alpha_t)}) \tag{C.2}$$

$$\left. + m'(\alpha_t)^2 v(\alpha_t)^2 \int_{-\infty}^{r} \frac{1}{\sqrt{2\pi v(\alpha_t)}} e^{-\left(\frac{\hat{r}_{(t,\alpha_t)} - m(\alpha_t)}{\sqrt{2v(\alpha_t)}}\right)^2} d(\hat{r}_{(t,\alpha_t)}) \right].$$

Which, by (2.64) from Section II.K, can be expressed as:

$$\frac{1}{[m'(\alpha_t)^2 v(\alpha_t)^2 + v(\alpha_t)v'(\alpha_t)^2]} \left[ \frac{v'(\alpha_t)^2 v(\alpha_t)}{2}\left(1 - (-1)^{\mathbf{1}_{(m(\alpha_t),\infty)}(r)} F_{g[\frac{3}{2},1]}\left(\left(\frac{r - m(\alpha_t)}{\sqrt{2v(\alpha_t)}}\right)^2\right)\right) \right.$$

$$- \frac{\sqrt{2}v'(\alpha_t)m'(\alpha_t)\sqrt{v(\alpha_t)}^3}{\sqrt{\pi}}\left(1 - F_{g[1,1]}\left(\left(\frac{r - m(\alpha_t)}{\sqrt{2v(\alpha_t)}}\right)^2\right)\right) \tag{C.3}$$

$$\left. + m'(\alpha_t)^2 v(\alpha_t)^2 \Phi\left(\frac{r - m(\alpha_t)}{\sqrt{v(\alpha_t)}}\right) \right]$$

$$= C_0 \left[ C_1 \left(1 - (-1)^{\mathbf{1}_{(m(\alpha_t),\infty)}(r)} F_{g[\frac{3}{2},1]}\left(\left(\frac{r - m(\alpha_t)}{\sqrt{2v(\alpha_t)}}\right)^2\right)\right) \right.$$

$$\left. + C_2 \left(1 - F_{g[1,1]}\left(\left(\frac{r - m(\alpha_t)}{\sqrt{2v(\alpha_t)}}\right)^2\right)\right) + C_3 \Phi\left(\frac{r - m(\alpha_t)}{\sqrt{v(\alpha_t)}}\right) \right], \tag{C.4}$$

where $C_0$, $C_1$, $C_2$, and $C_3$ are as defined in (4.16) and $\mathbf{1}_{(m(\alpha_t),\infty)}(r)$ is the indicator function which equals 1 when $m(\alpha_t) < r < \infty$ and equals 0 otherwise. Here, $\Phi(\cdot)$ is the CDF of a standard normal RV.

## Appendix D. Derivation of the Diagonal Elements $H_{t,t}$ of $\bar{\bar{H}}$ for $P_{NR}(\vec{\alpha}) = P(Ruin^C(\leq T_D))$

Each diagonal element $H_{t,t}$ of the Hessian matrix $\bar{\bar{H}}$ requires evaluation of the following 2$^{nd}$ order derivative:

$$\frac{\partial^2}{\partial \alpha_t^2}\left[\frac{1}{\sqrt{2\pi v(\alpha_t)}} e^{-\left(\frac{\hat{r}_{(t,\alpha_t)} - m(\alpha_t)}{\sqrt{2v(\alpha_t)}}\right)^2}\right]. \quad (D.1)$$

From (A.3), the 1$^{st}$ order derivative is given by:

$$\frac{\partial}{\partial \alpha_t}\left[\frac{1}{\sqrt{2\pi v(\alpha_t)}} e^{-\left(\frac{\hat{r}_{(t,\alpha_t)} - m(\alpha_t)}{\sqrt{2v(\alpha_t)}}\right)^2}\right] = f(\hat{r}_{(t,\alpha_t)}) * w(\hat{r}_{(t,\alpha_t)}) \quad (D.2)$$

where,

$$f(\hat{r}_{(t,\alpha_t)}) = \frac{1}{\sqrt{2\pi v(\alpha_t)}} e^{-\left(\frac{\hat{r}_{(t,\alpha_t)} - m(\alpha_t)}{\sqrt{2v(\alpha_t)}}\right)^2} \quad (D.3)$$

and,

$$w(\hat{r}_{(t,\alpha_t)}) = \left[\frac{v'(\alpha_t)\left(\hat{r}_{(t,\alpha_t)} - m(\alpha_t)\right)^2}{2v(\alpha_t)^2} + \frac{m'(\alpha_t)\left(\hat{r}_{(t,\alpha_t)} - m(\alpha_t)\right)}{v(\alpha_t)} - \frac{v'(\alpha_t)}{2v(\alpha_t)}\right] \quad (D.4)$$

so that,

$$\frac{\partial^2}{\partial \alpha_t^2}\left[\frac{1}{\sqrt{2\pi v(\alpha_t)}} e^{-\left(\frac{\hat{r}_{(t,\alpha_t)} - m(\alpha_t)}{\sqrt{2v(\alpha_t)}}\right)^2}\right] = f(\hat{r}_{(t,\alpha_t)}) * w'(\hat{r}_{(t,\alpha_t)}) + f'(\hat{r}_{(t,\alpha_t)}) * w(\hat{r}_{(t,\alpha_t)}) \quad (D.5)$$

$$= f(\hat{r}_{(t,\alpha_t)}) * w'(\hat{r}_{(t,\alpha_t)}) + f(\hat{r}_{(t,\alpha_t)}) * \left[w(\hat{r}_{(t,\alpha_t)})\right]^2, \quad (D.6)$$

since $f'(\hat{r}_{(t,\alpha_t)}) = f(\hat{r}_{(t,\alpha_t)}) * w(\hat{r}_{(t,\alpha_t)})$, from (D.2) above. Repeated applications of the product, quotient, and chain rules (Anton (1988)) to the term $w(\hat{r}_{(t,\alpha_t)})$ yields the following polynomial expression in $(\hat{r}_{(t,\alpha_t)} - m(\alpha_t))$ for $w'(\hat{r}_{(t,\alpha_t)})$:

$$w'(\hat{r}_{(t,\alpha_t)}) = \frac{v(\alpha_t)v''(\alpha_t) - 2v'(\alpha_t)^2}{2v(\alpha_t)^3} *$$

$$\left[\left(\hat{r}_{(t,\alpha_t)} - m(\alpha_t)\right)^2 - 2\left(\frac{2v(\alpha_t)v'(\alpha_t)m'(\alpha_t)}{v(\alpha_t)v''(\alpha_t) - 2v'(\alpha_t)^2}\right)\left(\hat{r}_{(t,\alpha_t)} - m(\alpha_t)\right) \right. \quad (D.7)$$

$$\left. - \left(\frac{v(\alpha_t)[v(\alpha_t)v''(\alpha_t) - v'(\alpha_t)^2 + 2v(\alpha_t)m'(\alpha_t)^2]}{v(\alpha_t)v''(\alpha_t) - 2v'(\alpha_t)^2}\right)\right].$$

Completing the square in $\left(\hat{r}_{(t,\alpha_t)} - m(\alpha_t)\right)$ gives:

$$w'\left(\hat{r}_{(t,\alpha_t)}\right) = \frac{v(\alpha_t)v''(\alpha_t) - 2v'(\alpha_t)^2}{2v(\alpha_t)^3} *$$

$$\left[\left(\hat{r}_{(t,\alpha_t)} - m(\alpha_t)\right)^2 - 2\left(\frac{2v(\alpha_t)v'(\alpha_t)m'(\alpha_t)}{v(\alpha_t)v''(\alpha_t) - 2v'(\alpha_t)^2}\right)\left(\hat{r}_{(t,\alpha_t)} - m(\alpha_t)\right)\right.$$

$$+ \left(\frac{2v(\alpha_t)v'(\alpha_t)m'(\alpha_t)}{v(\alpha_t)v''(\alpha_t) - 2v'(\alpha_t)^2}\right)^2$$

$$\left. - \left(\frac{2v(\alpha_t)v'(\alpha_t)m'(\alpha_t)}{v(\alpha_t)v''(\alpha_t) - 2v'(\alpha_t)^2}\right)^2 - \left(\frac{v(\alpha_t)[v(\alpha_t)v''(\alpha_t) - v'(\alpha_t)^2 + 2v(\alpha_t)m'(\alpha_t)^2]}{v(\alpha_t)v''(\alpha_t) - 2v'(\alpha_t)^2}\right)\right].$$

(D.8)

And, finally:

$$w'\left(\hat{r}_{(t,\alpha_t)}\right) = \frac{v(\alpha_t)v''(\alpha_t) - 2v'(\alpha_t)^2}{2v(\alpha_t)^3} * \left[\left(\hat{r}_{(t,\alpha_t)} - m(\alpha_t)\right) - \left(\frac{2v(\alpha_t)v'(\alpha_t)m'(\alpha_t)}{v(\alpha_t)v''(\alpha_t) - 2v'(\alpha_t)^2}\right)\right]^2$$

$$- \left(\frac{2v'(\alpha_t)^2 m'(\alpha_t)^2}{v(\alpha_t)[v(\alpha_t)v''(\alpha_t) - 2v'(\alpha_t)^2]}\right) - \left(\frac{v(\alpha_t)v''(\alpha_t) - v'(\alpha_t)^2 + 2v(\alpha_t)m'(\alpha_t)^2}{2v(\alpha_t)^2}\right)$$

(D.9)

The resulting 2$^{nd}$ derivative from (D.1) leaves the following 3 quantities to manage:

$$\frac{\partial^2}{\partial \alpha_t^2}\left[\frac{1}{\sqrt{2\pi v(\alpha_t)}} e^{-\left(\frac{\hat{r}_{(t,\alpha_t)} - m(\alpha_t)}{\sqrt{2v(\alpha_t)}}\right)^2}\right] =$$

$$\frac{\theta_t}{2v(\alpha_t)^3} * \left[\left(\hat{r}_{(t,\alpha_t)} - m(\alpha_t)\right) - \left(\frac{2v(\alpha_t)v'(\alpha_t)m'(\alpha_t)}{\theta_t}\right)\right]^2 f\left(\hat{r}_{(t,\alpha_t)}\right) \quad \text{(D.10a)}$$

$$+ \left(\frac{1}{v(\alpha_t)}\right)^2 \left[\frac{v'(\alpha_t)}{2v(\alpha_t)}\left(\hat{r}_{(t,\alpha_t)} - m(\alpha_t)\right)^2 + m'(\alpha_t)\left(\hat{r}_{(t,\alpha_t)} - m(\alpha_t)\right) - \frac{v'(\alpha_t)}{2}\right]^2 f\left(\hat{r}_{(t,\alpha_t)}\right) \quad \text{(D.10b)}$$

$$- \left[\left(\frac{2v'(\alpha_t)^2 m'(\alpha_t)^2}{v(\alpha_t) * \theta_t}\right) + \left(\frac{v(\alpha_t)v''(\alpha_t) - v'(\alpha_t)^2 + 2v(\alpha_t)m'(\alpha_t)^2}{2v(\alpha_t)^2}\right)\right] f\left(\hat{r}_{(t,\alpha_t)}\right), \quad \text{(D.10c)}$$

where, from (4.34),

$$\theta_t = v(\alpha_t)v''(\alpha_t) - 2v'(\alpha_t)^2. \tag{D.11}$$

We now multiply and divide the 1st quantity, (D.10a), by $v(\alpha_t) + \left[\frac{2v(\alpha_t)v'(\alpha_t)m'(\alpha_t)}{\theta_t}\right]^2$, and the 2nd quantity (D.10b) by $\left[\frac{2}{v'(\alpha_t)^2 + 2v(\alpha_t)m'(\alpha_t)^2}\right]$, which yields:

$$\frac{\partial^2}{\partial \alpha_t^2}\left[\frac{1}{\sqrt{2\pi v(\alpha_t)}}e^{-\left(\frac{\hat{r}_{(t,\alpha_t)} - m(\alpha_t)}{\sqrt{2v(\alpha_t)}}\right)^2}\right] =$$

$$\left[\frac{\theta_t}{2v(\alpha_t)^2} + \frac{2v'(\alpha_t)^2 m'(\alpha_t)^2}{\theta_t * v(\alpha_t)}\right] * \frac{\left[\left(\hat{r}_{(t,\alpha_t)} - m(\alpha_t)\right) - \left(\frac{2v(\alpha_t)v'(\alpha_t)m'(\alpha_t)}{\theta_t}\right)\right]^2}{v(\alpha_t) + \left[\frac{2v(\alpha_t)v'(\alpha_t)m'(\alpha_t)}{\theta_t}\right]^2} f\left(\hat{r}_{(t,\alpha_t)}\right)$$

$$+ \left(\frac{v'(\alpha_t)^2 + 2v(\alpha_t)m'(\alpha_t)^2}{2v(\alpha_t)^2}\right) * \qquad \text{(D.12)}$$

$$\frac{2\left[\frac{v'(\alpha_t)}{2v(\alpha_t)}\left(\hat{r}_{(t,\alpha_t)} - m(\alpha_t)\right)^2 + m'(\alpha_t)\left(\hat{r}_{(t,\alpha_t)} - m(\alpha_t)\right) - \frac{v'(\alpha_t)}{2}\right]^2}{v'(\alpha_t)^2 + 2v(\alpha_t)m'(\alpha_t)^2} f\left(\hat{r}_{(t,\alpha_t)}\right)$$

$$-\left[\left(\frac{2v'(\alpha_t)^2 m'(\alpha_t)^2}{v(\alpha_t) * \theta_t}\right) + \left(\frac{v(\alpha_t)v''(\alpha_t) - v'(\alpha_t)^2 + 2v(\alpha_t)m'(\alpha_t)^2}{2v(\alpha_t)^2}\right)\right] f\left(\hat{r}_{(t,\alpha_t)}\right).$$

Lastly, using the quantities $Q_t^{(1)}, Q_t^{(2)}$, and $Q_t^{(3)}$ introduced in (4.29), (4.31), and (4.33), along with $h_1(\hat{r}_{(t,\alpha_t)})$ and $h_2(\hat{r}_{(t,\alpha_t)})$ from (4.30) and (4.32), respectively, (D.1) is given by:

(D.13)
$$\frac{\partial^2}{\partial \alpha_t^2}\left[\frac{1}{\sqrt{2\pi v(\alpha_t)}}e^{-\left(\frac{\hat{r}_{(t,\alpha_t)} - m(\alpha_t)}{\sqrt{2v(\alpha_t)}}\right)^2}\right] = Q_t^{(1)} * h_1\left(\hat{r}_{(t,\alpha_t)}\right) + Q_t^{(2)} * h_2\left(\hat{r}_{(t,\alpha_t)}\right) + Q_t^{(3)} * f\left(\hat{r}_{(t,\alpha_t)}\right).$$

The final step, which is trivial and not shown here, is to insert this quantity back into (4.27) then pull terms that do not depend on the RVs, $\hat{r}_{(t,\alpha_t)}$, out as constants (that is, $Q_t^{(1)}, Q_t^{(2)}$, and $Q_t^{(3)}$). The result is given in (4.28[a-c]).

# Appendix E. Proof that $h_1(\hat{r}_{(t,\alpha_t)})$ is a Valid PDF

For $h_1(\hat{r}_{(t,\alpha_t)})$ to be a valid PDF it must be $\geq 0$ for $-\infty < \hat{r}_{(t,\alpha_t)} < \infty$ and integrate to 1.0 over its domain (Casella and Berger (1990)). Clearly for $-\infty < \hat{r}_{(t,\alpha_t)} < \infty$,

$$h_1(\hat{r}_{(t,\alpha_t)}) = \frac{\left[\left(\hat{r}_{(t,\alpha_t)} - m(\alpha_t)\right) - \left(\frac{2v(\alpha_t)v'(\alpha_t)m'(\alpha_t)}{\theta_t}\right)\right]^2}{v(\alpha_t) + \left[\frac{2v(\alpha_t)v'(\alpha_t)m'(\alpha_t)}{\theta_t}\right]^2} f(\hat{r}_{(t,\alpha_t)}) \tag{E.1}$$

is $\geq 0$ everywhere since all terms are non-negative. Further,

$$\int_{-\infty}^{\infty} \frac{\left[\left(\hat{r}_{(t,\alpha_t)} - m(\alpha_t)\right) - \left(\frac{2v(\alpha_t)v'(\alpha_t)m'(\alpha_t)}{\theta_t}\right)\right]^2}{v(\alpha_t) + \left[\frac{2v(\alpha_t)v'(\alpha_t)m'(\alpha_t)}{\theta_t}\right]^2} f(\hat{r}_{(t,\alpha_t)}) \, d(\hat{r}_{(t,\alpha_t)}) \tag{E.2}$$

$$= \frac{1}{v(\alpha_t) + \left[\frac{2v(\alpha_t)v'(\alpha_t)m'(\alpha_t)}{\theta_t}\right]^2} * \left[ \int_{-\infty}^{\infty} \left(\hat{r}_{(t,\alpha_t)} - m(\alpha_t)\right)^2 f(\hat{r}_{(t,\alpha_t)}) \, d(\hat{r}_{(t,\alpha_t)}) \right.$$

$$\left. - \frac{4v(\alpha_t)v'(\alpha_t)m'(\alpha_t)}{\theta_t} \int_{-\infty}^{\infty} \left(\hat{r}_{(t,\alpha_t)} - m(\alpha_t)\right) f(\hat{r}_{(t,\alpha_t)}) \, d(\hat{r}_{(t,\alpha_t)}) \right. \tag{E.3}$$

$$\left. + \left(\frac{2v(\alpha_t)v'(\alpha_t)m'(\alpha_t)}{\theta_t}\right)^2 \int_{-\infty}^{\infty} f(\hat{r}_{(t,\alpha_t)}) \, d(\hat{r}_{(t,\alpha_t)}) \right]$$

Using the definition of mean and variance, the bracketed term in (E.3) is $[\cdot] = [v(\alpha_t) - 0 + \left(\frac{2v(\alpha_t)v'(\alpha_t)m'(\alpha_t)}{\theta_t}\right)^2]$, so that (E.3) becomes:

$$\left(\frac{1}{v(\alpha_t) + \left[\frac{2v(\alpha_t)v'(\alpha_t)m'(\alpha_t)}{\theta_t}\right]^2}\right) * \left(v(\alpha_t) + \left[\frac{2v(\alpha_t)v'(\alpha_t)m'(\alpha_t)}{\theta_t}\right]^2\right) \tag{E.4}$$

$$= 1, \tag{E.5}$$

which completes the proof that $h_1(\hat{r}_{(t,\alpha_t)})$ from (4.30) is a valid PDF.

## Appendix F. Proof that $h_2(\hat{r}_{(t,\alpha_t)})$ is a Valid PDF

For $h_2(\hat{r}_{(t,\alpha_t)})$ to be a valid PDF it must be $\geq 0$ and integrate to 1.0 over its domain $-\infty < \hat{r}_{(t,\alpha_t)} < \infty$ (Casella and Berger (1990)). The PDF,

$$h_2(\hat{r}_{(t,\alpha_t)}) = \frac{2\left[\frac{v'(\alpha_t)}{2v(\alpha_t)}\left(\hat{r}_{(t,\alpha_t)} - m(\alpha_t)\right)^2 + m'(\alpha_t)\left(\hat{r}_{(t,\alpha_t)} - m(\alpha_t)\right) - \frac{v'(\alpha_t)}{2}\right]^2}{v'(\alpha_t)^2 + 2v(\alpha_t)m'(\alpha_t)^2} f(\hat{r}_{(t,\alpha_t)}), \quad \text{(F.1)}$$

is $\geq 0$ everywhere since all terms are non-negative. Also,

$$\int_{-\infty}^{\infty} \frac{2\left[\frac{v'(\alpha_t)}{2v(\alpha_t)}\left(\hat{r}_{(t,\alpha_t)} - m(\alpha_t)\right)^2 + m'(\alpha_t)\left(\hat{r}_{(t,\alpha_t)} - m(\alpha_t)\right) - \frac{v'(\alpha_t)}{2}\right]^2}{v'(\alpha_t)^2 + 2v(\alpha_t)m'(\alpha_t)^2} f(\hat{r}_{(t,\alpha_t)}) \, d(\hat{r}_{(t,\alpha_t)}) \quad \text{(F.2)}$$

$$= \frac{2}{v'(\alpha_t)^2 + 2v(\alpha_t)m'(\alpha_t)^2} * \left[\left(\frac{v'(\alpha_t)}{2v(\alpha_t)}\right)^2 \int_{-\infty}^{\infty} \left(\hat{r}_{(t,\alpha_t)} - m(\alpha_t)\right)^4 f(\hat{r}_{(t,\alpha_t)}) \, d(\hat{r}_{(t,\alpha_t)})\right.$$

$$+ \frac{m'(\alpha_t)v'(\alpha_t)}{v(\alpha_t)} \int_{-\infty}^{\infty} \left(\hat{r}_{(t,\alpha_t)} - m(\alpha_t)\right)^3 f(\hat{r}_{(t,\alpha_t)}) \, d(\hat{r}_{(t,\alpha_t)})$$

$$+ \left(m'(\alpha_t)^2 - \frac{v'(\alpha_t)^2}{2v(\alpha_t)}\right) \int_{-\infty}^{\infty} \left(\hat{r}_{(t,\alpha_t)} - m(\alpha_t)\right)^2 f(\hat{r}_{(t,\alpha_t)}) \, d(\hat{r}_{(t,\alpha_t)}) \quad \text{(F.3)}$$

$$- m'(\alpha_t)v'(\alpha_t) \int_{-\infty}^{\infty} \left(\hat{r}_{(t,\alpha_t)} - m(\alpha_t)\right) f(\hat{r}_{(t,\alpha_t)}) \, d(\hat{r}_{(t,\alpha_t)})$$

$$\left. + \left(\frac{v'(\alpha_t)}{2}\right)^2 \int_{-\infty}^{\infty} f(\hat{r}_{(t,\alpha_t)}) \, d(\hat{r}_{(t,\alpha_t)})\right]$$

$$= \left(\frac{2}{v'(\alpha_t)^2 + 2v(\alpha_t)m'(\alpha_t)^2}\right) *$$

$$\left[\frac{v'(\alpha_t)^2}{4v(\alpha_t)^2}\left(\frac{3v(\alpha_t)^2}{1}\right) + m'(\alpha_t)^2 v(\alpha_t) + \frac{v'(\alpha_t)^2}{4} - \frac{v'(\alpha_t)^2 v(\alpha_t)}{2v(\alpha_t)}\right] \quad \text{(F.4)}$$

$$= \left(\frac{2}{v'(\alpha_t)^2 + 2v(\alpha_t)m'(\alpha_t)^2}\right) * \left[\frac{3v'(\alpha_t)^2}{4} + m'(\alpha_t)^2 v(\alpha_t) + \frac{v'(\alpha_t)^2}{4} - \frac{2v'(\alpha_t)^2}{4}\right] \quad \text{(F.5)}$$

$$= \left(\frac{2}{v'(\alpha_t)^2 + 2v(\alpha_t)m'(\alpha_t)^2}\right) * \left[\frac{v'(\alpha_t)^2}{2} + \frac{2v(\alpha_t)m'(\alpha_t)^2}{2}\right] \tag{F.6}$$

$$= 1. \tag{F.7}$$

The derivations from (2.67) in Section II.L were applied when reducing (F.3). Note that when $f(\hat{r}_{(t,\alpha_t)})$ replaces $e^{-\left(\frac{x-\mu}{\sqrt{2\sigma^2}}\right)^2}$ in (2.66), the result from (2.67) must be divided by $\sqrt{2\pi\sigma^2}$. This completes the proof that $h_2(\hat{r}_{(t,\alpha_t)})$ from (4.32) is a valid PDF.

## Appendix G. Derivation of the CDF $H_1(\hat{r}_{(t,\alpha_t)})$ for $h_1(\hat{r}_{(t,\alpha_t)})$

Let $\hat{r}_{(t,\alpha_t)} \sim h_1(\hat{r}_{(t,\alpha_t)})$. The CDF $H_1(\hat{r}_{(t,\alpha_t)}) = P_{h_1}(\hat{r}_{(t,\alpha_t)} \leq r)$ of $h_1(\hat{r}_{(t,\alpha_t)})$ is defined as:

$$P_{h_1}(\hat{r}_{(t,\alpha_t)} \leq r) = \int_{-\infty}^{r} \frac{\left[\left(\hat{r}_{(t,\alpha_t)} - m(\alpha_t)\right) - \left(\frac{2v(\alpha_t)v'(\alpha_t)m'(\alpha_t)}{\theta_t}\right)\right]^2}{v(\alpha_t) + \left[\frac{2v(\alpha_t)v'(\alpha_t)m'(\alpha_t)}{\theta_t}\right]^2} f(\hat{r}_{(t,\alpha_t)}) \, d(\hat{r}_{(t,\alpha_t)}) \quad \text{(G.1)}$$

$$= \frac{1}{v(\alpha_t) + \left[\frac{2v(\alpha_t)v'(\alpha_t)m'(\alpha_t)}{\theta_t}\right]^2} * \left[ \int_{-\infty}^{r} \left(\hat{r}_{(t,\alpha_t)} - m(\alpha_t)\right)^2 f(\hat{r}_{(t,\alpha_t)}) \, d(\hat{r}_{(t,\alpha_t)}) \right.$$

$$- \frac{4v(\alpha_t)v'(\alpha_t)m'(\alpha_t)}{\theta_t} \int_{-\infty}^{r} \left(\hat{r}_{(t,\alpha_t)} - m(\alpha_t)\right) f(\hat{r}_{(t,\alpha_t)}) \, d(\hat{r}_{(t,\alpha_t)}) \quad \text{(G.2)}$$

$$\left. + \left(\frac{2v(\alpha_t)v'(\alpha_t)m'(\alpha_t)}{\theta_t}\right)^2 \int_{-\infty}^{r} f(\hat{r}_{(t,\alpha_t)}) \, d(\hat{r}_{(t,\alpha_t)}) \right]$$

$$= \left(\frac{1}{v(\alpha_t) + \left[\frac{2v(\alpha_t)v'(\alpha_t)m'(\alpha_t)}{\theta_t}\right]^2}\right)$$

$$* \left[ \frac{1}{\sqrt{2\pi v(\alpha_t)}} \int_{-\infty}^{r} \left(\hat{r}_{(t,\alpha_t)} - m(\alpha_t)\right)^2 e^{-\left(\frac{\hat{r}_{(t,\alpha_t)} - m(\alpha_t)}{\sqrt{2v(\alpha_t)}}\right)^2} d(\hat{r}_{(t,\alpha_t)}) \right.$$

$$- \frac{4v(\alpha_t)v'(\alpha_t)m'(\alpha_t)}{\theta_t \sqrt{2\pi v(\alpha_t)}} \int_{-\infty}^{r} \left(\hat{r}_{(t,\alpha_t)} - m(\alpha_t)\right) e^{-\left(\frac{\hat{r}_{(t,\alpha_t)} - m(\alpha_t)}{\sqrt{2v(\alpha_t)}}\right)^2} d(\hat{r}_{(t,\alpha_t)}) \quad \text{(G.3)}$$

$$\left. + \left(\frac{2v(\alpha_t)v'(\alpha_t)m'(\alpha_t)}{\theta_t}\right)^2 \int_{-\infty}^{r} \frac{1}{\sqrt{2\pi v(\alpha_t)}} e^{-\left(\frac{\hat{r}_{(t,\alpha_t)} - m(\alpha_t)}{\sqrt{2v(\alpha_t)}}\right)^2} d(\hat{r}_{(t,\alpha_t)}) \right]$$

$$= \left(\frac{1}{v(\alpha_t) + \left[\frac{2v(\alpha_t)v'(\alpha_t)m'(\alpha_t)}{\theta_t}\right]^2}\right) *$$

$$\left[\frac{v(\alpha_t)}{2}\left(1 - (-1)^{\mathbf{1}_{(m(\alpha_t),\infty)}(r)} F_{g[\frac{3}{2},1]}\left(\left(\frac{r - m(\alpha_t)}{\sqrt{2v(\alpha_t)}}\right)^2\right)\right)\right.$$

$$+ \left(\frac{\sqrt{2v(\alpha_t)}^3 v'(\alpha_t)m'(\alpha_t)}{\theta_t\sqrt{\pi}}\right)\left(1 - F_{g[1,1]}\left(\left(\frac{r - m(\alpha_t)}{\sqrt{2v(\alpha_t)}}\right)^2\right)\right)$$

$$\left. + \left(\frac{2v(\alpha_t)v'(\alpha_t)m'(\alpha_t)}{\theta_t}\right)^2 \Phi\left(\frac{r - m(\alpha_t)}{\sqrt{v(\alpha_t)}}\right)\right] \quad \text{(G.4)}$$

$$= H_{1,t}^{(0)} * \left[H_{1,t}^{(1)}\left(1 - (-1)^{\mathbf{1}_{(m(\alpha_t),\infty)}(r)} F_{g[\frac{3}{2},1]}\left(\left(\frac{r - m(\alpha_t)}{\sqrt{2v(\alpha_t)}}\right)^2\right)\right)\right.$$

$$\left. + H_{1,t}^{(2)}\left(1 - F_{g[1,1]}\left(\left(\frac{r - m(\alpha_t)}{\sqrt{2v(\alpha_t)}}\right)^2\right)\right) + H_{1,t}^{(3)}\Phi\left(\frac{r - m(\alpha_t)}{\sqrt{v(\alpha_t)}}\right)\right], \quad \text{(G.5)}$$

where $H_{1,t}^{(0)}$, $H_{1,t}^{(1)}$, $H_{1,t}^{(2)}$, and $H_{1,t}^{(3)}$ are as defined in (4.37), and the result from (2.67) was used to simplify the expressions.

## Appendix H. Derivation of the CDF $H_2(\hat{r}_{(t,\alpha_t)})$ for $h_2(\hat{r}_{(t,\alpha_t)})$

Let $\hat{r}_{(t,\alpha_t)} \sim h_2(\hat{r}_{(t,\alpha_t)})$. The CDF $H_2(\hat{r}_{(t,\alpha_t)}) = P_{h_2}(\hat{r}_{(t,\alpha_t)} \le r)$ of $h_2(\hat{r}_{(t,\alpha_t)})$ is defined as:

$$\int_{-\infty}^{r} \frac{2\left[\frac{v'(\alpha_t)}{2v(\alpha_t)}\left(\hat{r}_{(t,\alpha_t)} - m(\alpha_t)\right)^2 + m'(\alpha_t)\left(\hat{r}_{(t,\alpha_t)} - m(\alpha_t)\right) - \frac{v'(\alpha_t)}{2}\right]^2}{v'(\alpha_t)^2 + 2v(\alpha_t)m'(\alpha_t)^2} f(\hat{r}_{(t,\alpha_t)}) d(\hat{r}_{(t,\alpha_t)}) \quad (H.1)$$

$$= \left(\frac{2}{v'(\alpha_t)^2 + 2v(\alpha_t)m'(\alpha_t)^2}\right) *$$

$$\left[\left(\frac{v'(\alpha_t)^2}{4\sqrt{v(\alpha_t)}^5 \sqrt{2\pi}}\right) \int_{-\infty}^{r} \left(\hat{r}_{(t,\alpha_t)} - m(\alpha_t)\right)^4 e^{-\left(\frac{\hat{r}_{(t,\alpha_t)} - m(\alpha_t)}{\sqrt{2v(\alpha_t)}}\right)^2} d(\hat{r}_{(t,\alpha_t)})\right.$$

$$+ \frac{m'(\alpha_t)v'(\alpha_t)}{\sqrt{v(\alpha_t)}^3 \sqrt{2\pi}} \int_{-\infty}^{r} \left(\hat{r}_{(t,\alpha_t)} - m(\alpha_t)\right)^3 e^{-\left(\frac{\hat{r}_{(t,\alpha_t)} - m(\alpha_t)}{\sqrt{2v(\alpha_t)}}\right)^2} d(\hat{r}_{(t,\alpha_t)})$$

$$+ \left(\frac{2v(\alpha_t)m'(\alpha_t)^2 - v'(\alpha_t)^2}{2\sqrt{v(\alpha_t)}^3 \sqrt{2\pi}}\right) \int_{-\infty}^{r} \left(\hat{r}_{(t,\alpha_t)} - m(\alpha_t)\right)^2 e^{-\left(\frac{\hat{r}_{(t,\alpha_t)} - m(\alpha_t)}{\sqrt{2v(\alpha_t)}}\right)^2} d(\hat{r}_{(t,\alpha_t)}) \quad (H.2)$$

$$- \frac{m'(\alpha_t)v'(\alpha_t)}{\sqrt{2\pi v(\alpha_t)}} \int_{-\infty}^{r} \left(\hat{r}_{(t,\alpha_t)} - m(\alpha_t)\right) e^{-\left(\frac{\hat{r}_{(t,\alpha_t)} - m(\alpha_t)}{\sqrt{2v(\alpha_t)}}\right)^2} d(\hat{r}_{(t,\alpha_t)})$$

$$+ \left(\frac{v'(\alpha_t)}{2}\right)^2 \int_{-\infty}^{r} f(\hat{r}_{(t,\alpha_t)}) d(\hat{r}_{(t,\alpha_t)}) \Bigg]$$

Repeated applications of the identity in (2.67) to (H.2) yields $H_2(\hat{r}_{(t,\alpha_t)}) = P_{h_2}(\hat{r}_{(t,\alpha_t)} \le r)$ as:

$$\left(\frac{2}{v'(\alpha_t)^2 + 2v(\alpha_t)m'(\alpha_t)^2}\right) * \left[\left(\frac{3v'(\alpha_t)^2}{8}\right)\left(1 - (-1)^{\mathbf{1}_{(m(\alpha_t),\infty)}(r)} F_{g[\frac{5}{2},1]}\left(\left(\frac{r - m(\alpha_t)}{\sqrt{2v(\alpha_t)}}\right)^2\right)\right)\right.$$

$$- \frac{m'(\alpha_t)v'(\alpha_t)\sqrt{2v(\alpha_t)}}{\sqrt{\pi}}\left(1 - F_{g[2,1]}\left(\left(\frac{r - m(\alpha_t)}{\sqrt{2v(\alpha_t)}}\right)^2\right)\right) \quad (H.3a)$$

$$+ \left(\frac{2v(\alpha_t)m'(\alpha_t)^2 - v'(\alpha_t)^2}{4}\right)\left(1 - (-1)^{\mathbf{1}_{(m(\alpha_t),\infty)}(r)} F_{g[\frac{3}{2},1]}\left(\left(\frac{r - m(\alpha_t)}{\sqrt{2v(\alpha_t)}}\right)^2\right)\right)$$

$$+\frac{m'(\alpha_t)v'(\alpha_t)\sqrt{v(\alpha_t)}}{\sqrt{2\pi}}\left(1 - F_{g[1,1]}\left(\left(\frac{r - m(\alpha_t)}{\sqrt{2v(\alpha_t)}}\right)^2\right)\right)$$

(H.3b)

$$+\frac{v'(\alpha_t)^2}{4}\Phi\left(\frac{r - m(\alpha_t)}{\sqrt{v(\alpha_t)}}\right)\Bigg].$$

Using the quantities $H_{2,t}^{(0)}$, $H_{2,t}^{(1)}$, $H_{2,t}^{(2)}$, $H_{2,t}^{(3)}$, $H_{2,t}^{(4)}$, and $H_{2,t}^{(5)}$ as defined in (4.39) yields the following:

$$H2(\hat{r}_{(t,\alpha_t)}) = P_{h_2}(\hat{r}_{(t,\alpha_t)} \leq r) = H_{2,t}^{(0)}\Bigg[H_{2,t}^{(1)}\left(1 - (-1)^{\mathbf{1}_{(m(\alpha_t),\infty)}(r)}F_{g[\frac{5}{2},1]}\left(\left(\frac{r - m(\alpha_t)}{\sqrt{2v(\alpha_t)}}\right)^2\right)\right)$$

$$+ H_{2,t}^{(2)}\left(1 - F_{g[2,1]}\left(\left(\frac{r - m(\alpha_t)}{\sqrt{2v(\alpha_t)}}\right)^2\right)\right)$$

(H.4)

$$+ H_{2,t}^{(3)}\left(1 - (-1)^{\mathbf{1}_{(m(\alpha_t),\infty)}(r)}F_{g[\frac{3}{2},1]}\left(\left(\frac{r - m(\alpha_t)}{\sqrt{2v(\alpha_t)}}\right)^2\right)\right)$$

$$+ H_{2,t}^{(4)}\left(1 - F_{g[1,1]}\left(\left(\frac{r - m(\alpha_t)}{\sqrt{2v(\alpha_t)}}\right)^2\right)\right) + H_{2,t}^{(5)}\Phi\left(\frac{r - m(\alpha_t)}{\sqrt{v(\alpha_t)}}\right)\Bigg],$$

as was to be shown.

# Appendix I. Quasi-Concave Counter Example for $P_{NR}(\vec{\alpha})$ when $T_D=2$

Consider a retirement horizon of length $T_D=2$, and define glidepaths $\vec{\alpha}_1$ and $\vec{\alpha}_2$ as:

$$\vec{\alpha}_1 = \begin{pmatrix} \alpha_{11} \\ \alpha_{12} \end{pmatrix} = \begin{pmatrix} 0.439547 \\ 0.137059 \end{pmatrix}, \quad \vec{\alpha}_2 = \begin{pmatrix} \alpha_{21} \\ \alpha_{22} \end{pmatrix} = \begin{pmatrix} 0.140591 \\ 0.999999 \end{pmatrix} \tag{I.1}$$

Let $\lambda=0.688882$, so that $(1-\lambda)=0.311118$, and:

$$\vec{\alpha}_c = \begin{pmatrix} \alpha_{c1} \\ \alpha_{c2} \end{pmatrix} = \begin{pmatrix} \lambda\alpha_{11} + (1-\lambda)\alpha_{21} \\ \lambda\alpha_{12} + (1-\lambda)\alpha_{22} \end{pmatrix} = \begin{pmatrix} 0.346536 \\ 0.405535 \end{pmatrix}. \tag{I.2}$$

Finally, let $RF(0) = 0.586352$ ($\rightarrow W_R \approx 58.6\%$) recalling that this is a 2-period retirement horizon. Using simulation with sample size n=100 million we estimate the following success probabilities for these 3 glidepaths:

$$P_{NR}(\vec{\alpha}_1) = 0.158522 \tag{I.3a}$$

$$P_{NR}(\vec{\alpha}_2) = 0.190762 \tag{I.3b}$$

$$P_{NR}(\vec{\alpha}_c) = 0.148574 \tag{I.3c}$$

The differences of interest are $Z = P_{NR}(\vec{\alpha}_c) - P_{NR}(\vec{\alpha}_2)$ from the objective function (4.65) and $P_{NR}(\vec{\alpha}_1) - P_{NR}(\vec{\alpha}_c)$ from the constraint (4.66). Using the simulated probabilities in (I.3a), (I.3b), and (I.3c), these quantities are calculated as:

$$P_{NR}(\vec{\alpha}_c) - P_{NR}(\vec{\alpha}_2) = -0.042188 \tag{I.4a}$$

$$P_{NR}(\vec{\alpha}_1) - P_{NR}(\vec{\alpha}_c) = 0.009948 \tag{I.4b}$$

We have therefore identified glidepaths $\vec{\alpha}_1$, $\vec{\alpha}_2$, and a given convex combination $\vec{\alpha}_c$, such that $P_{NR}(\vec{\alpha}_c) < \text{Min}\{P_{NR}(\vec{\alpha}_1), P_{NR}(\vec{\alpha}_2)\}$. The NLP formulation approximated these quantities using the suggested grid approach with $Z_L=-13.1730$, $Z_U=13.1730$, and k=263,460 rectangles as:

$$\vec{P}^T * \vec{F}_{(C:2)} = -0.042246 \tag{I.5a}$$

$$\vec{P}^T * \vec{F}_{(1:C)} = 0.010014 \tag{I.5b}$$

The situation is depicted in Figure 8 below. Other quantities of interest relative to this figure are:

$$\begin{pmatrix} ZF_1(0) \\ ZF_2(0) \\ ZF_c(0) \end{pmatrix} = \begin{pmatrix} -4.545114 \\ -5.673628 \\ -5.056520 \end{pmatrix}, \text{ and } \begin{pmatrix} \frac{-m(\alpha_{12})}{\sqrt{v(\alpha_{12})}} \\ \frac{-m(\alpha_{22})}{\sqrt{v(\alpha_{22})}} \\ \frac{-m(\alpha_{c2})}{\sqrt{v(\alpha_{c2})}} \end{pmatrix} = \begin{pmatrix} -13.170457 \\ -5.394397 \\ -10.769658 \end{pmatrix}. \tag{I.6}$$

# Figure 8
## Quasi-Concave Counter Example for a Two-Period Retirement Horizon

This figure shows the quasi-concave counter example from Appendix I graphically for a 2-period retirement horizon. Here, $P_{NR}(\vec{\alpha}_c) < \text{Min}\{P_{NR}(\vec{\alpha}_1), P_{NR}(\vec{\alpha}_2)\}$ which implies that the function $P_{NR}(\vec{\alpha})$ as a surface has dips or valleys and is therefore not unimodal. Consequently, a hill climbing technique such as the one proposed within this research cannot guarantee that all local optimums are global optimums for reasonable $RF(0)$ (=$W_R$) and horizon lengths $T_D$. This is because there can be multiple hills. In this figure, retirement ruin is avoided when $(z_1, z_2)$ falls above and to the right of each curve which reflects the equation $z_2 = ZF_i(1)$ for $i=1,2,c$. The probability of avoiding ruin is the volume of the joint density $\phi(\cdot)$ above and to the right of each curve. The curve associated with the largest volume represents the glidepath with the greatest probability of avoiding financial ruin. In this figure it is visually reasonable that the curve associated with the convex combination glidepath $\vec{\alpha}_c$ has the lowest probability of avoiding financial ruin. The circles represent density contours out to 5 standard deviations and over 99.99% of the total probability is contained within the lightly drawn box around the largest circle. Further, if a box were drawn around the smallest circle it would contain over 68.2% of the total probability and this region is nearly missed by all 3 glidepaths. Note that when comparing 2 glidepaths it suffices to focus on the probability measure contained between their respective curves.

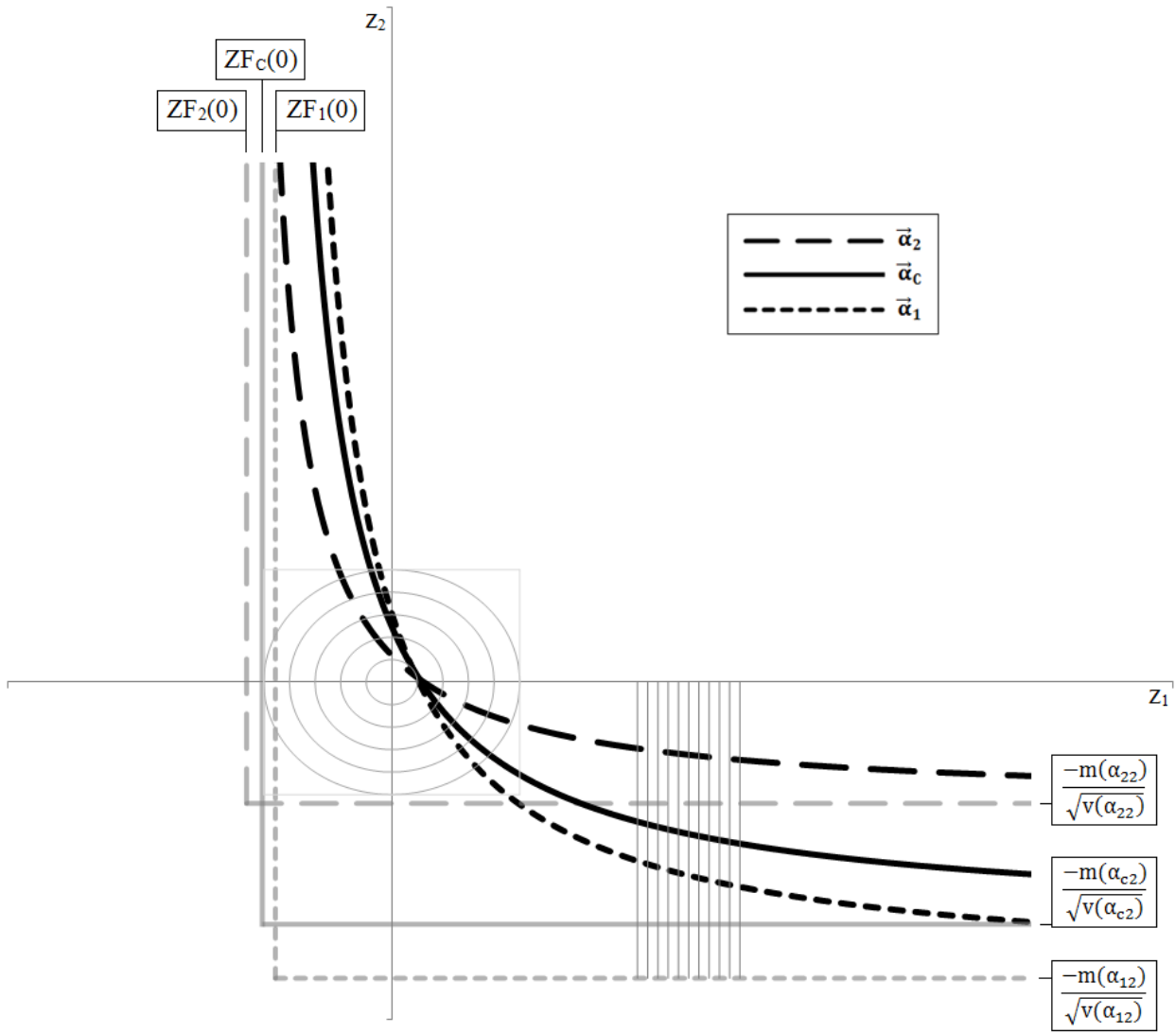

**Appendix J. Full C++ Implementation**

The source code provided consists of 13 functions and a header file that are compiled together into an executable (64-bit). It was used to generate the optimal glidepaths for Scenarios 1-8 and can be easily adapted to produce the optimal glidepaths in Scenarios 9 and 10 (using a simple looping mechanism). The user may submit one parameter to the executable which is the number of jobs to process concurrently. If the user supplies no value the code determines the number of concurrent processing units on the computer and exploits its full capacity. This is accomplished by multi-threading the calculations for estimates derived either via simulation or DP approximation. See the code for more details, specifically the functions ThrdPNRsim() and ThrdPNRdyn(). The freely available Boost® and Eigen® external libraries are used in this application and will need to be installed on the machine that compiles the source. Use caution if copying the code from this document as formatting issues can arise. For example, minus signs may be converted to hyphens. The code is being supplied under the open-source Apache License, Version 2.0.[1] Two files are mandatory and must exist in the directory specified by the user when launching the application: A control file ("control.txt") and a starting glidepath ("gp.txt"). The resulting optimal glidepath is written to a file ("output.txt") in the same directory. All details are displayed to the user on the screen as the program runs. In Section IV.E we discussed 4 ways to run the optimization, and examples of the control file for each arrangement are discussed next.

*A. Source Code Details*

We offer 4 different ways to perform the optimization. In most cases we recommend Newton's method with DP approximation since it generally converges faster than gradient ascent with DP approximation (2-4 iterations w/Hessian vs. 100-300 iterations w/out Hessian). When operating along the feasible region boundary we can define a related sub-problem, which is discussed below. The code will not automatically adjust the Hessian matrix to solve this sub-problem, but the existing gradient remains valid, pointing in the direction of steepest ascent. Therefore gradient ascent would be preferable along the boundary as was seen in Scenario #8 of Section IV.E. We do not recommend using simulation in general, however we do provide it as an option in the code. If simulation is used to estimate the quantities involved we supply 2 hypothesis test heuristics to achieve convergence faster. First, we allow the user to specify a Type I Error level for the hypothesis test that each gradient element equals 0. If this hypothesis is not rejected the gradient element is set to zero for that iteration, and we refer to the result as the *adjusted* gradient. This prevents climbing in the direction of sampling error which would need to be reversed at future iterations. We provide a 2$^{nd}$ test while climbing in the direction of the gradient, namely that the new probability is not inferior to the prior best probability. Adjusting the Type I Error level of this test allows the user to make good decisions during the climbing process, that is, decisions which take sampling error into account. As with noisy gradient estimates, poor decisions made while climbing need to be reversed at future iterations, therefore we benefit from avoiding them. We have experienced mixed results using these 2 heuristics in practice. Lastly, regardless of the optimization type chosen by the user, we always begin by taking a small number of steps (maximum of 50) in the direction of the gradient. This is a fast way to make good progress and it can also quickly move us off of the border when we select a poor starting glidepath. Details of 4 specific control file arrangements ("control.txt") are discussed next, followed by the input starting glidepath ("gp.txt"), and the output file containing the optimal glidepath ("output.txt").

---

[1] License details can be found at: http://www.apache.org/licenses/LICENSE-2.0. Contact the author with any questions/problems. <u>Reminder:</u> All code has bugs and the user shall assume responsibility for validation.

*A.1 Control File:* Newton's Method with Dynamic Programming Approximation

```
0.082509 0.0402696529 0.021409 0.0069605649 0.0007344180 0.000
30 0.05 0.00000000001
nr dp 5000 2.75
```

The 1$^{st}$ line of the control file contains the real stock and bond means, variances, covariance and the expense ratio in the following order: $\mu_s$, $\sigma^2_s$, $\mu_b$, $\sigma^2_b$, $\sigma_{(s,b)}$, and, $E_R$. This setup file uses historical returns. The 2$^{nd}$ line of the control file contains the horizon length, withdrawal rate, and epsilon convergence level as: $T_D$, $W_R$, and, $\varepsilon$. (<u>Note:</u> The first 2 lines will always contain the terms just described.) The 3$^{rd}$ line of the control file specifies the optimization scheme ("nr" vs. "ga"), the estimation/approximation method ("dp" vs. "sim"), followed by settings specific to the estimation/approximation method. When a DP is used the 2 settings are the discretization level ($P_R$) and the maximum ruin factor used during discretization ($RF_{Max}$). The control file shown here was used to produce the optimal glidepath in Scenario #1 of Section IV.E.

*A.2 Control File:  Gradient Ascent with Dynamic Programming Approximation*

```
0.055000 0.0428490000 0.017500 0.0042250000 0.0040365000 0.010
30 0.05 0.0000000013
ga dp 10000 2.75
```

The 1st and 2nd line of all control files will have terms specified in the same location and order.  The 3rd line of this control file specifies gradient ascent ("ga") as the optimization method and also uses DP approximation with $P_R$=10,000 and $RF_{Max}$=2.75.  This control file was used to produce the optimal glidepath in Scenario #8 of Section IV.E and assumes Evensky returns.

*A.3 Control File:  Newton's Method using Simulation Estimates*

```
0.082509 0.0402696529 0.021409 0.0069605649 0.0007344180 0.000
30 0.04 0.00005
nr sim 500000000 0.05 1.00
```

The 1st and 2nd line are the same as above and historical return assumptions are made here.  The 3rd line of this control file indicates that Newton's method will be used ("nr") along with simulation ("sim") as the estimation method.  When simulation is used there are 3 additional parameters to specify in the following order: N, alpha-level 1, alpha-level 2.  The sample size N provided becomes the base for the procedure which uses the following rules:  All $P_{NR}(\vec{\alpha})$ probabilities used in gradient calculations will have a sample size of 4*N.  Probabilities derived while climbing will use a sample size of 2*N, and all special probabilities will be derived using the sample size of N.  (Rejection sampling is computationally expensive.)

The alpha-level 1 setting contains the Type I Error probability for the non-inferiority hypothesis test used while climbing.  Probabilities estimated using simulation are subject to sampling error.  To account for this we test the hypothesis that the new probability is not inferior to the prior best probability while climbing in the direction of the gradient (see Section II.F).  Since we always start by taking a small number of steps in the direction of the gradient, this setting is required even when using Newton's method.  Setting alpha-level 1 = 0.5 will result in stepping only when the new probability exceeds the prior best probability, since $Z_{0.5} = 0$.  It thus reflects a way to turn this heuristic off if desired.  The non-inferiority test is increasingly forgiving as alpha-level 1 decreases, meaning that the test will rarely reject $H_o$ for very small alpha-levels which may result in over-climbing at each iteration.

The alpha-level 2 setting is used to adjust the gradient when estimates are produced using simulation. The null hypothesis tests 2 probabilities for equality (see Section II.E and the discussion at the end of Section IV.A). Gradient elements point in the direction of steepest ascent and noisy estimates lead us to climb in the wrong direction. We must backtrack and correct this during future iterations. If the gradient element is not significantly different from zero (that is, the test is not rejected) using Type I Error probability of alpha-level 2, we set it to zero for that iteration. The result is referred to as the *adjusted* gradient. Smaller alpha-levels are more forgiving and lead to more zeros in the adjusted gradient. Setting alpha-level 2 = 1.0 turns off this hypothesis test. If the user prefers estimation via simulation they can adjust these alpha levels as desired to achieve convergence faster for their particular arrangement.

*A.4 Control File:* Gradient Ascent using Simulation Estimates

```
0.055000 0.0428490000 0.017500 0.0042250000 0.0040365000 0.010
30 0.04 0.00005
ga sim 500000000 0.15 0.20
```

As noted, the 1$^{st}$ and 2$^{nd}$ lines always have the same structure, and this control file uses the Evensky return assumptions from Section IV.E. The 3$^{rd}$ line of this control file indicates that gradient ascent will be used with simulation ("sim") as the estimation method. The meaning of the 3 additional simulation parameters were given in A.3 and refer to N, alpha-level 1, and alpha-level 2, respectively. Here, the glidepath test for non-inferiority while climbing would use an alpha-level of 0.15. As this value increases the number of steps taken will generally decrease. The goal is not to take a large or small number of steps but the right number of steps at each iteration. Using this control file, the test of $H_o$: $g_t = 0$ vs. $H_a$: $g_t \neq 0$ for t=1, 2, …, 30 would use an alpha-level of 0.20. As this value increases the number of gradient elements set to zero decreases.

*A.5 Input File: Initial Glidepath*

The procedure requires an initial glidepath be specified and optimization begins at this point. If the surface being optimized has multiple local optimums then the starting glidepath determines where the procedure ends. In Section IV.E we found that all 5 starting glidepaths converge to the same optimum glidepath in Scenarios 1-8. The same was found for Scenarios 9 and 10 in Section V, using different starting glidepaths. Contents of "gp.txt" using Random Glidepath #1 from Figure 5 is shown next. We do not include blank lines or extra spaces in this file, and we start on line #1. Therefore this file has exactly 30 lines.

```
0.636
0.214
0.193
0.637
0.626
0.597
0.943
0.877
0.254
0.823
0.903
0.294
0.444
0.513
0.529
0.160
0.564
0.293
0.698
0.228
0.311
0.776
0.689
0.764
0.596
0.793
0.911
0.624
0.709
0.205
```

*A.5 Output File: Optimal Glidepath*

A single output file named "output.txt" is written to the directory that contains the control file and the initial glidepath. A sample of this file is shown below. GP[00] is $\alpha_1$, GP[01] is $\alpha_2$, etc... The first equity ratio is set at time t=0 and the corresponding return is observed at time t=1. This particular output file is from Scenario #7 of Section IV.E.

```
--> Success probability for this Glide-Path = 0.527952155270
GP[00]=+0.8235272966  GP[06]=+0.8590020928  GP[12]=+0.7307821069  GP[18]=+0.6235703362  GP[24]=+0.5436024264
GP[01]=+0.8617503896  GP[07]=+0.8386754281  GP[13]=+0.7107993614  GP[19]=+0.6085468596  GP[25]=+0.5323875470
GP[02]=+0.8850732670  GP[08]=+0.8170174723  GP[14]=+0.6916598009  GP[20]=+0.5942528104  GP[26]=+0.5216839028
GP[03]=+0.8931568162  GP[09]=+0.7949412393  GP[15]=+0.6733802825  GP[21]=+0.5806493950  GP[27]=+0.5114608960
GP[04]=+0.8890061069  GP[10]=+0.7730074379  GP[16]=+0.6559544071  GP[22]=+0.5676980327  GP[28]=+0.5016899088
GP[05]=+0.8765350143  GP[11]=+0.7515547956  GP[17]=+0.6393611079  GP[23]=+0.5553611039  GP[29]=+0.4923442815
```

*B. Implementation/Source Code Miscellanea*

*B.1 Operating Along the Boundary*

The methods proposed here are mainly designed to find interior point optimums, not those on the feasible region boundary. Consider the feasible region as defined in (3.3) along with a very high initial withdrawal rate relative to the horizon length $T_D$. The gradient will quickly drive some equity ratios to 1.0 and, remaining unsatisfied, it will point higher. Since we do not allow borrowing or shorting, the constraints in (3.3) prohibit equity ratios from exceeding 1.0. Therefore large gradient elements go unreduced while climbing and this can prevent convergence. For this reason, we only consider the effective gradient elements when determining convergence. That is, the magnitude that will not drive the solution outside of the feasible region. This situation was encountered in Scenario #8 from Section IV.E. The optimal solution requires equity ratios greater than 1.0 during the first decade of a 30-year retirement. Since the constraints will not allow this we become stuck at a sub-optimal solution and cannot drive all gradient elements to zero. When this occurs we can define a related sub-problem that assumes these equity ratios are 1.0 and removes them as parameters in the optimization. For example, consider the following sub-problem for Scenario #8:

Maximize: $\quad Z = P_{NR}(\vec{\alpha}) = P_{NR}(\vec{\alpha}, 30) = P(\text{Ruin}^C(\leq 30))$ \hfill (J.1)

Subject to: $\quad \alpha_t = 1.0 \quad\quad\quad\quad$ for t = 1, 2, ..., 10 \hfill (J.2)

$\quad\quad\quad\quad\quad MV_t(\alpha) + \varepsilon \leq \alpha_t \leq 1.0 \quad$ for t = 11, 12, ..., 30 \hfill (J.3)

This sub-problem has 20 parameters, not 30, and the corresponding gradient would be a 20 element vector having elements identical to $g_{11} - g_{30}$ as derived in (3.7) for the 30-dimensional problem with $\alpha_t = 1.0$, for t=1, 2, ..., 10. The point is that the gradient element for each dimension of the sub-problem matches the corresponding gradient element for the full problem and we can still use it to climb. The code will do this automatically at the border as was seen in Scenario #8. We can think of the solution as an interior-point optimum for the sub-problem that assumes the first 10 equity ratios are constants (= 1.0). The corresponding 20x20 Hessian matrix for the sub-problem would be required for Newton's method, and as written, the code will not automatically construct it. For this reason, Newton's method should be avoided when operating along the boundary with this implementation.

*B.2 Newton's Method can Converge to a Minimum*

Newton's method attempts to drive the gradient to zero which can reflect a local/global maximum or minimum. The procedure may attempt to find a minimum when very large or small withdrawal rates are used, relative to the horizon length. This reflects the fact that our procedure is designed to "climb a hill". If the surface being optimized resembles nothing shaped like a hill or section of it, then the procedure is unlikely to be effective. We recommend constant inspection of the largest and smallest eigenvalues of the Hessian matrix during optimization when using Newton's method as it indicates when the procedure is working as intended, and also when it isn't.

*B.3 Rejection Sampling Bounds*

Rejection sampling is used to simulate observations from unknown PDFs (Von Neumann (1951)). We can envision this approach as drawing a very tight box around the density function such that it includes nearly all area. Two uniform random values are then generated using domains that match the box length and height. One is for the horizontal axis (length) and one is for the vertical axis (height). If the 2$^{nd}$ uniform random value is less than the density evaluated at the first uniform random value we keep the observation. If not, we discard it. The observation is the value of the 1$^{st}$ uniform random variable. This is the approach used to simulate observations from the special gradient density, $g(\hat{r}_{(t,\alpha_t)})$, and the 2 special Hessian densities, $h_1(\hat{r}_{(t,\alpha_t)})$ and $h_2(\hat{r}_{(t,\alpha_t)})$. Note that these PDFs are functions of the stock and bond means, variances, and covariance from (4.1) and (4.2). We have not sized these boxes dynamically to work with generic distributional assumptions and the user should confirm they are sized properly if using simulation with different return assumptions. The limits on these boxes are set in the function PNRsim( ).

```
/*
/ Copyright (c) 2015 Chris Rook
/
/ Licensed under the Apache License, Version 2.0 (the "License");  you may not use this file except in compliance with the License.  You may obtain a copy of
/ the License at http://www.apache.org/licenses/LICENSE-2.0.  Unless required by applicable law or agreed to in writing, software distributed under the License
/ is distributed on an "AS IS" BASIS, WITHOUT WARRANTIES OR CONDITIONS OF ANY KIND, either express or implied.  See the License for the specific language
/ governing permissions and limitations under the License.
/
/ Filename:  stdafx.h  (Include this header at the top of all other code files.)
/
/ Summary:
/
/        This file includes headers that define prototypes for standard functions.  Various constants, inline functions and prototypes for functions
/        created within this application are also defined.  (See details below.)
/
/ Inline functions defined in this header file:
/
/        1.) m()   = The real/expense-adjusted return mean as a function of the equity ratio at a given time-point.
/        2.) v()   = The real/expense-adjusted return variance as a function of the equity ratio at a given time-point.
/        3.) mp()  = The derivative of m() with respect to the equity ratio at a given time-point.
/        4.) vp()  = The derivative of v() with respect to the equity ratio at a given time-point.
/        5.) vpp() = The 2nd derivative of v() with respect to the equity ratio at a given time-point.
/        6.) mva() = The equity ratio that reflects the minimum variance portfolio given the stock/bond return variance & covariance.
/        7.) kh1() = A quantity used in defining the special Hessian density h1().
/        8.) f()   = The density that represents real/expense-adjusted returns (historical distribution).
/        9.) g()   = The special density used for computing the gradient at each time-point, and also used to construct the Hessian.
/       10.) h1()  = The 1st special density used for computing the diagonal Hessian elements at each time-point.
/       11.) h2()  = The 2nd special density used for computing the diagonal Hessian elements at each time-point.
/
/ Function prototypes defined in this header file:
/
/        1.) GetPNR()    = Prototype for the function that returns the probability of avoiding ruin (i.e., success probability).
/        2.) GetConst()  = Prototype for the function that returns the vector of constants used to build any CDF call within this application.
/        3.) GetGamma()  = Prototype for the function that returns the vector of gamma RVs used to build any CDF call within this application.
/        4.) GetCDF()    = Prototype for the function that returns any CDF value needed within this application using the vectors of constants and gamma RVs.
/        5.) ThrdPNRsim()= Prototype for function that threads the simulation ruin computation across CPU cores to reduce processing time.
/        6.) ThrdPNRdyn()= Prototype for function that threads the dynamic programming ruin computation across CPU cores to reduce processing time.
/        7.) PNRsim()    = Prototype for function that derives the probability of no ruin (using simulation) for a given glide-path/withdrawal rate.
/        8.) PNRdyn()    = Prototype for function that derives the probability of no ruin (using dynamic programming) for a given glide-path/withdrawal rate.
/        9.) BldGrad()   = Prototype for function that accepts an empty array and populates it with the gradient vector elements for a given glide-path.
/       10.) Climb()     = Prototype for function that climbs in the direction of the gradient and stops when no further progress can be made.
/       11.) DrvHess()   = Prototype for function that derives and returns the Hessian matrix for a given glide-path.
/       12.) WrtAry()    = Prototype for function that writes an array (either the glide-path or the gradient) to standard output or a file as a block.
/-----------------------------------------------------------------------------------------------------------------------------------------------------*/
#pragma once

// Include header files.

#include "targetver.h"
#include <stdio.h>
#include <iomanip>
#include <string>
#include <tchar.h>
#include <math.h>
#include <stdlib.h>
#include <iostream>
#include <fstream>
```

```cpp
#include <random>
#include <string>
#include <algorithm>
#include <boost/math/distributions.hpp>
#include <boost/thread/thread.hpp>
#include <boost/algorithm/string.hpp>
#include <Eigen/Dense>
#include <Eigen/Eigenvalues>

using namespace std;

// Define constants.

const string paramfile="control.txt";       // Name of parameter control input file.
const string initgpfile="gp.txt";           // Name of input glidepath file.
const string outfile="output.txt";          // Name of output file.
const double pi = 3.1415926535897932384;    // Constant representation of PI.

// Define inline functions.

inline long double m(const long double prms[6], const long double a) {return (1.00-prms[5])*(1 + a*prms[0] + (1.00-a)*prms[2]);}
inline long double v(const long double prms[6], const long double a) {return pow(1.00-prms[5],2)*(a*a*prms[1] + (1.00-a)*(1.00-a)*prms[3] + 2.00*a*(1.00-a)*prms[4]);}
inline long double mp(const long double prms[6]) {return (1.00-prms[5])*(prms[0] - prms[2]);}
inline long double vp(const long double prms[6], const long double a) {return pow(1.00-prms[5],2)*(2.00*a*prms[1] - 2.00*(1.00-a)*prms[3] + (2.00 - 4.00*a)*prms[4]);}
inline long double vpp(const long double prms[6]) {return pow(1.00-prms[5],2)*(2.00*prms[1] + 2.00*prms[3] - 4.00*prms[4]);}
inline long double mva(const long double prms[6]) {return (prms[3] - prms[4])/(prms[1] + prms[3] - 2.00*prms[4]);}
inline long double kh1(const long double prms[6], const long double a) {return (-2.00*vp(prms,a)*mp(prms)*v(prms,a))/(v(prms,a)*vpp(prms) - 2.00*pow(vp(prms,a),2));}
inline long double f(const long double prms[6], const long double a, const long double r) {return (1/sqrt(2.00*pi*v(prms,a)))*exp(-pow(r-m(prms,a),2)/(2.00*v(prms,a)));}
inline long double g(const long double prms[6], const long double a, const long double r) {return f(prms,a,r)*pow(vp(prms,a)*(r-m(prms,a)) + mp(prms)*v(prms,a),2)/(pow(mp(prms),2)*pow(v(prms,a),2) + v(prms,a)*pow(vp(prms,a),2));}
inline long double h1(const long double prms[6], const long double a, const long double r) {return f(prms,a,r)*pow((r - m(prms,a) + kh1(prms,a)),2)/(v(prms,a) + pow(kh1(prms,a),2));}
inline long double h2(const long double prms[6], const long double a, const long double r) {return f(prms,a,r)*2.00*pow(pow(r-m(prms,a),2)*vp(prms,a)/(2.00*v(prms,a)) + mp(prms)*(r-m(prms,a)) - vp(prms,a)/2.00,2)/(pow(vp(prms,a),2) + 2.00*v(prms,a)*pow(mp(prms),2));}

// Define function prototypes.

long double GetPNR(const string type, const long double prms[6], long double * a, const int fxTD, const double RF0, const long long int n,
                   const long int nbuckets, const long prec, const int partls[4], const int plproc);
vector<long double> GetConst(const long double mts[7], const int partls[4], const int y);
vector<boost::math::gamma_distribution<>> GetGamma(const int partls[4], const int y);
long double GetCDF(const long double mts[7], const vector<long double> & C, const boost::math::normal n, const vector<boost::math::gamma_distribution<>> & G,
                  const double x);
long double ThrdPNRsim(const long double prms[6], long double * a, const int fxTD, const double RF0, const long long int n, const int partls[4],
                      const int plproc);
long double ThrdPNRdyn(const long double prms[6], long double * a, const int fxTD, const double RF0, const long int nbuckets, const int partls[4],
                      const int plproc, const long prec);
void PNRsim(const long double prms[6], const long double *a, const int fxTD, const double RF0, const long long int n, const int tnum, long double * probs,
            const int partls[4]);
void PNRdyn(const long double mts[7], const long pr, const long bkts[3], const long double * Vp, long double * V, const vector<long> & PrB,
            const vector<long double> & C, const vector<boost::math::gamma_distribution<>> & G);
long double BldGrad(const string type, const long double prms[6], long double * a, const int fxTD, const double RF0, const long long int n,
                   const long int nbuckets, const long prec, const long double stdpnr, const long long int npnr, const double alpha, const int plproc,
                   long double * gradvctr);
```

```cpp
long double Climb(const string type, const long double prms[6], long double * gpath, const int fxTD, const double RF0, const long long int n,
                  const long int nbuckets, const long prec, const int plproc, const long double mxgrd, const long long int npnr,
                  const long double * grdnt, const double alpha, const int nitrs=0);
Eigen::MatrixXd DrvHess(const string type, const long double prms[6], long double * a, const int fxTD, const double RF0, const long long int n,
                        const long int nbuckets, const long prec, const int plproc, const long double * gradvctr, const long double stdpnr);
void WrtAry(const long double stdpnr, const long double * a, const string lbl, const int fxTD, const string filenm=" ");
```

### Code File: StaticGP.cpp

```
/*
/ Copyright (c) 2015 Chris Rook
/
/ Licensed under the Apache License, Version 2.0 (the "License");  you may not use this file except in compliance with the License.  You may obtain a copy of
/ the License at http://www.apache.org/licenses/LICENSE-2.0.  Unless required by applicable law or agreed to in writing, software distributed under the License
/ is distributed on an "AS IS" BASIS, WITHOUT WARRANTIES OR CONDITIONS OF ANY KIND, either express or implied.  See the License for the specific language
/ governing permissions and limitations under the License.
/
/ Filename:  StaticGP.cpp  (Defines the entry point for the console application.)
/
/ Function:  main()
/
/ Summary:
/
/       After launching the executable the user is prompted for the location of a directory where the control file and initial glide-path is stored in files
/       named "control.txt" and "gp.txt".  The file names can be changed in the header.  The control file contains the means/variances/covariance for stock and
/       bond returns (real) which are assumed normally distributed and constant with respect to time for this application.  (Neither is a requirement for the
/       proposed method, only for this particular implementation.)  The expense ratio, retirement horizon length, fixed inflation-adjusted withdrawal rate,
/       optimization method (gradient ascent/Newton-Raphson), convergence threshold, estimation method (dynamic program/simulation), base simulation sample
/       size and testing alphas, dynamic program discretization size and maximum ruin factor are also included in the control file.  The glide-path file holds
/       the initial glide-path in a single column which is the starting point for the optimization.  When using gradient ascent, the direction of steepest
/       ascent is computed for the current glide-path and climbing begins in that direction until no further progress can be made.  When estimating gradient
/       elements using simulation, no progress is made when the null hypothesis that the new probability of success is >= the old probability of success is
/       rejected (using the 1st alpha from the control file).  (Note that the probabilities being compared are subject to sampling error when estimating
/       probabilities using simulation which is why a hypothesis test is used to determine when to stop climbing.)  When using dynamic programming to estimate
/       success probabilities we use a direct comparison, not a hypothesis test.  When climbing progress ends we recompute the new gradient and proceed to
/       climb in the new direction until no further progress is made.  This procedure is repeated until the largest absolute effective value of the gradient
/       vector is below the convergence threshold specified in the control file.  By driving the gradient vector to zero in all dimensions we move to a point
/       on the surface that reflects a local/global maximum, which is true when the region is concave (i.e., Hessian matrix is negative semi-definite <=> all
/       its eigenvalues are non-positive.)  If all initial glide-paths exist in a concave region and lead to the same local optimum we consider it an empirical
/       interior point optimum and argue against the need for a meta-heuristic.  All elements of the gradient vector and Hessian matrix can be expressed as
/       probabilities (or differences/linear combinations of) and these are estimated using either dynamic programming or simulation.  When climbing using
/       simulation a sample size of 2X the base sample size is used since probabilities are compared.  Once no further progress is made we recompute the
/       success probability for that glide-path to remove any built-in upward sampling bias using a simulation sample size of 4X the base sample size.
/       Gradient and Hessian computations use the base sample size.  This probability is used when computing the subsequent gradient.  When using gradient
/       ascent with simulation we test each element of the gradient for equality with zero and set it to zero if the hypothesis test is not rejected (using the
/       2nd alpha from the control file).  Since these quantities are subject to sampling error we do not want to climb in the direction of that sampling error
/       if it can be avoided.  The resulting vector is referred to as the adjusted gradient and only applies when simulation is the estimation method.  The
/       Newton-Raphson optimization method is also available and uses a first-order Taylor expansion to estimate the gradient in the neighborhood of a given
/       point.  We know that the optimal value is the point where the gradient equals zero and this approximation is set to zero and solved yielding a new
/       point.  The new point becomes the old point and the process is repeated iteratively until the convergence criteria is met.  Convergence is achieved for
/       both optimization methods when the largest absolute gradient entry is smaller than the value supplied in the control file.  The user thus has 2 options
/       for optimization (gradient ascent or Newton-Raphson) and 2 options for estimating success probabilities (simulation or dynamic programming).  The final
/       glide-path is written to the file output.txt located in the directory specified by the user.  The Hessian and its eigenvalues are calculated at the
/       final glidepath to confirm that the region is concave.
/
/ Parameters:
/
/       The executable takes either one or no arguments when it is launched.  This value reflects the number of concurrent processes to utilize when running
/       the program.  If the user does not supply a value then the program will determine the maximum number allowed on the PC running it and use this value.
```

```cpp
/
/ Input Files:
/
/       1.) Control file named "control.txt" in the directory that the user provides.  Change the name in the header file if desired.  This value is a constant
/           set in the header to the constant paramfile.
/       2.) Initial glide-path which is in the file specified by the header constant initgpfile (default "gp.txt").  This file should contain exactly TD values
/           between MVA and 1.00 where MVA is the minimum variance alpha for this arrangement.  We do not use alphas below the MVA.  This file is also located
/           in the directory specified by the user.
/
/ Output Files:
/
/       1.) The file "output.txt" is written to the directory specified by the user and contains the final optimal glide-path, assuming convergence is
/           achieved.
/
/ Return Value:
/
/       A value of 0 (for success) or an appropriate error code (for failure).
/----------------------------------------------------------------------------------------------------------------------------------------------------------*/
#include "stdafx.h"
int main(int argc, char *argv[])
{
        // Local variables.
        //===================
        long double params[6], *grad=NULL, probnr, strtpnr, *GP=NULL, *OrigGP=NULL, maxgrad, stpthrsh, maxeigen, mineigen;
        double WR, alpha[2]={0.50, 1.00}, RFMax=2.75;
        int TD, pllproc=0, prtls[4]={-1,-1,-1,-1}, iterint=1;
        long long int SN=(long long int) pow(10.0,7);
        long int prec=(long int) pow(10.0,4);
        string rootdir, type, alg;

        // Get the directory location where setup files are stored, strip any leading/trailing blanks.
        //=============================================================================================
        cout << "Enter the directory where the setup files reside (eg, c:\\staticgp\\): " << endl;
        cin >> rootdir; cin.get();
        boost::algorithm::trim(rootdir);
        cout << endl;

        // Read setup details from control file.
        //=======================================
        ifstream getparams(rootdir+paramfile);
        if (getparams.is_open())
        {
                getparams >> params[0] >> params[1] >> params[2] >> params[3] >> params[4] >> params[5];
                getparams >> TD >> WR >> stpthrsh;
                getparams >> alg >> type;
                if (type=="sim")
                        getparams >> SN >> alpha[0] >> alpha[1];
                else if (type=="dp")
                        getparams >> prec >> RFMax;
                getparams.close();
        }
        else
        {
                cout << "ERROR: Could not read file: " << rootdir + paramfile << endl;
                cout << "EXITING...main()..." << endl; cin.get();
                exit (EXIT_FAILURE);
        }
```

```cpp
// Read initial glide-path from file.
//====================================
GP = new long double [TD];
ifstream getgp(rootdir+initgpfile);
if (getgp.is_open())
{
        int g=0;
        while (!getgp.eof() && g < TD)
        {
                getgp >> GP[g];
                g=g+1;
        }
        getgp.close();
        if (g < TD)
        {
                cout << "ERROR: File: " << rootdir + initgpfile << " needs " << TD << " initial asset allocations, but has fewer." << endl;
                cout << "EXITING...main()..." << endl; cin.get();
                exit (EXIT_FAILURE);
        }
}
else
{
        cout << "ERROR: Could not read file: " << rootdir + initgpfile << endl;
        cout << "EXITING...main()..." << endl; cin.get();
        exit (EXIT_FAILURE);
}

// Parse arguments to main().
// Pllproc is optional, default is optimal # of threads on the machine running the application.
//===========================================================================================
if (argc == 2)
        pllproc = stoi(argv[1]);
else if (argc != 1)
{
        cout << "ERROR: Parameter misspecification.  Incorrect # of parameters to the executable (expecting zero or one...)." << endl;
        cout << "EXITING...main()..." << endl; cin.get();
        exit (EXIT_FAILURE);
}

// When pllproc==0, replace with the # of independent processing units on the computer running the application.
// (Increase by a multiple of 4 when using a DP.  Some efficiencies are run sequentially.)
//============================================================================================================
if (pllproc == 0)
        pllproc = boost::thread::hardware_concurrency();

// Display optimization algorithm.
//================================
cout << "===>  Optimization algorithm:  ";
if (alg=="nr")
        cout << "Newton's Method" << endl;
else if (alg=="ga")
        cout << "Gradient Ascent" << endl;

// Display estimation method.
//===========================
cout << endl << "===>  Estimation method:  ";
if (type=="sim")
        cout << "Simulation" << endl;
```

```cpp
        else if (type=="dp")
        {
                cout << "Dynamic Program" << endl;
                pllproc = (int) 4*pllproc;
        }

        // Declare variables that depend on data read from the control file for sizing.
        //================================================================================
        Eigen::MatrixXd hess(TD,TD);
        Eigen::VectorXd sol(TD), ngrdnt(TD);
        Eigen::EigenSolver<Eigen::MatrixXd> hevals;
        grad = new long double[TD];

        // Take some steps (1 full iteration but no more than 50 steps) in the direction of steepest ascent.  This can move us off
        // the boundary region where computations may be unstable (infinite), especially when constructing the Hessian for Newton's method.
        // Also, this initial stepping usually makes improvements very quickly before proceeding with the optimization routine.
        //=====================================================================================================================
        probnr=GetPNR(type, params, GP, TD, WR, (long long int) 4*SN, (long int) (RFMax*prec), prec, prtls, pllproc);
        cout << endl << "Initial Glide-Path (w/Success Probability): " << endl;
        WrtAry(probnr, GP, "GP", TD);
        for (int s=1; s<=2; ++s)
        {
                maxgrad=BldGrad(type, params, GP, TD, WR, (long long int) SN, (long int) (RFMax*prec), prec, probnr, (long long int) 4*SN, alpha[1], pllproc,
                                grad);
                if (maxgrad <= stpthrsh)
                {
                        cout << "The glide-path supplied satisfies the EPSILON convergence criteria: " << maxgrad << " vs. " << stpthrsh << endl;
                        s=s+1;
                }
                else if (s != 2)
                {
                        probnr=Climb(type, params, GP, TD, WR, (long long int) 2*SN, (long int) (RFMax*prec), prec, pllproc, maxgrad, probnr,
                                    (long long int) 4*SN, grad, alpha[0], 50);
                        cout << endl << "New (Post Initial Climb) Glide-Path (w/Success Probability): " << endl;
                        WrtAry(probnr, GP, "GP", TD);
                }
                else if (maxgrad <= stpthrsh)
                        cout << "The glide-path supplied satisfies the EPSILON convergence criteria after intial climb without iterating: " << maxgrad <<
                                " vs. " << stpthrsh << endl;
        }

        // Negate the gradient if using NR method.
        //=========================================
        if (alg=="nr")
        {
                for (int y=0; y<TD; ++y)
                        ngrdnt[y]=-1.00*grad[y];
        }

        // If convergence is not achieved after initial climb then launch into full iteration mode.
        //==========================================================================================
        while (maxgrad > stpthrsh)
        {
                cout << noshowpos << endl << string(25, '=') << endl << "Start Iteration #" << (iterint) << endl << string(25, '=') << endl;
                if (alg=="nr")
                {
                        // Record the probability before iterating.
                        //==========================================
```

```cpp
                    strtpnr=probnr;

                    // Build the Hessian matrix for this glide-path and derive its eigenvalues.  (Display the largest & smallest value.)
                    // This is required when method=nr.  When either procedure ends with convergence we recompute the Hessian matrix to
                    // ensure we are at a local/global maximum (done below after convergence).
                    //=================================================================================================================
                    hess = DrvHess(type, params, GP, TD, WR, SN, (long int) (RFMax*prec), prec, pllproc, grad, probnr);
                    hevals.compute(hess, false);
                    maxeigen=-999.00; mineigen=999.00;
                    for (int y=0; y<TD; ++y)
                    {
                            complex<double> eval=hevals.eigenvalues()[y];

                            // Maximum eigenvalue.
                            //=====================
                            if (eval.real() > maxeigen)
                                    maxeigen=eval.real();

                            // Minimum eigenvalue.
                            //=====================
                            if (eval.real() < mineigen)
                                    mineigen=eval.real();
                    }

                    // Display the smallest/largest eigenvalues.
                    //==========================================
                    cout.setf(ios_base::fixed, ios_base::floatfield); cout.precision(30);
                    cout << endl << "Min Hessian eigenvalue for this iteration (>=0.00 --> convex region): " << mineigen << endl;
                    cout << endl << "Max Hessian eigenvalue for this iteration (<=0.00 --> concave region): " << maxeigen << endl;

                    // Update the glidepath and recompute the probability using the new glidepath.
                    //=============================================================================
                    sol=hess.colPivHouseholderQr().solve(ngrdnt);
                    for (int y=0; y<TD; ++y)
                            GP[y]=sol[y]+GP[y];

                    probnr=GetPNR(type, params, GP, TD, WR, (long long int) 4*SN, (long int) (RFMax*prec), prec, prtls, pllproc);

                    // If success probability has worsened alert the user.
                    //====================================================
                    if (probnr < strtpnr)
                    {
                      cout << endl << "NOTE:  The success probability has worsened during the last iteration.  This could happen for different reasons:";
                      cout << endl << "       1.) The difference in probabilities is beyond the system's ability to measure accurately (i.e., beyond 15 significant digits).";
                      cout << endl << "       2.) The difference is due to estimation/approximation error.";
                      cout << endl << "       3.) You may be operating along the boundary region.  In general the procedure is not well defined on the boundaries.  (Try gradient ascent.)";
                      cout << endl;
                    }
            }
            else if (alg=="ga")
            {
                    // Update the glide-path and recompute the probability using the new glide-path.
                    //===============================================================================
                    probnr=Climb(type, params, GP, TD, WR, (long long int) 2*SN, (long int) (RFMax*prec), prec, pllproc, maxgrad, probnr,
                                 (long long int) 4*SN, grad, alpha[0]);
            }
```

```cpp
        // Display the new glide-path.
        //===============================
        cout << endl << "New Glide-Path:";
        WrtAry(probnr, GP, "GP", TD);

        // Rebuild the gradient and negate it when using NR.
        //===================================================
        maxgrad=BldGrad(type, params, GP, TD, WR, (long long int) 1*SN, (long int) (RFMax*prec), prec, probnr, (long long int) 4*SN, alpha[1], pllproc,
                    grad);
        if (alg=="nr")
        {
                for (int y=0; y<TD; ++y)
                        ngrdnt[y]=-1.00*grad[y];
        }

        // Report the convergence status.
        //================================
        cout.setf(ios_base::fixed, ios_base::floatfield); cout.precision(20);
        cout << endl << "EPSILON Convergence Criteria: " << maxgrad << " vs. " << stpthrsh << endl;
        if (maxgrad <= stpthrsh)
                cout << endl << "==========>  EPSILON Convergence criteria satisfied.  <==========" << endl;

        cout << noshowpos << endl << string(25, '=') << endl << "End Iteration #" << (iterint) << endl << string(25, '=') << endl;
        iterint=iterint+1;
}

// Build Hessian and confirm we are at a maximum, not a saddle-point or plateau for example.
//===========================================================================================
cout << endl << "Convergence Achieved:  Final step is to confirm we are at a local/global maximum.  Hessian is being built." << endl;
hess = DrvHess(type, params, GP, TD, WR, SN, (long int) (RFMax*prec), prec, pllproc, grad, probnr);
hevals.compute(hess, false);
maxeigen=-999.00; mineigen=999.00;
for (int y=0; y<TD; ++y)
{
        complex<double> eval=hevals.eigenvalues()[y];

        // Maximum eigenvalue.
        //=====================
        if (eval.real() > maxeigen)
                maxeigen=eval.real();

        // Minimum eigenvalue.
        //=====================
        if (eval.real() < mineigen)
                mineigen=eval.real();
}

// Display the smallest/largest eigenvalues.
//===========================================
cout.setf(ios_base::fixed, ios_base::floatfield); cout.precision(30);
cout << endl << "Min Hessian eigenvalue at solution [>=0.00 --> convex region --> (local/global) minimum]: " << mineigen << endl;
cout << endl << "Max Hessian eigenvalue at solution [<=0.00 --> concave region --> (local/global) maximum]: " << maxeigen << endl;

// Write final GP to the output file.
//====================================
cout << endl;
if (maxeigen <= 0 || mineigen >= 0)
```

```cpp
				cout << "(Local/Global) Optimal ";
			cout << "Glide-Path:" << endl;
			WrtAry(probnr, GP, "GP", TD, rootdir+outfile);

			// Free up dynamic memory allocations.
			//====================================
			delete [] GP; GP=nullptr;
			delete [] grad; grad=nullptr;
			delete [] OrigGP; OrigGP=nullptr;
			cout << endl << "Done (hit return to exit)."; cin.get();
			return 0;
}
```

## Code File: BldGrad.cpp

```
/*
/ Copyright (c) 2015 Chris Rook
/
/ Licensed under the Apache License, Version 2.0 (the "License");  you may not use this file except in compliance with the License.  You may obtain a copy of
/ the License at http://www.apache.org/licenses/LICENSE-2.0.  Unless required by applicable law or agreed to in writing, software distributed under the License
/ is distributed on an "AS IS" BASIS, WITHOUT WARRANTIES OR CONDITIONS OF ANY KIND, either express or implied.  See the License for the specific language
/ governing permissions and limitations under the License.
/
/ Filename:  BldGrad.cpp
/
/ Function:  BldGrad()
/
/ Summary:
/
/        This function computes each gradient element and when using simulation tests the null hypothesis (Ho) that it equals zero using the type I error alpha
/        set in the control file (2nd alpha).  If simulation is used and this hypothesis test is not rejected then the element is set to zero in a 2nd vector
/        termed the adjusted gradient.  When simulation is used, each gradient element is subject to sampling error and if the noise is retained as the
/        direction of steepest ascent then we will move in this direction when climbing which will need to be reversed during future iterations.  By setting the
/        gradient element to zero when we are at the optimal allocation for a given time point during a given iteration we do not modify that equity ratio when
/        climbing.  The incoming appropriately sized array is populated with the gradient (when estimating via dp) or adjusted gradient (when estimating via
/        simulation) entries and the maximum absolute effective value of this array is the return value for the function.  The effective gradient entry is the
/        value for each time point that will not drive the equity ratio outside of the feasible region.  This maximum effective gradient value can be used to
/        determine convergence.  Note that when we are operating along the border region, the gradient will continue to point in the direction of steepest
/        ascent even if we cannot climb any further in that direction.  Therefore, without using the maximum effective gradient value, the procedure would fail
/        to converge.  The goal is to drive each element of the gradient to a value of zero since this implies that there is no direction we can move in to
/        improve the success probability, thus we are at a local optimum.  Both the adjusted (when using simulation) and unadjusted gradient vectors are printed
/        to the screen in block form.  (Note that the goal is to drive the unadjusted gradient vector to zero in all dimensions and any technique that achieves
/        this goal faster is valid as long as it works, which is the reason for testing each element against zero and eliminating values that are not
/        significantly different from zero.)  When this function is invoked, an estimation method is provided and determines how each gradient element is
/        estimated.  The two estimation choices are dynamic program (dp) or simulation (sim).  When simulation is used each quantity estimated is subject to
/        sampling error, and when dynamic programming is used each quantity is subject to approximation error.
/
/ Parameters:
/
/        1.) Estimation type:  Either dynamic programming (dp) or simulation (sim).
/        2.) Six member long double array of parameter settings for stock mean, stock variance, bond mean, bond variance, stock-bond covariance, expense ratio.
/        3.) Array of long doubles holding the glide-path that is the basis for computing the gradient (as pointer).  The corresponding array must have exactly
/            TD elements.
/        4.) Number of time-points for this implementation.  This represents a fixed number of years to consider for the retirement horizon (i.e., 30).
/        5.) The fixed inflation-adjusted withdrawal rate that the retiree is using during decumulation (i.e., 0.04 for a 4% withdrawal rate).
/        6.) The sample size to use when simulating the ruin/success probability.  (Note that the ruin probability is the number of times ruin occurs divided
/            by the number of trial runs.  The success probability is 1 minus the ruin probability.)  Here the probability will use the special density g() at
/            the given time point since we are building gradient elements.  (This parameter is only used when type="sim".)
/        7.) The number of buckets to use when discretizing the ruin factor dimension.  This value equals RFMax*(Discretization Precision), where these 2 values
/            are set by the user in the control file.  (This parameter is only used when type="dp".)
```

```
/           8.)  The number of buckets to use when discretizing each ruin factor unit value.  For example, if this value=1000 then each ruin factor unit is
/                represented by 1000 buckets.  This value is specified by the user in the control file when type="dp".
/           9.)  The success probability (i.e., probability of no ruin) for the incoming glide-path.  (Use a large sample size when estimating this if type="sim"
/                since it is subject to sampling error and used when deriving each gradient value.  A poor representation will negatively impact each gradient
/                element.)
/          10.)  The sample size used to simulate the success probability passed in with the above parameter.  (We need to know the sample size when testing the
/                hypothesis that each gradient element equals zero.  (This parameter is only used when type="sim".)
/          11.)  The alpha value used to test the hypothesis that the gradient element equals zero.  If this hypothesis holds then the value can be considered
/                sampling noise and we set it to zero when climbing.  (The reason for doing this is to achieve convergence faster.  This value is set by the user in
/                the control file, and it is only used when type="sim".)
/          12.)  The number of independent processing units on the computer running the application or alternatively the number of parallel processes to use during
/                execution as specified by the user.  (If user specified it is set as a parameter when invoking the application.)
/          13.)  A pointer to an empty array of appropriate size (i.e., TD elements) to populate with the gradient values.
/
/ Return Value:
/
/         This function populates an empty array passed to it with the gradient elements and it returns the maximum absolute effective gradient value.  (The
/         gradient element with largest effective magnitude can be used to determine convergence (i.e., stopping criteria).)  Effective here means it takes into
/         account the boundary values since we have a constrained optimization problem.
/-----------------------------------------------------------------------------------------------------------------------------------------------------------*/
#include "stdafx.h"
long double BldGrad(const string type, const long double prms[6], long double * a, const int fxTD, const double RF0, const long long int n,
                    const long int nbuckets, const long prec, const long double stdpnr, const long long int npnr, const double alpha, const int plproc,
                    long double * gradvctr)
{
        // Declare local variables.
        //========================
        int prtls[4] = {-1,-1,-1,-1};
        long double *K=new long double[fxTD], grdpnr, maxval=0, cmbpnr, ts, pval;
        boost::math::normal normdist = boost::math::normal(0.00,1.00);

        // Iterate over each time point deriving each gradient entry.
        //===========================================================
        cout << endl << "Building gradient ";
        for (int g=0; g<fxTD; ++g)
        {
                // Construct the constant needed for gradient entries.
                //====================================================
                K[g] = vp(prms,a[g])/(2.00*v(prms,a[g])) + pow(mp(prms),2)/(2.00*vp(prms,a[g]));

                // Populate the gradient vector for this time point.
                //==================================================
                prtls[0]=g;
                grdpnr = GetPNR(type, prms, a, fxTD, RF0, n, nbuckets, prec, prtls, plproc);
                gradvctr[g] = K[g]*(grdpnr - stdpnr);
                cout << ".";

                // Maximum effective absolute value of this gradient vector.
                //==========================================================
                if (a[g]+gradvctr[g] > 1.00)
                {
                        if (1.00-a[g] > maxval)
                                maxval=1.00-a[g];
                }
                else if (a[g]+gradvctr[g] < mva(prms)+0.0001)
                {
                        if (a[g]-(mva(prms)+0.0001) > maxval)
                                maxval=a[g]-(mva(prms)+0.0001);
```

```cpp
                    }
                    else if (abs(gradvctr[g]) > maxval)
                            maxval=abs(gradvctr[g]);
            }
            cout << " (Done)" << endl;

            // Print the unadjusted gradient vector entries.
            //=================================================
            cout << endl << "Gradient (no adjustment):";
            WrtAry(-1, gradvctr, "Grd", fxTD);

            // If using simulation, test each element for equality with zero.  If test holds set the element to zero.
            //=========================================================================================================
            if (type == "sim" && alpha < 1.00)
            {
                    maxval=0;
                    for (int g=0; g<fxTD; ++g)
                    {
                            cmbpnr = (((long double) n*(stdpnr + gradvctr[g]/K[g]) + (long double) npnr*stdpnr)/((long double) (n + npnr)));
                            ts = ((stdpnr + gradvctr[g]/K[g]) - stdpnr)/sqrt((cmbpnr)*(1.00 - cmbpnr)*(1.00/n + 1.00/npnr));
                            pval = 2.00*min(boost::math::cdf(normdist, ts), 1.00 - boost::math::cdf(normdist, ts));
                            if (pval > alpha)
                                    gradvctr[g] = 0.00;         // The element is not different from zero at significance level alpha.

                            // Maximum effective absolute value of this gradient vector.
                            //============================================================
                            if (a[g]+gradvctr[g] > 1.00)
                            {
                                    if (1.00-a[g] > maxval)
                                            maxval=1.00-a[g];
                            }
                            else if (a[g]+gradvctr[g] < mva(prms)+0.0001)
                            {
                                    if (a[g]-(mva(prms)+0.0001) > maxval)
                                            maxval=a[g]-(mva(prms)+0.0001);
                            }
                            else if (abs(gradvctr[g]) > maxval)
                                    maxval=abs(gradvctr[g]);
                    }

                    // Print the adjusted gradient vector entries.
                    //=============================================
                    cout << endl << "ADJUSTED gradient:";
                    WrtAry(-1, gradvctr, "Adj-Grd", fxTD);
            }
            // Display the maximum effective absolute value of the gradient vector (used to determine convergence).
            //=======================================================================================================
            cout << endl << "Maximum effective absolute value of this gradient: " << maxval << endl;

            // Free up temporary memory allocations.
            //=======================================
            delete [] K; K=nullptr;

            // This function returns the maximum absolute value of the gradient elements.
            // (Which is used to define the stopping/convergence criteria.)
            //=============================================================================
            return maxval;
}
```

# Code File: Climb.cpp

```
/*
/ Copyright (c) 2015 Chris Rook
/
/ Licensed under the Apache License, Version 2.0 (the "License");  you may not use this file except in compliance with the License.  You may obtain a copy of
/ the License at http://www.apache.org/licenses/LICENSE-2.0.  Unless required by applicable law or agreed to in writing, software distributed under the License
/ is distributed on an "AS IS" BASIS, WITHOUT WARRANTIES OR CONDITIONS OF ANY KIND, either express or implied.  See the License for the specific language
/ governing permissions and limitations under the License.
/
/ Filename: Climb.cpp
/
/ Function:  Climb()
/
/ Summary:
/
/       This function steps in the direction of the gradient until no further progress can be made.  When using simulation as the estimation type, the ruin
/       probabilities are subject to sampling error and we test the hypothesis that the updated GP is not inferior to the best performing GP encountered thus
/       far in the iteration.  When using simulation, climbing ends when this test is rejected at an alpha level specified by the user (the first alpha in the
/       control file).  Since the test is for non-inferiority and alpha is the probability of making a Type I Error (reject Ho when it is true), a small alpha
/       is more likely to result in climbing past the true stopping point at each iteration, whereas a large alpha will do the exact opposite, namely result in
/       stopping too soon.  Since the gradient is an expensive computation (in terms of runtime) the alpha decision is an important one.  The hypothesis test
/       is:
/              Ho:  Probability of ruin using new GP >= Probability of ruin using base GP    (base GP is the best performing thus far in the iteration)
/              Ha:  Probability of ruin using new GP <  Probability of ruin using base GP    (base GP is the best performing thus far in the iteration)
/
/       and only applies when simulation is the estimation technique.  Simulated probabilities are subject to sampling error which is a function of the sample
/       size, which is taken into account by the test above.  Quantities that are estimated using dynamic programming (dp) are subject to approximation error
/       and climbing proceeds until the probability declines from the last value.  No hypothesis test is used when the estimation type is dynamic programming.
/       The step size has been set as a function of the largest gradient (absolute) value.  The user may archive better performance by changing the step size
/       rules.  A small step size leads to longer run times, whereas a larger step size leads to imprecise stopping criteria which must be corrected at a
/       future iteration.
/
/ Parameters:
/
/       1.) Estimation type:  Either dynamic programming (dp) or simulation (sim).
/       2.) Six member long double array of parameter settings for stock mean, stock variance, bond mean, bond variance, stock-bond covariance, expense ratio.
/       3.) Array of long doubles holding the glide-path to start climbing at (as pointer).  The corresponding array must have exactly TD elements and is
/           updated with new values when the hypothesis test above is not rejected or climbing continues when using DP approximation.
/       4.) Number of time-points for this implementation.  This represents a fixed number of years to consider for the retirement horizon (i.e., 30).
/       5.) The fixed inflation-adjusted withdrawal rate that the retiree is using during decumulation (i.e., 0.04 for a 4% withdrawal rate).
/       6.) The sample size to use when simulating the ruin/success probability.  (Note that the ruin probability is the number of times ruin occurs divided
/           by the number of trial runs.  The success probability is 1 minus the ruin probability.)  This parameter is only used when type="sim".
/       7.) The number of buckets to use when discretizing the ruin factor dimension.  This value equals RFMax*(Discretization Precision), where these 2 values
/           are set by the user in the control file.  (This parameter is only used when type="dp".)
/       8.) The number of buckets to use when discretizing each ruin factor unit value.  For example, if this value=1000 then each ruin factor unit is
/           represented by 1000 buckets.  This value is specified by the user in the control file when type="dp".
/       9.) The number of independent processing units on the computer running the application or alternatively the number of parallel processes to use during
/           execution as specified by the user.  (If user specified it is set as a parameter when invoking the application.)
/       10.) The largest absolute gradient value used to determine the step size.  Larger gradient entries result in smaller step sizes and vice-versa.
/       11.) The success probability for the glide-path provided which is the starting point for climbing.  To proceed in the direction of the gradient the new
/            success probability must be non-inferior to the current value (using the hypothesis test above) when using type="sim" or just greater than when
/            using type="dp".
/       12.) The sample size used to simulate the success probability passed via the previous parameter.  (This parameter is only used when type="sim".)
/       13.) The gradient vector as a pointer to an array of exactly TD elements.  This is the direction of steepest ascent.
/       14.) The alpha value used to test the hypothesis that the new GP is non-inferior to the best prior GP.  We continue climbing until this Ho is rejected.
/            (This parameter is only used when type="sim".  This is the first alpha level set in the control file.)
/       15.) The maximum number of climbing iterations before returning to the invoking program.  (For example, use with initial stepping to avoid a lengthy
/            climb that can occur with a poorly selected starting point.)
```

```cpp
/
/  Return Value:
/
/       This function returns the optimal success probability achieved while climbing.  The updated glide-path is placed directly into the glide-path array
/       that is passed to this function as a pointer in argument #3.
/----------------------------------------------------------------------------------------------------------------------------------------------------------*/
#include "stdafx.h"
long double Climb(const string type, const long double prms[6], long double * gpath, const int fxTD, const double RF0, const long long int n,
                 const long int nbuckets, const long prec, const int plproc, const long double mxgrd, const long double stdpnr, const long long int npnr,
                 const long double * grdnt, const double alpha, const int nitrs)
{
        // Declare/initialize local variables.
        //====================================
        long double newpnr=stdpnr, maxpnr=stdpnr, ts, cmbvar, pval=1.00, iter=1.00, *OrigGP=NULL, *PrevGP=NULL;
        const int prtls[4] = {-1,-1,-1,-1};
        long long int maxn=npnr, prevn=npnr;
        int cont=1, fstimpr=0, iindx, tryup=0, origindx=0;
        boost::math::normal normdist = boost::math::normal(0.00,1.00);

        // Get initial glidepath provided, assign to both original GP array and prior GP array.
        //=====================================================================================
        PrevGP = new long double [fxTD];
        OrigGP = new long double [fxTD];
        for (int y=0; y<fxTD; ++y)
                OrigGP[y]=PrevGP[y]=gpath[y];

        // Define the step size for climbing.  The step size depends on the largest gradient element and grows exponentially.
        // This is a heuristic that has worked well and can be modified if desired.  Better step sizes can reduce runtimes.
        //====================================================================================================================
        for (int i=1; i<=10; ++i)
                if ((mxgrd >= 1.00/(10.00*pow(10.00, i))) && (mxgrd < 1.00/(10.00*pow(10.00, i-1))))
                {
                        iindx = i;
                        iter = (double) pow(exp(log(5.00)/4.00),iindx);
                }

                // Output details for the current iteration.
                //==========================================
                cout.setf(ios_base::fixed, ios_base::floatfield); cout.precision(10);
                cout << endl << noshowpos << string(70, '=') << endl << "Iteration step size = " << iter << endl; cout.precision(20);
                cout << "Trying to improve on success probability = " << stdpnr << endl << string(70, '=') << endl;

                // Climb in the direction of the gradient.
                //========================================
                for (int t=0; (cont==1 || fstimpr==0); ++t)                            // Iterate until no more progress is made
                {
                        if (cont==0 && fstimpr==0)                                     // If no progress is made reduce step size and try again.
                        {
                                // Set the original index value when entering this problematic scenario.
                                //======================================================================
                                if (origindx == 0)
                                        origindx=iindx;

                                // Adjust glidepath back one iteration since it failed to improve the probability.
                                //================================================================================
                                for (int y=0; y<fxTD; ++y)
                                        gpath[y] = PrevGP[y];                          // Reverse final update, since no progress was made.
```

```cpp
            if (iindx != 0)
                    cout.precision(10); cout << endl << "No Progress Made:  Iteration step size changed from " << iter;
            if (iindx==0)
            {
                    cout << " to 0.00.  (No additional climbing attempts will be made.)" << endl;
                    cout.setf(ios_base::fixed, ios_base::floatfield); cout.precision(10);
                    cout << endl << endl ;
                    cout << "ERROR: No progress can be made, the procedure is stuck.  (Step size has been reduced to 0.)" << endl;
                    cout << "       You may be operating along the boundary where the process is not well defined or your" << endl;
                    cout << "       estimation/approximation precision level is not adequate for your epsilon level." << endl << endl;
                    cout << endl << "Current Glide-Path: ";
                    WrtAry(maxpnr, gpath, "GP", fxTD);
                    cout << endl << "EXITING...Climb()..." << endl; cin.get();
                    exit (EXIT_FAILURE);
            }
            else if (iindx==1 && tryup==5)
            {
                    iindx=iindx-1;
                    iter = (double) iter/2.00;
            }
            else if (iindx > 1 && tryup==5)
            {
                    if (iindx == origindx+5 )
                            iindx=origindx-1;
                    else
                            iindx=iindx-1;
                    iter = (double) pow(exp(log(5.00)/4.00),iindx);
            }
            else if (iindx > 1 && tryup < 5)
            {
                    iindx=iindx+1;
                    iter = (double) pow(exp(log(5.00)/4.00),iindx);
                    tryup=tryup+1;
            }
            cout << " to " << iter << ".  (Attempting to climb again.)" << endl;
            cont=1;
    }
    for (int y=0; y<fxTD; ++y)                                  // Iterate over glide-path and update it
    {
            PrevGP[y] = gpath[y];                                       // Reset the previous glide-path element
            gpath[y] = gpath[y] + (iter)*grdnt[y];               // Update each individual glide-path element

            if (gpath[y] < mva(prms)+0.0001)
                    gpath[y] = mva(prms)+0.0001;                 // Stay above MVA and consistent with ThrdPNRdyn() and ThrdPNRsim().
            else if (gpath[y] > 1.00)
                    gpath[y] = 1.00;                             // Consistent with ThrdPNRdyn() and ThrdPNRsim().
    }
    newpnr = GetPNR(type, prms, gpath, fxTD, RF0, n, nbuckets, prec, prtls, plproc);     // Derive new probability and display old vs new.

    cout.setf(ios_base::fixed, ios_base::floatfield); cout.precision(20); cout.imbue(std::locale(""));
    cout << endl << "Base Prob(NR)  = " << maxpnr;
    if (type=="sim")
            cout << " (N=" << maxn << ")";
    cout << endl << "New Prob(NR)   = " << newpnr;
    if (type=="sim")
            cout << " (N=" << n << ")";
    else if (newpnr > maxpnr)
            cout << "  (Better, CONTINUE climbing ...)";
```

```cpp
                else
                        cout << "  (Worse, STOP climbing ...)";
                cout << endl;

                // If using simulation, conduct a non-inferiority test of the new vs max base GP.
                // =====> Continue to climb if the new GP is at least as good as the max base GP.
                // Otherwise, compare new probability with old and climb while making progress.
                //======================================================================
                if (type=="sim")
                {
                        cmbvar = maxpnr*(1.00 - maxpnr)/maxn + newpnr*(1.00 - newpnr)/n;
                        ts = (newpnr - maxpnr)/sqrt(cmbvar);
                        pval = boost::math::cdf(normdist, ts);
                        cout.setf(ios_base::fixed, ios_base::floatfield); cout.precision(5);
                        cout << "Test Statistic = " << ts << endl;
                        cout << "P-Value        = " << pval << "     (Alpha="<< alpha << ")" << endl;
                        if (pval > alpha)
                        {
                                fstimpr=1; cout << "=====> Accept Ho (non-inferiority), CONTINUE climbing ..." << endl;
                        }
                        else
                        {
                                cont=0; cout << "=====> Reject Ho (non-inferiority), STOP climbing ..." << endl;
                        }

                        // Update PNR and sample size for base GP.
                        //=========================================
                        if (newpnr > maxpnr)
                        {
                                maxpnr=newpnr; maxn=n;
                        }
                }
                else if (type=="dp")
                {
                        if (newpnr > maxpnr)
                        {
                                fstimpr=1; maxpnr=newpnr;
                        }
                        else
                                cont=0;
                }

                // For lengthy climbing, display the current glidepath at 100 iteration intervals.
                //======================================================================
                if (((t+1) % 100) == 0 )
                {
                        cout << endl << "Current Glide-Path at Iteration: " << t+1;
                        WrtAry(newpnr, gpath, "GP", fxTD);
                }

                // Stop when maximum number of iterations has been reached, if specified.
                //==================================================================
                if (nitrs > 0 && t+1 == nitrs && cont==1)
                {
                        cout << endl << "Climbing limit reached at " << nitrs << " iterations." << endl;
                        cont=2;
                }
        }
}
```

```cpp
                // Adjust glidepath back one iteration since it failed to improve the probability.
                // (This is only done when climbing failed to improve, not when limit is reached.)
                //=================================================================================
                if (cont != 2)
                        for (int y=0; y<fxTD; ++y)
                                gpath[y] = PrevGP[y];

                if (type=="sim")
                {
                        cout << endl << "Resetting the probability (to remove any built-in upward sampling bias) ..." << endl;
                        maxpnr=ThrdPNRsim(prms, gpath, fxTD, RF0, (long long int) 2*n, prtls, plproc);
                }

                // Release temporary memory allocations.
                //======================================
                delete [] OrigGP; OrigGP=nullptr;
                delete [] PrevGP; PrevGP=nullptr;

                // Return the max success probability.
                //====================================
                return maxpnr;
}
```

## Code File: DrvHess.cpp

```
/*
/ Copyright (c) 2015 Chris Rook
/
/ Licensed under the Apache License, Version 2.0 (the "License");  you may not use this file except in compliance with the License.  You may obtain a copy of
/ the License at http://www.apache.org/licenses/LICENSE-2.0.  Unless required by applicable law or agreed to in writing, software distributed under the License
/ is distributed on an "AS IS" BASIS, WITHOUT WARRANTIES OR CONDITIONS OF ANY KIND, either express or implied.  See the License for the specific language
/ governing permissions and limitations under the License.
/
/ Filename:  DrvHess.cpp
/
/ Function:  DrvHess()
/
/ Summary:
/
/       This function computes the Hessian values and populates, then returns, a square TDxTD symmetric matrix with these values.  Each element of the Hessian
/       matrix can be expressed as a linear combination of ruin probabilities.  The form of this linear combination differs if the element is on the diagonal
/       or not.  The user selects an estimation method of simulation (sim) or dynamic programming (dp) for estimating these ruin probabilities.  The off-
/       diagonal elements use first order partial derivatives in both the i,j dimension and the diagonal elements use 2nd order partial derivatives in the i-i
/       dimension.  The off-diagonal elements can make use of the existing gradient vector and success probability therefore only requires one new ruin
/       probability be estimated which uses the gradient special density g() in both the i and j dimensions.  The diagonal elements can make use of the success
/       probability only and requires 2 new ruin probabilities be estimated with either simulation or a dynamic program.  The diagonal elements make use of the
/       special densities h1() and h2(), or their corresponding CDFs.
/
/ Parameters:
/
/       1.) Estimation type:  Either dynamic programming (dp) or simulation (sim).
/       2.) Six member long double array of parameter settings for stock mean, stock variance, bond mean, bond variance, stock-bond covariance, expense ratio.
/       3.) Array of long doubles holding the glide-path to use for this calculation (as pointer).  The corresponding array must have exactly TD elements.
/       4.) Number of time-points for this implementation.  This represents a fixed number of years to consider for the retirement horizon (i.e., 30).
/       5.) The fixed inflation-adjusted withdrawal rate that the retiree is using during decumulation (i.e., 0.04 for a 4% withdrawal rate).
/       6.) The sample size to use when simulating the ruin/success probability.  (Note that the ruin probability is the number of times ruin occurs divided
/           by the number of trial runs.  The success probability is 1 minus the ruin probability.  (Here the probability will use the special densities h1()
/           and h2() when building diagonal Hessian elements and will use the gradient density g() for off-diagonal elements.)  This parameter is only used
/           when estimation type is simulation.
```

```
/            7.) The number of buckets to use when discretizing the ruin factor dimension.  This value equals RFMax*(Discretization Precision), where these 2 values
/                are set by the user in the control file.  (This parameter is only used when type="dp".)
/            8.) The number of buckets to use when discretizing each ruin factor unit value.  For example, if this value=1000 then each ruin factor unit is
/                represented by 1000 buckets.  This value is specified by the user in the control file when type="dp".
/            9.) The number of independent processing units on the computer running the application or alternatively the number of parallel processes to use during
/                execution as specified by the user.  (If user specified it is set as a parameter when invoking the application.)
/           10.) A pointer to the gradient vector for this glide-path.  The off-diagonal Hessian elements include ruin calculations that can be derived directly
/                from the existing gradient values and these are used instead of recalculating them.
/           11.) The success probability for the given glide-path.  This value is also used for computing off-diagonal Hessian elements and is not recomputed each
/                time.
/
/ Return Value:
/
/         This function returns a square TDxTD symmetric matrix representing the Hessian of the glide-path supplied.
/----------------------------------------------------------------------------------------------------------------------------------------------------------------*/
#include "stdafx.h"
Eigen::MatrixXd DrvHess(const string type, const long double prms[6], long double * a, const int fxTD, const double RF0, const long long int n,
                        const long int nbuckets, const long prec, const int plproc, const long double * gradvctr, const long double stdpnr)
{
        // Declare local variables.
        //===========================
        Eigen::MatrixXd hess(fxTD, fxTD);
        string blnks;

        // Derive the Hessian matrix.
        //===========================
        cout << endl << "Building Hessian ";
        for (int i=0; i<fxTD; ++i)
        {
                blnks=string(17*(i>0)+i,' '); cout << blnks;
                for (int j=0; j<fxTD; ++j)
                {
                        if (j >= i)
                        {
                                cout << ".";
                                if (i < 0 || j < 0 || i >= fxTD || j >= fxTD)
                                {
                                        cout << "ERROR:  Both i and j must be integers between 0 and " << (fxTD-1) << ", i=" << i << " and j=" << j << endl;
                                        cout << "EXITING...DrvHess()..." << endl; cin.get();
                                        exit (EXIT_FAILURE);
                                }
                                else if (i != j)        // ***** Off-diagonal elements ***** //
                                {
                                        // Define needed quantities and return the off-diagonal Hessian element.
                                        //======================================================================
                                        long double K1, K2;
                                        K1 = vp(prms,a[i])/(2.00*v(prms,a[i])) + pow(mp(prms),2)/(2.00*vp(prms,a[i]));
                                        K2 = vp(prms,a[j])/(2.00*v(prms,a[j])) + pow(mp(prms),2)/(2.00*vp(prms,a[j]));
                                        int prtls[4] = {i,j,-1,-1};
                                        hess(i,j) = K1*K2*(GetPNR(type, prms, a, fxTD, RF0, n, nbuckets, prec, prtls, plproc) - gradvctr[i]/K1 -
                                                           gradvctr[j]/K2 - stdpnr);
                                }
                                else                    // ***** Diagonal elements ***** //
                                {
                                        // Define needed quantities and return the diagonal Hessian element.
                                        // [Note #1:  K1 is for h1, K2 is for h2, and K3 is for f(). And the diagonal term is K1*h1() + K2*h2() + K3*f().]
                                        // [Note #2:  The Hessian diagonals will divide by zero for one alpha, check for that value and exit if encountered.]
                                        //==================================================================================================================
```

```cpp
                                        if (v(prms,a[i])*vpp(prms) - 2.00*pow(vp(prms,a[i]),2) == 0.00)
                                        {
                                                cout.setf(ios_base::fixed, ios_base::floatfield); cout.precision(10);
                                                cout << "ERROR:  Hessian diagonal element does not exist for alpha value=" << a[i] <<
                                                        " encountered at time point t=" << i << "." << endl;
                                                cout << "EXITING...DrvHess()..." << endl; cin.get();
                                                exit (EXIT_FAILURE);
                                        }
                                        else
                                        {
                                                long double K1, K2, K3, H1, H2;
                                                K1=(v(prms,a[i]) + pow(kh1(prms,a[i]),2))*(v(prms,a[i])*vpp(prms) -
                                                        2.00*pow(vp(prms,a[i]),2))/(2.00*pow(v(prms,a[i]),3));
                                                K2=(pow(vp(prms,a[i]),2) + 2.00*v(prms,a[i])*pow(mp(prms),2))/(2.00*pow(v(prms,a[i]),2));
                                                K3=-((vpp(prms)*v(prms,a[i]) - pow(vp(prms,a[i]),2) +
                                                        2.00*v(prms,a[i])*pow(mp(prms),2))/(2.00*pow(v(prms,a[i]),2))
                                                        + (2.00*pow(vp(prms,a[i]),2)*pow(mp(prms),2))/(pow(v(prms,a[i]),2)*vpp(prms) -
                                                        2.00*pow(vp(prms,a[i]),2)*v(prms,a[i])));
                                                int h1prtls[4] = {-1,-1,i,-1};
                                                H1=GetPNR(type, prms, a, fxTD, RF0, n, nbuckets, prec, h1prtls, plproc);
                                                int h2prtls[4] = {-1,-1,-1,i};
                                                H2=GetPNR(type, prms, a, fxTD, RF0, n, nbuckets, prec, h2prtls, plproc);
                                                hess(i,j) = K1*H1 + K2*H2 + K3*stdpnr;
                                        }
                                }
                        }
                        else
                                hess(i,j) = hess(j,i);
                }
                if (i == (fxTD-1)) cout << " (Done)";
                cout << endl;
        }
        cout << endl;

        // Return the Hessian matrix.
        //===========================
        return hess;
}
```

### Code File:  GetCDF.cpp

```cpp
/*
/ Copyright (c) 2015 Chris Rook
/
/ Licensed under the Apache License, Version 2.0 (the "License");  you may not use this file except in compliance with the License.  You may obtain a copy of
/ the License at http://www.apache.org/licenses/LICENSE-2.0.  Unless required by applicable law or agreed to in writing, software distributed under the License
/ is distributed on an "AS IS" BASIS, WITHOUT WARRANTIES OR CONDITIONS OF ANY KIND, either express or implied.  See the License for the specific language
/ governing permissions and limitations under the License.
/
/ Filename:  GetCDF.cpp
/
/ Function:  GetCDF()
/
/ Summary:  Single function that returns any CDF value required by this application.  All CDF calls derived in this application can be expressed as a linear
/           combination of CDF calls from normal and gamma random variables.  The constants and random variables are passed as parameters and this function
/           combines them and computes then returns the CDF value for the value x (set in last argument).  CDF calls are only needed when estimation is via
/           dynamic programming.  When simulation is the estimation method the PDFs are used.  Having one function return any CDF value that is required
/           allows for only one function to compute needed probabilities via dynamic programming.
/
/ Parameters:
```

```
/
/       1.) Array of precomputed moments & constants for the given time-point and asset allocation (to save processing time): m, mp, v, vp, vpp, mva, kh1.
/       2.) Vector of constants.  All CDFs in this application can be expressed as a linear combination of known CDF calls.  This is the constant vector for
/           that linear combination.
/       3.) A normal random variable (class type=boost distribution) reflecting the inflation/expense-adjusted return for this time-point and asset allocation.
/       4.) A vector of gamma distributed random variables (class type=boost distribution) needed to build the CDF for a given time-point and asset allocation.
/           The CDF G() of g(), and the CDF H1() of h1() both use 2 gamma RVs.  The CDF H2() of h2() uses 4 gamma RVs.
/       5.) A value on the axis of the random variable of interest.  The value returned is the probability of being to the left of this value.
/
/ Return Value:
/
/       This function returns the CDF of x, where the CDF is a linear combination (specific form) of the constants and random variables passed as arguments.
/       This function can return CDF values for F() of f(), G() of g(), H1() of h1(), and H2() of h2().
/----------------------------------------------------------------------------------------------------------------------------------------------------*/
#include "stdafx.h"
long double GetCDF(const long double mts[7], const vector<long double> & C, const boost::math::normal n, const vector<boost::math::gamma_distribution<>> & G,
                const double x)
{
        // Declare local variables.
        //==========================
        long double *F=NULL, retval=0, sgn=1.00*(x <= mts[0]) - 1.00*(x > mts[0]);
        int hival=1;

        // Array to hold Gamma CDF values.
        //=================================
        F = new long double [(int) G.size() + 1];
        F[0]=0.00;

        // Derive & return the CDF value for the given parameters.
        //=========================================================
        for (int i=1; i<=(int) G.size(); ++i)
        {
                F[i] = (1.00 - pow(sgn,i)*(cdf(G[i-1], pow((x - mts[0])/sqrt(2.00*mts[2]),2))));
                hival=hival*(F[i]==1.0+pow(-1.0,i+1));
                retval = retval + C[i]*F[i];
        }

        // Free up dynamic memory allocations.
        //=====================================
        delete [] F; F=nullptr;

        // Return the CDF value, making sure the highest possible value is 1.00.
        //=======================================================================
        if (hival==1 && cdf(n,x)==1.00)
                return 1.00;
        else
                return C[0]*(retval + C[(int) C.size()-1]*cdf(n, x));
}
```

# Code File: GetConst.cpp

```cpp
/*
/ Copyright (c) 2015 Chris Rook
/
/ Licensed under the Apache License, Version 2.0 (the "License");  you may not use this file except in compliance with the License.  You may obtain a copy of
/ the License at http://www.apache.org/licenses/LICENSE-2.0.  Unless required by applicable law or agreed to in writing, software distributed under the License
/ is distributed on an "AS IS" BASIS, WITHOUT WARRANTIES OR CONDITIONS OF ANY KIND, either express or implied.  See the License for the specific language
/ governing permissions and limitations under the License.
/
/ Filename:  GetConst.cpp
/
/ Function:  GetConst()
/
/ Summary:
/
/       Each CDF call used in this application can be expressed as a linear combination of CDF calls to known random variables.  The random variables are
/       normal and gamma distributed.  The value of the constants in the linear combination depend on whether or not special densities are used to represent
/       inflation/expense-adjusted returns at each time-point.  This function takes the information for the given time-point, including the moments, the time-
/       point, and the special density indicator array and builds the appropriately sized vector of constants.  A similar function will build the appropriately
/       sized vector of random variables, and the GetCDF() function combines the constants and random variables to return the needed CDF value.
/
/ Parameters:
/
/       1.) Array of precomputed moments & constants for the given time-point and asset allocation (to save processing time): m, mp, v, vp, vpp, mva, kh1.
/       2.) A 4-element indicator array that specifies which densities should represent inflation/expense-adjusted returns at the given time-point.
/           The settings for this array are:
/               a.) prtls[] = {-1,-1,-1,-1} for all standard real/expense adjusted return densities.
/               b.) prtls[] = {i,-1,-1,-1} for gradient element "i" which uses gradient density g() for the real/expense-adjusted return at the i-th
/                   time-point.
/               c.) prtls[] = {i,j,-1,-1} for the off-diagonal Hessian elements which uses the gradient density g() at both time-points "i" and "j".
/               d.) prtls[] = {-1,-1,i,-1} for the diagonal Hessian elements which uses h1() as the real/expense-adjusted return at time-point "i".
/               e.) prtls[] = {-1,-1,-1,i} for the diagonal Hessian elements which uses h2() as the real/expense-adjusted return at time-point "i".
/       3.) The time-point currently being processed.  This value is compared with the indicator array passed in argument #2 to determine which
/           constants are needed for the CDF call.
/
/ Return Value:
/
/       This function returns the vector of constants needed to build the CDF for a given asset allocation and time-point.  The size of the vector and
/       constant values returned depend on the density being used to represent inflation/expense-adjusted returns for the given time-point.
/-----------------------------------------------------------------------------------------------------------------------------------------------------------*/
#include "stdafx.h"
vector<long double> GetConst(const long double mts[7], const int partls[4], const int y)
{
        // Declare local variables.
        //===========================
        vector<long double> C;

        // Build vector of constants.
        //===========================
        if (partls[0] != y && partls[1] != y && partls[2] != y && partls[3] != y)
                C.push_back(1.00);
        else if (partls[0]==y || partls[1]==y)
        {
                C.push_back(1.00/(pow(mts[1],2)*pow(mts[2],2) + mts[2]*pow(mts[3],2)));
                C.push_back(pow(mts[3],2)*mts[2]/2.00);
                C.push_back(-mts[3]*mts[1]*mts[2]*sqrt(2.00*mts[2]/pi));
                C.push_back(pow(mts[1],2)*pow(mts[2],2));
        }
```

```cpp
        else if (partls[2]==y)
        {
                C.push_back(1.00/(mts[2] + pow(mts[6],2)));
                C.push_back(mts[2]/2.00);
                C.push_back(-sqrt((2.00*mts[2])/pi)*mts[6]);
                C.push_back(pow(mts[6],2));
        }
        else if (partls[3]==y)
        {
                C.push_back(2.00/(pow(mts[3],2) + 2.00*mts[2]*pow(mts[1],2)));
                C.push_back(3.00*pow(mts[3],2)/8.00);
                C.push_back(-sqrt(2.00*mts[2]/pi)*mts[3]*mts[1]);
                C.push_back((2.00*pow(mts[1],2)*mts[2] - pow(mts[3],2))/4.00);
                C.push_back(sqrt(mts[2]/(2.00*pi))*mts[3]*mts[1]);
                C.push_back(pow(mts[3],2)/4.00);
        }
        return C;
}
```

## Code File: GetGamma.cpp

```cpp
/*
/ Copyright (c) 2015 Chris Rook
/
/ Licensed under the Apache License, Version 2.0 (the "License");  you may not use this file except in compliance with the License.  You may obtain a copy of
/ the License at http://www.apache.org/licenses/LICENSE-2.0.  Unless required by applicable law or agreed to in writing, software distributed under the License
/ is distributed on an "AS IS" BASIS, WITHOUT WARRANTIES OR CONDITIONS OF ANY KIND, either express or implied.  See the License for the specific language
/ governing permissions and limitations under the License.
/
/ Filename:  GetGamma.cpp
/
/ Function:  GetGamma()
/
/ Summary:
/
/       Each CDF call needed by this application can be expressed as a linear combination of known CDF calls to normal and gamma distributed random variables.
/       The constants and random variables needed to build the CDF call depend on the current time-point and the current special-density indicator array.  Once
/       these are known the vector of gamma random variables can be built.  This is combined with the appropriate normal random variable and required constants
/       in the function GetCDF().
/
/ Parameters:
/
/       1.) A 4-element indicator array that specifies which densities should represent inflation/expense-adjusted returns at the given time-point.  The
/           settings for this array are:
/              a.) prtls[] = {-1,-1,-1,-1} for all standard real/expense adjusted return densities.
/              b.) prtls[] = {i,-1,-1,-1} for gradient element "i" which uses gradient density g() for the real/expense-adjusted return at the i-th time-point.
/              c.) prtls[] = {i,j,-1,-1} for the off-diagonal Hessian elements which uses the gradient density g() at both time-points "i" and "j".
/              d.) prtls[] = {-1,-1,i,-1} for the diagonal Hessian elements which uses h1() as the real/expense-adjusted return at time-point "i".
/              e.) prtls[] = {-1,-1,-1,i} for the diagonal Hessian elements which uses h2() as the real/expense-adjusted return at time-point "i".
/       2.) The time-point currently being processed.  This value is compared with the indicator array passed in argument #1 to determine which gamma random
/           variables are needed for the CDF call.
/
/ Return Value:
/
/       This function returns a vector of gamma random variables (with appropriate shape/scale settings) to use when building the CDF required at a specific
/       time-point and special-density setting.
/------------------------------------------------------------------------------------------------------------------------------------------------------------*/
#include "stdafx.h"
vector<boost::math::gamma_distribution<>> GetGamma(const int partls[4], const int y)
{
```

```cpp
        // Declare local variables.
        //==========================
        vector<boost::math::gamma_distribution<>> G;

        // Build vector of random variables.
        //==================================
        if (partls[3]==y)
        {
                G.push_back(boost::math::gamma_distribution<> (2.50, 1.00));
                G.push_back(boost::math::gamma_distribution<> (2.00, 1.00));
        }
        if (partls[0]==y || partls[1]==y || partls[2]==y || partls[3]==y)
        {
                G.push_back(boost::math::gamma_distribution<> (1.50, 1.00));
                G.push_back(boost::math::gamma_distribution<> (1.00, 1.00));
        }
        return G;
}
```

### Code File: GetPNR.cpp

```
/*
/ Copyright (c) 2015 Chris Rook
/
/ Licensed under the Apache License, Version 2.0 (the "License");  you may not use this file except in compliance with the License.  You may obtain a copy of
/ the License at http://www.apache.org/licenses/LICENSE-2.0.  Unless required by applicable law or agreed to in writing, software distributed under the License
/ is distributed on an "AS IS" BASIS, WITHOUT WARRANTIES OR CONDITIONS OF ANY KIND, either express or implied.  See the License for the specific language
/ governing permissions and limitations under the License.
/
/ Filename:  GetPNR.cpp
/
/ Function:  GetPNR()
/
/ Summary:
/
/       This function computes a single probability of avoiding ruin by passing needed parameters to either the simulation specific or dynamic programming
/       specific functions that perform and return the actual calculation.  The estimation type (simulation or dynamic programming) is specified by the user
/       in the control file.
/
/ Parameters:
/
/       1.) Estimation type:  Either dynamic programming (dp) or simulation (sim).
/       2.) Six member long double array of parameter settings for stock mean, stock variance, bond mean, bond variance, stock-bond covariance, expense ratio.
/       3.) Array of long doubles holding the glide-path that is the basis for the probability being requested.
/       4.) Number of time-points for this application.  This represents a fixed number of years to consider for the retirement horizon (i.e., 30).
/       5.) The fixed inflation-adjusted withdrawal rate that the retiree is using during decumulation (i.e., 0.04 for a 4% withdrawal rate).
/       6.) The sample size to use when simulating the ruin/success probability.  (Note that the simulated ruin probability is the number of times ruin occurs
/           divided by the number of trial runs.  The success probability is 1 minus the ruin probability.)  (This parameter is only used when type="sim".)
/       7.) The number of buckets to use when discretizing the ruin factor dimension.  This value equals RFMax*(Discretization Precision), where these 2 values
/           are set by the user in the control file.  (This parameter is only used when type="dp".)
/       8.) The number of buckets to use when discretizing each ruin factor unit value.  For example, if this value=1000 then each ruin factor unit is
/           represented by 1000 buckets.  This value is specified by the user in the control file when type="dp".
/       9.) A 4-element indicator array that specifies which densities should represent inflation/expense-adjusted returns at the given time-point. The
/           settings for this array are:
/               a.) prtls[] = {-1,-1,-1,-1} for all standard real/expense adjusted return densities.
/               b.) prtls[] = {i,-1,-1,-1} for gradient element "i" which uses gradient density g() for the real/expense-adjusted return at the i-th time-point.
/               c.) prtls[] = {i,j,-1,-1} for the off-diagonal Hessian elements which uses the gradient density g() at both time-points "i" and "j".
/               d.) prtls[] = {-1,-1,i,-1} for the diagonal Hessian elements which uses h1() as the real/expense-adjusted return at time-point "i".
/               e.) prtls[] = {-1,-1,-1,i} for the diagonal Hessian elements which uses h2() as the real/expense-adjusted return at time-point "i".
/      10.) The number of independent processing units on the computer running the application or alternatively the number of parallel processes to use during
/           execution as specified by the user.  (If user specified it is set as a parameter when invoking the application.)
```

```
/
/ Return Value:
/
/       This function returns the computed probability.
/----------------------------------------------------------------------------------------------------------------------*/
#include "stdafx.h"
long double GetPNR(const string type, const long double prms[6], long double * a, const int fxTD, const double RF0, const long long int n,
                  const long int nbuckets, const long prec, const int partls[4], const int plproc)
{
        long double retvar;
        if (type=="sim")
                retvar = ThrdPNRsim(prms, a, fxTD, RF0, n, partls, plproc);
        else if (type=="dp")
                retvar = ThrdPNRdyn(prms, a, fxTD, RF0, nbuckets, partls, plproc, prec);
        return retvar;
}
```

## Code File: PNRdyn.cpp

```
/*
/ Copyright (c) 2015 Chris Rook
/
/ Licensed under the Apache License, Version 2.0 (the "License");  you may not use this file except in compliance with the License.  You may obtain a copy of
/ the License at http://www.apache.org/licenses/LICENSE-2.0.  Unless required by applicable law or agreed to in writing, software distributed under the License
/ is distributed on an "AS IS" BASIS, WITHOUT WARRANTIES OR CONDITIONS OF ANY KIND, either express or implied.  See the License for the specific language
/ governing permissions and limitations under the License.
/
/ Filename:  PNRdyn.cpp
/
/ Function:  PNRdyn()
/
/ Summary:
/
/       This function computes the probability of ruin between the two RF(t) buckets provided using a dynamic program.  The function accepts an array of
/       probabilities from the prior time-point along with a vector of bucket numbers that point to the buckets that represent unique probabilities (the last
/       bucket in the sequence).  The probabilities computed are entered into the array passed to this function as a pointer.  If this array already holds
/       values they will be replaced by the new computations, it is not required that the array be empty.  The array of prior probabilities is named Vp[] and
/       the array of current time-point probabilities is V[].  To compute probabilities using a dynamic program only the CDF of the random variables
/       representing inflation/expense-adjusted returns is needed.  When computing a standard ruin probability the inflation/expense-adjusted returns are
/       assumed normally distributed for this implementation with means/variances/covariances of the corresponding stock/bond returns specified in the control
/       file.  When computing gradient entries the special density g() with CDF G() is used to represent the inflation/expense-adjusted return at the current
/       time-point.  When computing diagonal Hessian entries the densities h1() and h2() with CDFs H1() and H2() are used to represent inflation/expense-
/       adjusted returns.  Off-diagonal Hessian entries use the gradient special density g() with CDF G() at both time-points represented by the diagonal
/       indices.
/
/ Parameters:
/
/       1.) Array of precomputed moments & constants for the given time-point and equity ratio (to save processing time): m, mp, v, vp, vpp, mva, kh1.
/       2.) The number of buckets to use when discretizing each ruin factor unit value.  For example, if this value=1000 then each ruin factor unit is
/           represented by 1000 buckets.  This value is specified by the user in the control file when type="dp".
/       3.) A 3-value array holding the start ruin factor bucket, end ruin factor bucket, and total # buckets in the discretization.  The probability of ruin
/           is computed and entered into the incoming array for all buckets between the start and end buckets (inclusive).  Allowing for bucket-specific calls
/           enables multi-threading when this function is invoked from a wrapper.
/       4.) An array (as pointer) of long doubles holding the probabilities of ruin for each bucket at the prior time-point.  The size of this array is the
/           total number of buckets in the discretization (i.e., RFMax*Prec).
/       5.) An array (as pointer) of long doubles that will be populated between the buckets given in parameter #3.  The size of this array is the total number
/           of buckets in the discretization (i.e., RFMax*Prec).
/       6.) A vector containing the bucket numbers from the prior time-point where unique ruin probabilities reside.  If there is a string of consecutive
/           buckets with the same ruin probabilities they can be treated as a single larger bucket to save processing time at the current time-point.
/       7.) Vector of constants.  All CDFs in this application can be expressed as a linear combination of known CDF calls.  This is the constant vector for
```

```cpp
/                that linear combination.  This vector along with the vector of gamma random variables in the next parameter are passed to the GetCDF() function.
/        8.) A vector of gamma distributed random variables (class type=boost distribution) needed to build the CDF for a given time-point and equity ratio.
/            The CDF G() of g(), and the CDF H1() of h1() both use 2 gamma RVs.  The CDF H2() of h2() uses 4 gamma RVs.  This vector along with the constant
/            vector from the previous argument are passed to the GetCDF() function.
/
/ Return Value:
/
/                This function returns no value but populates the array supplied in the 5th argument with the corresponding ruin probabilities between the two
/                RF(t) buckets specified in the 3rd argument.
/----------------------------------------------------------------------------------------------------------------------------------------------------------*/
#include "stdafx.h"
void PNRdyn(const long double mts[7], const long prec, const long bkts[3], const long double * Vp, long double * V, const vector<long> & PrB,
            const vector<long double> & C, const vector<boost::math::gamma_distribution<>> & G)
{
        // Declare local variables.
        //==========================
        const double tiethresh=0.5;
        const long nuqbkts=(long) PrB.size();
        int ties=0, cont=1;
        double rf;
        long double cdfval, eprob, rhs_cdf, lhs_cdf, pruin;

        // Define needed distributions.
        //=============================
        boost::math::normal normdist=boost::math::normal(mts[0],sqrt(mts[2]));

        // Iterate over buckets until a probability of 1.00 is encountered.
        //=================================================================
        for (long b=bkts[0]; cont==1 && b<=bkts[1]; ++b)
        {
                rf = (double) b / prec;          // Derive ruin factor from bucket # and precision.
                pruin = 99.00;                   // Set to unrealistic value.

                // Derive P(Ruin) for this bucket, time point, and asset allocation.
                //==================================================================
                cdfval = GetCDF(mts, C, normdist, G, rf);
                if (cdfval >= 1.00)
                        eprob = Vp[bkts[2]-1];
                else
                {
                        rhs_cdf = 1.00;
                        lhs_cdf = GetCDF(mts, C, normdist, G, rf*(1 + (prec/1.5)));
                        eprob = (rhs_cdf - lhs_cdf)*Vp[0];                                  // First bucket, unique processing.
                        rhs_cdf = lhs_cdf;
                        for (long pb=2; pb <= nuqbkts; ++pb)                                // All others but last bucket, standard processing for unique probs only.
                        {
                                lhs_cdf = GetCDF(mts, C, normdist, G, rf*(1.0 + prec/(PrB[pb-1] + 0.5)));
                                eprob = eprob + (rhs_cdf - lhs_cdf)*Vp[PrB[pb-1]-1];
                                rhs_cdf = lhs_cdf;
                        }
                        eprob = (eprob + (rhs_cdf - cdfval)*Vp[bkts[2]-1])/(1.00 - cdfval);   // Last bucket, unique processing, make it conditional.
                }

                // Deal with numerical instability near zero and one.
                //===================================================
                if (ties == 0)
                {
                        pruin = (cdfval + eprob - (cdfval*eprob));
```

```cpp
                        if (pruin > tiethresh)
                        {
                                pruin = (1.00) - (1.00-cdfval)*(1.00-eprob);
                                ties = 1;
                        }
                }
                else
                        pruin = (1.00) - (1.00-cdfval)*(1.00-eprob);

                // Load ruin probability into array.
                //==================================
                V[b-1] = pruin;

                // Pruning:  Once we hit probability of 1.00, set the remaining buckets manually.
                //================================================================================
                if (V[b-1] >= 1.00)
                {
                        for (long bb=b+1; bb<=bkts[1]; ++bb)
                                V[bb-1]=1.00;
                        cont=0;
                }
        }
        return;
}
```

### Code File:  PNRsim.cpp

```
/*
/ Copyright (c) 2015 Chris Rook
/
/ Licensed under the Apache License, Version 2.0 (the "License");  you may not use this file except in compliance with the License.  You may obtain a copy of
/ the License at http://www.apache.org/licenses/LICENSE-2.0.  Unless required by applicable law or agreed to in writing, software distributed under the License
/ is distributed on an "AS IS" BASIS, WITHOUT WARRANTIES OR CONDITIONS OF ANY KIND, either express or implied.  See the License for the specific language
/ governing permissions and limitations under the License.
/
/ Filename:  PNRsim.cpp
/
/ Function:  PNRsim()
/
/ Summary:
/
/       This function derives the probability of no ruin (PNR), i.e. the success probability for a given glide-path using simulation.  We assume that
/       stock/bond real returns are iid normal RVs in this implementation.  This is not a requirement for the model but an assumption that we make.  To
/       derive an element of the gradient we need a success probability where the normal RV is replaced by an RV that follows a different distribution and
/       this is specified by the parameter prtls[4].  When deriving the elements of the Hessian matrix other success probabilities are needed and those also
/       are specified using the array prtls[4].  The array prtls[4] uses indicators to request a specific success probability (more below).  This function
/       accepts parameters that set the stock/bond means, variances, covariance along with the withdrawal rate, sample size, and glide-path.  At each time-
/       point we update the ruin factor RF(t).  Ruin occurs if-and-only-if RF(t) becomes negative.  Note that the ruin probability is not returned by this
/       function but placed into a long double array that is supplied to the function.  This enables multi-threading of the ruin calculation to save processing
/       time, which is done in the wrapper function ThrdPNRsim().
/
/ Parameters:
/
/       1.) Six member long double array of parameter settings for stock mean, stock variance, bond mean, bond variance, stock-bond covariance, expense ratio.
/       2.) Array of long doubles holding the glide-path to use for this ruin calculation (as pointer).  The corresponding array must have exactly TD elements.
/       3.) Number of time-points for this implementation.  This represents a fixed number of years to consider for the retirement horizon (i.e., 30).
/       4.) The fixed inflation-adjusted withdrawal rate that the retiree is using during decumulation (i.e., 0.04 for a 4% withdrawal rate).
/       5.) The sample size to use when simulating the ruin/success probability.  (Note that the ruin probability is the number of times ruin occurs divided
/               by the number of trial runs.  The success probability is 1 minus the ruin probability.)
/       6.) The thread ID.  PNRsim() is invoked from a wrapper function that determines the number of independent processing units on the PC running the
```

```
/              application and splits the ruin derivation into smaller jobs then runs them concurrently in separate threads and combines the results when
/              finished.
/        7.) The array to return the calculated success probability in using the thread ID as the array element index.
/        8.) A 4-element indicator array that specifies exactly which densities to use for the requested ruin calculation.  The settings for this array are:
/              a.) prtls[] = {-1,-1,-1,-1} for all standard real/expense adjusted return densities.
/              b.) prtls[] = {i,-1,-1,-1} for gradient element "i" which uses the gradient density g() for the real/expense-adjusted return at the i-th time-
/                   point.
/              c.) prtls[] = {i,j,-1,-1} for the off-diagonal Hessian elements which uses the gradient density g() at both time-points "i" and "j".
/              d.) prtls[] = {-1,-1,i,-1} for the diagonal Hessian elements which uses h1() as the real/expense-adjusted return at time-point "i".
/              e.) prtls[] = {-1,-1,-1,i} for the diagonal Hessian elements which uses h2() as the real/expense-adjusted return at time-point "i".
/
/ Return Value:
/
/       This function returns the success probability in the array supplied as probs[tnum-1].
/----------------------------------------------------------------------------------------------------------------------------------------------------------*/
#include "stdafx.h"
void PNRsim(const long double prms[6], const long double * a, const int fxTD, const double RF0, const long long int n, const int tnum, long double * probs,
            const int partls[4])
{
        // Declare local variables.
        //=========================
        long double *mn=NULL, *std=NULL, RF, rtrn, unx, uny, xlow, xhigh, yhigh;
        bool ruin;
        long long int cntr=0;
        std::random_device rd;
        std::default_random_engine gen(rd());
        std::normal_distribution<long double> * rrtrn=NULL;   // Array
        std::uniform_real_distribution<long double> urtrn1;   // Scalar
        std::uniform_real_distribution<long double> urtrn2;   // Scalar

        // Get means/standard deviations for each timepoint and build the normal RV generator.
        //====================================================================================
        mn = new long double [fxTD];
        std = new long double [fxTD];
        rrtrn = new std::normal_distribution<long double> [fxTD];

        // Check that no value of the partls[] array is equal to or greater than fxTD.
        //============================================================================
        for (int i=0; i<4 && partls[i]>=fxTD; ++i)
        {
                cout.setf(ios_base::fixed, ios_base::floatfield);
                cout << "ERROR: Invalid (>=TD) partial derivative specification partls[" << i << "]=" << partls[i] << "." << endl;
                cout << "EXITING...PNRsim()..." << endl; cin.get();
                exit (EXIT_FAILURE);
        }

        // Iterate over valid time points and set parameters needed during simulation.
        //============================================================================
        for (int y=0; y<fxTD; ++y)
        {
                // Check that only one Hessian special density is specified.
                // Set the lower/upper range and density bounds for all special densities.
                //=======================================================================
                if (partls[2] != -1 && partls[3] != -1)
                {
                        cout.setf(ios_base::fixed, ios_base::floatfield); cout.precision(10);
                        cout << "ERROR:  Only 1 Hessian special density should be used per simulation." << endl;
                        cout << "EXITING...PNRsim()..." << endl; cin.get();
```

```cpp
                exit (EXIT_FAILURE);
        }
        else if (partls[0] != -1 || partls[1] != -1)        // Gradient density g() ranges.  (Confirmed for our distributional assumptions.)
        {
                xlow = -0.15; xhigh = 2.20; yhigh = 6.20;
        }
        else if (partls[2] != -1)                           // Hessian density h1() ranges.  (Confirmed for our distributional assumptions.)
        {
                xlow = -0.10; xhigh = 2.30; yhigh = 5.70;
        }
        else if (partls[3] != -1 )                          // Hessian density h2() ranges.  (Confirmed for our distributional assumptions.)
        {
                xlow = -0.15; xhigh = 2.40; yhigh = 5.85;
        }

        // Build arrays for means/variances based on the incoming glidepath.
        //======================================================================
        mn[y] = m(prms, a[y]);
        std[y] = sqrt(v(prms, a[y]));
        rrtrn[y] = std::normal_distribution<long double> (mn[y], std[y]);
}

// Estimate the probability.
//===========================
if (partls[0]==-1 && partls[1]==-1 && partls[2]==-1 && partls[3]==-1)    // Processing for no special densities.
{
        for (long long int i=1; i<=n; ++i)
        {
                RF = (long double) RF0;
                ruin = 0;
                for (int y=0; y<fxTD && ruin==0; ++y)
                {
                        rtrn = rrtrn[y](gen);                           // All returns from real/expense adjusted PDF.
                        if (RF > 0 && rtrn > RF)
                                RF = RF/(rtrn - RF);
                        else
                                ruin = 1;
                }
                cntr = cntr + (long long int) ruin;
        }
}
else                                                                    // Processing for special densities.
{
        for (long long int i=1; i<=n; ++i)
        {
                RF = (long double) RF0;
                ruin = 0;
                for (int y=0; y<fxTD && ruin==0; ++y)
                {
                        if (partls[0] != y && partls[1] != y && partls[2] != y && partls[3] != y)
                                rtrn = rrtrn[y](gen);                                           // Regular density at this time point.
                        else
                        {
                                rtrn = -999;
                                do
                                {
                                        unx = (xhigh - xlow)*urtrn1(gen) + xlow;                // Set domain for special density.
                                        uny = (yhigh)*urtrn2(gen);                              // Set maximum value for special density.
```

```cpp
                                        if (((partls[0]==y || partls[1]==y) && uny <= g(prms,a[y],unx))      // Generate RV from special density.
                                         || (partls[2]==y && uny <= h1(prms,a[y],unx))
                                         || (partls[3]==y && uny <= h2(prms,a[y],unx)))
                                            rtrn = unx;
                                    } while (rtrn == -999);
                                }
                                if (RF > 0 && rtrn > RF)
                                    RF = RF/(rtrn - RF);
                                else
                                    ruin = 1;
                            }
                            cntr = cntr + (long long int) ruin;
                        }
                    }

                    // Get the probability of no ruin and add it to the incoming array.
                    //============================================================
                    probs[tnum-1] = 1.00 - (long double) cntr/n;

                    // Free temporary memory.
                    //========================
                    delete [] mn;     mn=nullptr;
                    delete [] std;    std=nullptr;
                    delete [] rrtrn;  rrtrn=nullptr;

                    return;
                }
```

### Code File:  ThrdPNRdyn.cpp

```
/*
/ Copyright (c) 2015 Chris Rook
/
/ Licensed under the Apache License, Version 2.0 (the "License");  you may not use this file except in compliance with the License.  You may obtain a copy of
/ the License at http://www.apache.org/licenses/LICENSE-2.0.  Unless required by applicable law or agreed to in writing, software distributed under the License
/ is distributed on an "AS IS" BASIS, WITHOUT WARRANTIES OR CONDITIONS OF ANY KIND, either express or implied.  See the License for the specific language
/ governing permissions and limitations under the License.
/
/ Filename:  ThrdPNRdyn.cpp
/
/ Function:  ThrdPNRdyn()
/
/ Summary:
/
/       This function serves as a wrapper for the PNRdyn() function and it splits the job of computing a ruin probability with a dynamic program across the
/       various independent processing units on the computer running the application.  This function is passed the number of independent processing units on
/       the computer and it divides the DP discretization by this number and launches simultaneous calls to PNRdyn() to process specific sets of buckets
/       concurrently populating a pre-defined array at these bucket indices.  When invoking this function expect a single success probability to be returned.
/
/ Parameters:
/
/       1.) Six member long double array of parameter settings for stock mean, stock variance, bond mean, bond variance, stock-bond covariance, expense ratio.
/       2.) Array of long doubles holding the glide-path to use for this ruin calculation (as pointer).  The corresponding array must have exactly TD elements.
/          (Note that we only consider alphas between the low-volatility portfolio and 1.00.  If any alpha provided in this array is not between the low-
/           volatility portfolio and 1.00 it is forced to the nearest acceptable value.)
/       3.) Number of time-points for this implementation.  This represents a fixed number of years to consider for the retirement horizon (i.e., 30).
/       4.) The fixed inflation-adjusted withdrawal rate that the retiree is using during decumulation (i.e., 0.04 for a 4% withdrawal rate).
/       5.) The number of buckets to use when discretizing the ruin factor dimension.  This value equals RFMax*(Discretization Precision), where these 2 values
/           are set by the user in the control file.
/       6.) A 4-element indicator array that specifies which densities should represent inflation/expense-adjusted returns at the given time-point.
```

```
/                The settings for this array are:
/                    a.) prtls[] = {-1,-1,-1,-1} for all standard real/expense adjusted return densities.
/                    b.) prtls[] = {i,-1,-1,-1} for gradient element "i" which uses gradient density g() for the real/expense-adjusted return at the i-th
/                        time-point.
/                    c.) prtls[] = {i,j,-1,-1} for the off-diagonal Hessian elements which uses the gradient density g() at both time-points "i" and "j".
/                    d.) prtls[] = {-1,-1,i,-1} for the diagonal Hessian elements which uses h1() as the real/expense-adjusted return at time-point "i".
/                    e.) prtls[] = {-1,-1,-1,i} for the diagonal Hessian elements which uses h2() as the real/expense-adjusted return at time-point "i".
/         7.) The number of independent processing units on the computer running the application or alternatively the number of parallel processes to use during
/             execution as specified by the user.  (If user specified it is set as a parameter when invoking the application.)
/         8.) The number of buckets to use when discretizing each ruin factor unit value.  For example, if this value=1000 then each ruin factor unit is
/             represented by 1000 buckets.  This value is specified by the user in the control file when type="dp".
/
/ Return Value:
/
/         This function returns the success probability derived using a DP.
/-------------------------------------------------------------------------------------------------------------------------------------------*/
#include "stdafx.h"
long double ThrdPNRdyn(const long double prms[6], long double * a, const int fxTD, const double RF0, const long int nbuckets, const int partls[4],
                       const int plproc, const long prec)
{
        // Declare local variables.
        //===========================
        long double *Vp=new long double[nbuckets], *V=new long double[nbuckets], prevprob=0.00, mts[7];
        long int bktsprun, prnbkt=nbuckets, **bktarys=NULL;
        int cont, pwr;
        vector<long> uBkts; uBkts.push_back(1); uBkts.push_back(nbuckets);
        vector<long double> C;
        vector<boost::math::gamma_distribution<>> G;
        boost::thread *t=NULL;

        // Initiate prior timepoint's probabilities with zeros.
        //=====================================================
        for (long b=1; b<=nbuckets; ++b)
                Vp[b-1] = 0.00;

        // Check that no value of the partls[] array is equal to or greater than fxTD.
        //============================================================================
        for (int i=0; i<4 && partls[i]>=fxTD; ++i)
        {
                cout.setf(ios_base::fixed, ios_base::floatfield);
                cout << "ERROR: Invalid (>=TD) partial derivative specification partls[" << i << "]=" << partls[i] << "." << endl;
                cout << "EXITING...PNRsim()..." << endl; cin.get();
                exit (EXIT_FAILURE);
        }

        // Check that all equity ratios are between MVA and 1.00.
        //=======================================================
        for (int y=0; y<fxTD; ++y)
        {
                if (a[y] < mva(prms)+0.0001 || a[y] > 1.00)
                {
                        if (a[y] > 1.00)
                                a[y]=1.00;
                        else if (a[y] < mva(prms)+0.0001)
                                a[y]=mva(prms)+0.0001;
                }
        }
```

```cpp
// Iterate over all time points, launching separate threads to process equal sized collections of
// buckets concurrently within each time point.
//=================================================================================================
for (int y=fxTD-1; y>=1; y--)
{
        // Populate moments array for this time point.
        //============================================
        mts[0]=m(prms,a[y]); mts[1]=mp(prms);
        mts[2]=v(prms,a[y]); mts[3]=vp(prms,a[y]); mts[4]=vpp(prms);
        mts[5]=mva(prms); mts[6]=kh1(prms,a[y]);

        // Process collections of buckets concurrently.
        //=============================================
        t = new boost::thread[plproc];

        // Define needed constants.
        //=========================
        C=GetConst(mts, partls, y);

        // Define needed distributions.
        //=============================
        G=GetGamma(partls, y);

        // Update prnbkt and derive the new # of buckets to process per run for the next iteration.
        // (Buckets with trivial assignments are handled separately.)
        //=========================================================================================
        cont=1; pwr=1;
        prnbkt=nbuckets;
        if (y < (fxTD - (int) (fxTD/6)))
        {
                prnbkt=1;
                pwr=2;
        }

        for (long b=prnbkt + (long) pow(-1.00, pwr)*(2*plproc); (cont==1 && b>=1 && b<=nbuckets); b=b + (long) pow(-1.00, pwr)*(2*plproc))
        {
                long int prnbkts[3] = {b, b, nbuckets};
                PNRdyn(mts, prec, prnbkts, Vp, V, uBkts, C, G);

                if (pwr==1 && V[b-1] >= 1.00)
                        prnbkt=b;
                else if (pwr==2 && V[b-1] < 1.00)
                        prnbkt=min(b + (long) pow(-1.00, pwr)*(2*plproc),nbuckets);
                else
                        cont=0;
        }

        // The value of prnbkt should be ge plproc and le nbuckets.
        //=========================================================
        if (prnbkt > nbuckets)
                prnbkt = nbuckets;
        else if (prnbkt < plproc)
                prnbkt = plproc;

        bktsprun = (long int) (prnbkt/plproc + 1);

        if (bktsprun*plproc > nbuckets)
                bktsprun = (long int) (nbuckets/plproc);
```

```cpp
        // If # buckets/run is still not right, exit with an error and fix.
        //========================================================================
        if (bktsprun*plproc > nbuckets || bktsprun < 1)
        {
                cout.setf(ios_base::fixed, ios_base::floatfield); cout.precision(50);
                cout << endl << "The # of buckets per run is not being derived correctly, must fix:" << endl;
                cout << "# of buckets per run = " << bktsprun << endl;
                cout << "Total # of buckets = " << nbuckets << endl;
                cout << "# of concurrent processes being used = " << plproc << endl;
                cout << "EXITING...ThrdPNRdyn()..." << endl; cin.get();
                exit(EXIT_FAILURE);
        }

        // An array of size 3 is used to specify the start/end/total buckets for each thread.
        //==================================================================================
        bktarys = new long * [plproc];
        for (int i=1; i<=plproc; ++i)
        {
                bktarys[i-1] = new long[3];
                bktarys[i-1][0]=bktsprun*(i-1) + 1;
                bktarys[i-1][1]=min(bktsprun*(i),nbuckets);
                bktarys[i-1][2]=nbuckets;
                t[i-1] = boost::thread(PNRdyn, boost::cref(mts), prec, boost::cref(bktarys[i-1]), boost::cref(Vp), boost::ref(V), boost::cref(uBkts),
                        boost::cref(C), boost::cref(G));
        }

        // Assign trivial (known) values.
        //===============================
        if (bktarys[plproc-1][1] < nbuckets)
                for (long int b=bktsprun*(plproc)+1; b<=nbuckets; ++b)
                        V[b-1]=1.00;

        // Wait for all threads to finish, then proceed.
        //==============================================
        for (int i=0; i<plproc; ++i)
                t[i].join();

        // Free temporary memory allocations and reused containers.
        //=========================================================
        for (int i=0; i<plproc; ++i)
        {
                delete [] bktarys[i]; bktarys[i]=nullptr;
        }
        delete [] bktarys; bktarys=nullptr;
        delete [] t; t=nullptr;
        uBkts.clear(); uBkts.shrink_to_fit();
        C.clear(); C.shrink_to_fit();
        G.clear(); C.shrink_to_fit();

        // Update quantities needed for next iteration.
        // Reset the prior year's probabilities in Vp[] and derive the unique bucket quantities.
        //======================================================================================
        cont=1; prevprob=0.00;
        for (long b=1; b<=nbuckets; ++b)
        {
                Vp[b-1] = V[b-1];
```

```cpp
                        // Verify the integrity of the probabilities derived during this iteration.
                        //===========================================================================
                        if ( (Vp[b-1] < (prevprob - pow(0.1,15))) || (Vp[b-1] > 1.00 + 2.00*pow(0.1,16)))
                        {
                                cout.setf(ios_base::fixed, ios_base::floatfield); cout.precision(50);
                                cout << endl << "There is an issue with the probabilities derived at this timepoint (t=" << y << "), see below:" << endl;
                                if (b > 1)
                                        cout << "Vp[" << (b-2) << "] = " << Vp[b-2] << endl;
                                cout << "Vp[" << (b-1) << "] = " << Vp[b-1] << endl;
                                cout << "EXITING...ThrdPNRdyn()..." << endl; cin.get();
                                exit(EXIT_FAILURE);
                        }
                        prevprob = Vp[b-1];

                        // Derive the new vector of bucket #'s with unique probabilities at the prior timepoint.
                        //=====================================================================================
                        if ((b == 1) || (b != nbuckets && V[b-1] != V[b] && cont==1) || (b == nbuckets))
                                uBkts.push_back(b);

                        // Once PRuin=1 stop collecting unique buckets.
                        //=============================================
                        if (cont==1 && Vp[b-1] >= 1.00)
                                cont=0;
                }

                // Check that the last bucket at this time point has a PRuin of 1.00.
                // (Otherwise RFMax needs to be increased.  This must hold when using special densities also.)
                //===========================================================================================
                if (Vp[nbuckets-1] < 1.00)
                {
                        cout.setf(ios_base::fixed, ios_base::floatfield); cout.precision(50);
                        cout << "Timepoint (t=" << y << "), has V[" << nbuckets-1 << "]=" << V[nbuckets-1] << ", which is < 1.00." << endl;
                        cout << "(Increase RFMax.)" << endl;
                        cout << "EXITING...ThrdPNRdyn()..." << endl; cin.get();
                        exit(EXIT_FAILURE);
                }
        }

        // Process final timepoint only for the given RF0.
        //================================================
        mts[0]=m(prms,a[0]); mts[1]=mp(prms);
        mts[2]=v(prms,a[0]); mts[3]=vp(prms,a[0]); mts[4]=vpp(prms);
        mts[5]=mva(prms); mts[6]=kh1(prms,a[0]);
        long int RF0bkt = (long int) (RF0*prec + 0.5);
        C=GetConst(mts, partls, 0);
        G=GetGamma(partls, 0);
        long int fnlbkts[3] = {RF0bkt, RF0bkt, nbuckets};
        PNRdyn(mts, prec, fnlbkts, Vp, V, uBkts, C, G);

        // Retrieve the probability to return.
        //====================================
        long double rtprob=1.00 - V[RF0bkt-1];

        // Delete temporary memory allocations.
        //=====================================
        delete [] Vp; Vp=nullptr;
        delete [] V; V=nullptr;
```

```
        // The single PNR derived using a DP is returned.
        //================================================
        return rtprob;
}
```

### Code File: ThrdPNRsim.cpp

```
/*
/ Copyright (c) 2015 Chris Rook
/
/ Licensed under the Apache License, Version 2.0 (the "License");  you may not use this file except in compliance with the License.  You may obtain a copy of
/ the License at http://www.apache.org/licenses/LICENSE-2.0.  Unless required by applicable law or agreed to in writing, software distributed under the License
/ is distributed on an "AS IS" BASIS, WITHOUT WARRANTIES OR CONDITIONS OF ANY KIND, either express or implied.  See the License for the specific language
/ governing permissions and limitations under the License.
/
/ Filename:  ThrdPNRsim.cpp
/
/ Function:  ThrdPNRsim()
/
/ Summary:
/
/       This function serves as a wrapper for the PNRsim() function and it splits the job of simulating a ruin probability across the various independent
/       processing units on the computer running the application.  This function is passed the number of independent processing units on the computer and it
/       divides the simulation sample size by this number and launches simultaneous calls to PNRsim() then averages the resulting probability across calls for
/       a final value that is returned to the calling program.  When invoking this function expect a single success probability to be returned.
/
/ Parameters:
/
/       1.) Six member long double array of parameter settings for stock mean, stock variance, bond mean, bond variance, stock-bond covariance, expense ratio.
/       2.) Array of long doubles holding the glide-path to use for this ruin calculation (as pointer).  The corresponding array must have exactly TD elements.
/           (Note that we only consider alphas between the low-volatility portfolio and 1.00.  If any alpha provided in this array is not between the low-
/            volatility portfolio and 1.00 it is forced to the nearest acceptable value.)
/       3.) Number of time-points for this implementation.  This represents a fixed number of years to consider for the retirement horizon (i.e., 30).
/       4.) The fixed inflation-adjusted withdrawal rate that the retiree is using during decumulation (i.e., 0.04 for a 4% withdrawal rate).
/       5.) The sample size to use when simulating the ruin/success probability.  (Note that the ruin probability is the number of times ruin occurs divided
/           by the number of trial runs.  The success probability is 1 minus the ruin probability.)
/       6.) A 4-element indicator array that specifies exactly which densities to use for the requested ruin calculation.  The settings for this array are:
/           a.) prtls[] = {-1,-1,-1,-1} for all standard real/expense adjusted return densities.
/           b.) prtls[] = {i,-1,-1,-1} for gradient element "i" which uses the gradient density g() for the real/expense-adjusted rtrn at the ith time point.
/           c.) prtls[] = {i,j,-1,-1} for the off-diagonal Hessian elements which uses the gradient density g() at both time-points "i" and "j".
/           d.) prtls[] = {-1,-1,i,-1} for the diagonal Hessian elements which uses h1() as the real/expense-adjusted return at time-point "i".
/           e.) prtls[] = {-1,-1,-1,i} for the diagonal Hessian elements which uses h2() as the real/expense-adjusted return at time-point "i".
/       7.) The number of independent processing units on the computer running the application or alternatively the number of parallel processes to use during
/           execution as specified by the user.  (If user specified it is set as a parameter when invoking the application.)
/
/ Return Value:
/
/       This function returns the simulated success probability.
/----------------------------------------------------------------------------------------------------------------------------------------------------------*/
#include "stdafx.h"
long double ThrdPNRsim(const long double prms[6], long double * a, const int fxTD, const double RF0, const long long int n, const int partls[4],
                       const int plproc)
{
        // Define local variables.
        //=========================
        long double *probvals=NULL, probnr=0;
        boost::thread * t=NULL;

        // Check that all equity allocations are between MVA and 1.00.
        //============================================================
        for (int y=0; y<fxTD; ++y)
        {
```

```cpp
            if (a[y] < mva(prms)+0.0001 || a[y] > 1.00)
            {
                    if (a[y] > 1.00)
                            a[y]=1.00;
                    else if (a[y] < mva(prms)+0.0001)
                            a[y]=mva(prms)+0.0001;
            }
    }

    // Split call to PNR() by threads to speed up processing.
    //======================================================
    probvals = new long double[plproc];
    t = new boost::thread[plproc];

    // Separate processing when using simulation vs DP to determine PNR.
    //=================================================================
    for (int i=1; i<=plproc; ++i)
            t[i-1] = boost::thread(PNRsim, boost::cref(prms), boost::cref(a), fxTD, RF0, (long long int) n/plproc, i, boost::ref(probvals),
                                   boost::cref(partls));

    // Wait for all threads to finish, then proceed.  Average probabilities across threads.
    //====================================================================================
    for (int i=0; i<plproc; ++i)
    {
            t[i].join();
            probnr = probnr + probvals[i]/plproc;
    }

    // Delete temporary memory allocations.
    //=====================================
    delete [] t;   t = nullptr;
    delete [] probvals;   probvals = nullptr;

    // The single simulated PNR is returned.
    //======================================
    return probnr;
}
```

### Code File:  WrtAry.cpp

```
/*
/ Copyright (c) 2015 Chris Rook
/
/ Licensed under the Apache License, Version 2.0 (the "License");  you may not use this file except in compliance with the License.  You may obtain a copy of
/ the License at http://www.apache.org/licenses/LICENSE-2.0.  Unless required by applicable law or agreed to in writing, software distributed under the License
/ is distributed on an "AS IS" BASIS, WITHOUT WARRANTIES OR CONDITIONS OF ANY KIND, either express or implied.  See the License for the specific language
/ governing permissions and limitations under the License.
/
/ Filename:  WrtAry.cpp
/
/ Function:  WrtAry()
/
/ Summary:
/
/       This function will write the contents of an array as a block to either standard output or to standard output and a file specified by the user.
/
/ Parameters:
/
/       1.) The success probability for the glide-path being printed, if it applies.  (Set to something not between 0.00 and 1.00 when printing gradient.)
/       2.) The array being printed to standard output or a file specified as a pointer.  (Will be either the glide-path or the gradient.)
/       3.) The text to use for labeling each element of the array.  (For example, "GP" for GP[] or "Grd" for Grd[].)
/       4.) The number of elements in the array to be printed.  (This will be the # of time-points for the arrangement being analyzed.)
/       5.) An optional filename.  If specified the array is printed the file specified, which is placed in the directory specified by the user.
```

```cpp
/
/ Return Value:
/
/       This function has no return value.
/----------------------------------------------------------------------------------------------------------------------------*/
#include "stdafx.h"
void WrtAry(const long double stdpnr, const long double * a, const string lbl, const int fxTD, const string filenm)
{
        // Declare and initialize local variables.
        //==============================================
        int nrows, maxcols=5;
        if (fxTD % maxcols == 0)
                nrows=fxTD/maxcols;
        else
                nrows=(fxTD/maxcols)+1;

        // Write glide-path to output window by default.
        //==============================================
        cout.setf(ios_base::fixed, ios_base::floatfield); cout << endl;
        if (stdpnr >= -0.000001 && stdpnr <= 1.000001)
        {
                cout.precision(20);
                cout << noshowpos << "--> Success probability for this Glide-Path=" << stdpnr << endl;
        }
        cout.precision(10);
        for (int r=0; r<nrows; ++r)
        {
                for (int c=0; c<maxcols; ++c)
                        if (r + nrows*c < fxTD)
                                cout << lbl + "[" << setfill('0') << setw(2) << noshowpos << r + nrows*c << "]=" << right << setw(10) << showpos
                                        << a[r + nrows*c] << "  ";
                cout << endl << noshowpos;
        }

        // If a filename is given, write to file.
        //========================================
        if (filenm != " ")
        {
                ofstream fout(filenm, ios::out);
                fout.setf(ios_base::fixed, ios_base::floatfield); fout << endl;
                if (stdpnr >= -0.000001 && stdpnr <= 1.000001)
                {
                        fout.precision(12);
                        fout << noshowpos << "--> Success probability for this Glide-Path = " << stdpnr << endl;
                }
                fout.precision(10);
                for (int r=0; r<nrows; ++r)
                {
                        for (int c=0; c<maxcols; ++c)
                                if (r + nrows*c < fxTD)
                                        fout << lbl + "[" << setfill('0') << setw(2) << noshowpos << r + nrows*c << "]=" << right << setw(10) << showpos
                                                << a[r + nrows*c] << "  ";
                        fout << endl << noshowpos;
                }
                fout.close();
        }
        return;
}
```